\documentclass[12pt]{iopart}
\usepackage{amsfonts}   
\usepackage{xcolor}
\usepackage{amssymb, graphicx}

\begin{document}

\title[perturbation theory]{A perturbation theory for the Anderson model}

\author{J. Kern}

\address{Formerly: Institut f{\"u}r Theoretische Physik, Universit{\"a}t Regensburg, 
93040 Regensburg, Germany}
\ead{johannes.kern@physik.uni-regensburg.de}
\begin{abstract}
Within the diagrammatic real time approach \cite{König96, Schoeller97}, the current 
across a quantum dot which is tunnel coupled to two leads at different chemical potentials 
is calculated by the use of two objects referred to as kernels. The stationary reduced 
density matrix of the quantum dot is determined by the use of the {\em density matrix kernel}, 
while the {\em current kernel} is used in a second step to determine the stationary current 
across the dot. 
If the tunneling Hamiltonian is multiplied by a coupling parameter ``$w$``, then everything, 
including the kernels, the stationary density matrix as well as the stationary current,  
can be viewed as a function of $w$. In the time space, and at every single and fixed time $t$, 
the kernels have the clear structure of a convergent power series in $w$.
Refer to the coefficients of these power series as the {\em orders} of the kernels. 
It is intuitive to truncate the kernels at some finite order and to perform remaining 
calculations by the use of the corresponding approximate kernels. However, the quantities 
which actually appear in the calculations are not the kernels as a function of time but rather 
their Laplace transforms, and here only in the limit $\lambda \to 0$. The statement that 
even in this limit the structure of a convergent power series in the coupling parameter
is still conserved is shown in the text. The statement that the  
stationary density matrix and current are still analytic in the coupling parameter is shown, 
assuming the quantum dot is the single impurity Anderson model (SIAM). 
Finally, results for the kernels up to sixth order, neglecting 
the doubly occupied state and assuming equal energies $E_\uparrow= E_\downarrow$, 
are presented and discussed. In case the degenerate level lies below the Fermi level, a zero
bias resonance, getting more and more pronounced with decreasing temperature, is expected.                   
\end{abstract}


\maketitle

\section{Introduction}
The SIAM quantum dot \cite{Bruus} is a quantum dot with only four possible states: 
It can be empty (state $0$), occupied by an electron with spin $\sigma$ (state $\sigma$) 
or be in the doubly occupied state $2$, i.e., filled with two electrons of opposite spin. 
If the quantum dot is tunnel coupled to contacts at different chemical 
potentials, then a current can flow between them. Within the diagrammatic real time 
approach \cite{König96, Schoeller97} it is given by the equation 
 \begin{equation} \label{curent kernel}
 I = Tr \left\lbrace K_c (\lambda = 0) \rho \right\rbrace,
\end{equation}
where $K_c$ is the current kernel and $\rho $ is the 
stationary reduced density matrix \cite{Blum} of the dot. The latter contains information 
about the stationary probabilities of finding the dot in the possible states. 
The current kernel is originally obtained as a function of time. The object 
of Eq. (\ref{curent kernel}) is the Laplace transform of this function of time. 
The map  
\begin{equation} \label{transformation}
 (...)(t) \mapsto \int_0^\infty dt (...)(t)  e^{-\lambda t}  
\end{equation}
is applied to $K_c (t)$, where $\lambda$ is a positive number, and the 
limit $\lambda \to 0$ is taken.

The stationary reduced density matrix is determined by solving the quantum master 
equation in the stationary limit:
\begin{equation} \label{quantum master equation}
 0 = \frac{i}{\hbar} \left[ \rho, H_\odot \right]  + K(\lambda = 0) 
\rho.
\end{equation}
The operator $H_\odot$ is the Hamiltonian of the isolated dot. It is 
diagonal in the introduced states of the quantum dot: $H_\odot (a) = E_a  a$, 
where $E_a$ is the eigenenergy of the state $a = 0, \sigma, 2$. The object 
$K(\lambda = 0)$ is the density matrix kernel; its structure is analogous  
to that of the current kernel and, at first, it is obtained as a function of the time; 
application of the Laplace transform to $K(t)$ and the limit $\lambda \to 0$ 
yield $K(\lambda = 0)$.

A possible approach to the problem of determining the current is 
perturbation theory: Introduce a coupling parameter which expresses the 
strength of the tunneling coupling and consider the kernels and finally the 
current $I(w)$ as a function of this parameter. Calculate the 
Taylor series of $I(w)$ in $w = 0$ up to an order $z$ as high as possible. 
The resulting polynomial $I^{(z)}(w)$ of degree $z$ can be expected to be a  
good approximation for $I(w)$ in the case of small values of $w$. 
Since the current is calculated via the kernels, it is natural to try to obtain the 
Taylor series of the current by calculation of the 
Taylor expansions of the kernels. Indeed, the kernels are analytic in the 
coupling parameter $w$ around $w=0$, and the Taylor expansion of the current is 
obtained from the corresponding expansions of the kernels.


The same basic theory has been applied in Ref.
\cite{Kern epjb} in a non-perturbative way. All contributions to the kernels
can be represented by diagrams. Although diagrams of all orders have been calculated, 
the summation remained incomplete,
since only diagrams within a selection called the {\em dressed second order} (DSO)
were taken into account. The diagram selection has also been discussed in Ref. 
\cite{Koller_Diss, Koller12}. Moreover, and much earlier, e.g. in Ref. \cite{König96},
another diagram selection called the {\em resonant tunneling approximation} (RTA) 
has been used. All DSO diagrams are contained in the RTA selection.

\section{Hamiltonian}
The Hamilton operator of the SIAM quantum dot can be written as 
\begin{equation} \label{dot Hamiltonian}
 H_\odot = U d_\uparrow^\dagger d_\uparrow  d_\downarrow^\dagger d_\downarrow  + 
 \sum_\sigma 
   \epsilon_\sigma d_\sigma^\dagger d_\sigma,   
\end{equation}
where $d_\sigma^\dagger$ ($d_\sigma$) is the creation (annihilation) operator of the 
one electron level with spin $\sigma$. The operator acts on the four dimensional 
complex vector space spanned by the subsets of the set of the two one electron 
levels, $\left\lbrace \uparrow, \downarrow \right\rbrace$, referred to as $0$, 
$\sigma $, and $2$.
The eigenvalues of $H_\odot$ are $ E_0 = 0, E_\sigma = \epsilon_\sigma,
E_2 = U +  \sum_\sigma  \epsilon_\sigma $.  


The contribution of the contacts to the total Hamiltonian is assumed to be given 
by    
\begin{displaymath}
 H_R = \sum_{l\sigma {\bf k}} \varepsilon_{l \sigma {\bf k}}  
c_{l \sigma {\bf k}}^\dagger c_{l \sigma {\bf k}}.
\end{displaymath}
This is the usual choice of the Hamiltonian of the leads. The electrons in the leads 
are assumed to be noninteracting, Ref. \cite{Ashcroft}.

Finally, there is the tunneling Hamiltonian, which expresses the possibility that 
electrons can tunnel from the leads to the quantum dot or vice versa 
\cite{Gottlieb}. The conventional tunneling Hamiltonian is: 
\begin{equation} \label{tunneling Hamiltonian}
 H_T = \sum_{ l \sigma {\bf k}} T_{l{\bf k}\sigma}  d_{\sigma}^\dagger 
c_{l{\bf k}\sigma} + 
		      \mbox{h. c. (hermitian conjugate)},
\end{equation}
where $c_{l{\bf k}\sigma}^\dagger$ ($c_{l{\bf k}\sigma}$) is the creation- 
(annihilation) operator of the electron level in the lead $l$ with wave vector 
${\bf k} $ and spin $\sigma$. For simplicity, I will assume that the coefficients 
$ T_{l{\bf k}\sigma}$ of the tunneling Hamiltonian are independent of spin, 
$ T_{l{\bf k}\sigma} =  T_{l{\bf k}}$.

\section{Taylor expansion of the kernels}

\subsection{The kernels in the time space}
The kernels are obtained by taking the sum of all diagrams, the diagrammatic 
series is infinite. A possible way of writing down the density matrix kernel in 
the Laplace space is explained in Ref. \cite{Kern10}. It is derived from the 
conventional diagrammatic language (e.g. Ref. \cite{Schoeller97}) by grouping the 
diagrams according to their topology \cite{Koller_Diss}. 
For any linear operator $x$ acting on the vector space spanned by the 
quantum dot states one obtains: 

\begin{eqnarray*}
 \int_0^t K(t') dt'x =  
&&  1/\hbar \sum_{n=1}^\infty \sum_{k=0}^{2n} \quad \sum_{\beta_0, \dots, \beta_k, 
\alpha_k, \dots \alpha_{2n}} 
     \quad \sum_{l_1 \sigma_1, \dots , l_{n} \sigma_{n} } \quad    \\
&& \sum_{(p,q)} \quad \sum_{\mbox{J: consistent with (p,q)}}   \\
&& (-1)^{(n+k)}   sign(p,q) \quad |\beta_0><\alpha_{2n}| \quad  
<\beta_k|x|\alpha_k>    \\
\end{eqnarray*}

\begin{eqnarray*}
&&    \int d {\bf k_1} Z_{l_1} \dots 
\int d {\bf k_n} Z_{l_n}   
\end{eqnarray*}
\begin{eqnarray*}
&&  \prod_{i: p_i < q_i \le k} f_{l_i}^{(v_{p_i})}(\varepsilon_{l_i \sigma_i 
{\bf k_i}})
    < \beta_{p_{i-1}} | D_{l_i \sigma_i { \bf k_i  }}^{(v_{p_i})} |\beta_{p_i} >
    < \beta_{q_{i-1}} | D_{l_i \sigma_i { \bf k_i  }}^{(-v_{p_i})} |\beta_{q_i} >   \\ 	
&&  \prod_{i: k+1 \le p_i < q_i} f_{l_{i}}^{(v_{p_i})}(\varepsilon_{l_i \sigma_i 
{\bf k_i}})
    < \alpha_{p_{i-1}} | D_{l_i \sigma_i { \bf k_i  }}^{(v_{p_i})} |\alpha_{p_i} > 
    < \alpha_{q_{i-1}} | D_{l_i \sigma_i { \bf k_i  }}^{(-v_{p_i})} |\alpha_{q_i} > \\ 	
&&  \prod_{i: p_i \le k, k+1 \le q_i} f_{l_{i}}^{(-v_{p_i})}(\varepsilon_{l_i 
\sigma_i {\bf k_i}})
    < \beta_{p_{i-1}} | D_{l_i \sigma_i { \bf k_i  }}^{(v_{p_i})} |\beta_{p_i} > 
    < \alpha_{q_{i-1}} | D_{l_i \sigma_i { \bf k_i  }}^{(-v_{p_i})} |\alpha_{q_i} >   
\\ 	
\end{eqnarray*}

\begin{eqnarray*}
 && {\int \int \dots \int }_{\tau_1 + \tau_2 + \dots + \tau_{2n-1} \le 
\frac{t}{\hbar}} 
\quad  \prod_{j=1}^{2n-1}
\end{eqnarray*}
\begin{displaymath}
 \exp \left\lbrace    - i \tau_j  \left( \Delta E_j  
    + \sum_{i: p_i  \in \tilde{J} (j), q_i \not\in \tilde{J} (j)}^{}   v_{p_i} 
    \varepsilon_{l_i {\bf k_i} \sigma_i} - \sum_{i: p_i  \not\in \tilde{J} (j), 
q_i \in \tilde{J} (j)}   v_{p_i} 
    \varepsilon_{l_i {\bf k_i} \sigma_i}                 \right) \right\rbrace .
\end{displaymath}

\begin{figure}[h]\centering 
\includegraphics[width = 0.5\textwidth]{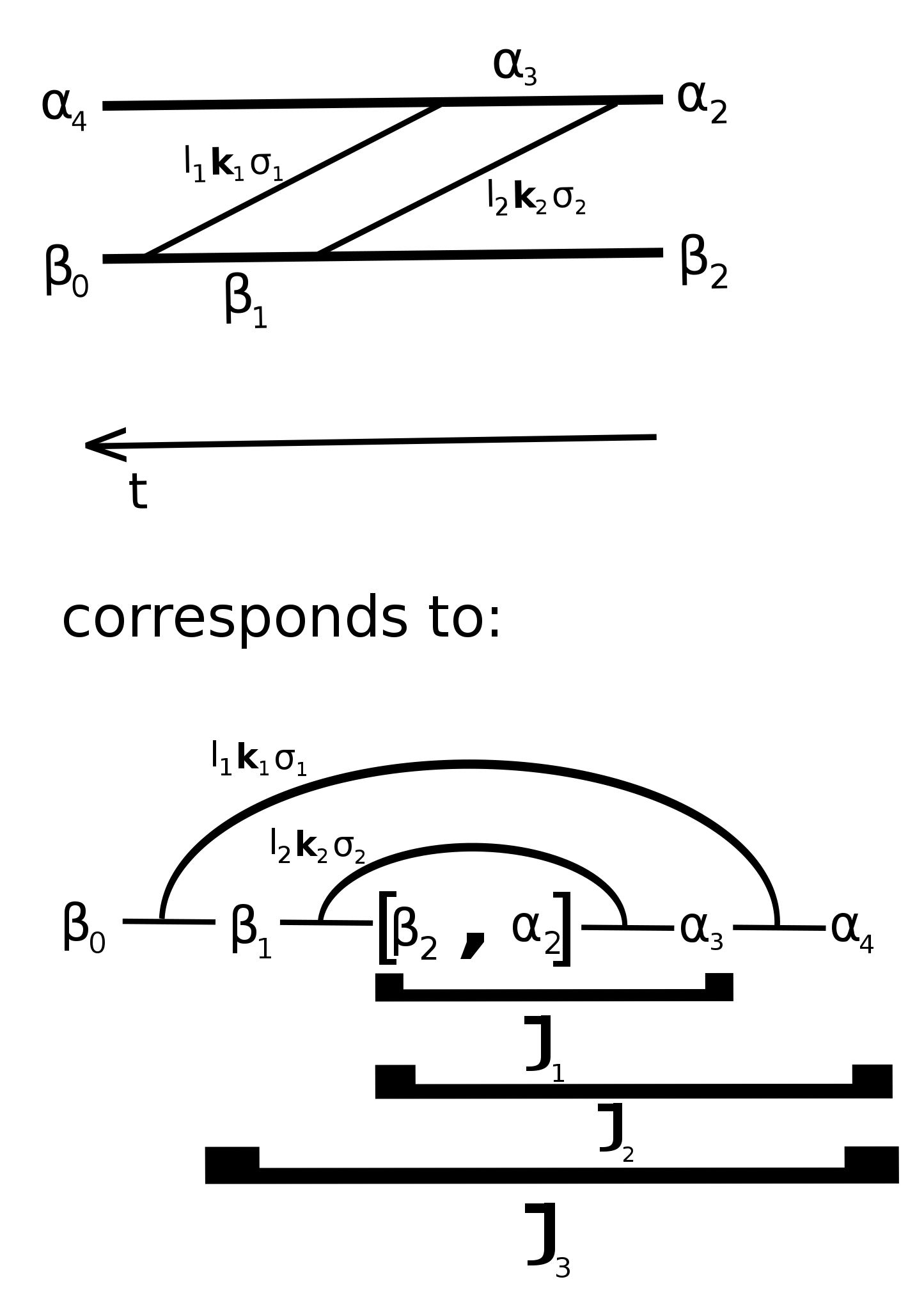}
\caption{\small An example of a fourth-order diagram, written in the conventional
(above) and an alternative (below) form. $2n = 4$ in this diagram, moreover, 
$k = 2$; the pairs $(p_i, q_i)$ appear in the lower version of the diagram as pairs of 
horizontal lines, i.e., pairs of numbers between $1$ and $4$ once these are counted
from the left to the right. The pairs in the example are $(1,4),(2,3)$. The sequence 
of intervals is given by 
$J(1) = \left\lbrace 2,3 \right\rbrace, J(2) = \left\lbrace 2,3,4 \right\rbrace , 
 J(3) = \left\lbrace 1,2,3,4 \right\rbrace $. The pair formation and the 
sequence of intervals are consistent in the sense that the diagram is irreducible. }
\label{topology}
\end{figure}

The meaning of the appearing objects is defined in Ref. \cite{Kern10}, 
but I briefly recall the terminology here, and included Fig. \ref{topology}: 
$2n$ is the order of the 
diagram, $n$ the number of its tunneling lines; $k$ is the number of quantum dot 
states appearing on the left-hand side of the diagram; the $\beta_i, \alpha_i$ are 
arbitrary quantum dot states; I used the bra-ket notation. 
$l_i, \sigma_i$ is the lead- and the spin index of the
$i$-th tunneling line; $(p,q) $ is a set of $n$ pairs of numbers between one and $2n$ 
and ''is'' the set of the tunneling lines; always $p_i < q_i$. 
$J$ is an increasing sequence of 
intervals of integers $J(1) \subset J(2) \dots \subset J(2n-1) \subset 
\left\lbrace 0,1, \dots ,2n \right\rbrace $
and defines the time ordering of the diagram; $\tilde{J}(j) := J(j) \setminus 
min J(j)$. The pair formations $(p,q)$ originate from Wick's theorem 
\cite{Bruus} and are required to be consistent with the sequence $J$ of intervals
in such a way that the diagram is {\em irreducible}: For every $j \in \left\lbrace
1, \dots , 2n-1 \right\rbrace $ there must be at least one $i \in \left\lbrace 
1, \dots , n \right\rbrace $ such that $p_i \in \tilde{J}(j), 
q_i \notin \tilde{J}(j)$ or vice versa.
This condition becomes graphically manifest in the property of Fig. \ref{topology} 
that for any horizontal square bracket representing one of the intervals $J(j)$
there is at least one tunneling line which leaves the region enclosed by the 
square bracket.

${\bf k}_i$ is the wave 
vector of the electron level attached to the tunneling line $i$, $Z_{l_i}$ the number 
of allowed wave vecors  per volume in the first Brillouin zone of lead $l_i$. The 
integrations go over these. By $f_{l}^{(\pm)}$ the Fermi-Dirac distribution/ 
one minus the Fermi-Dirac distribution of lead $l$ is denoted.
The   operators  $ D_{l\sigma {\bf k}} :=  T_{l {\bf k}}^\ast d_{\sigma}  $ contain 
the tunneling coupling; the integral with respect to the time is performed over the 
set 
\begin{displaymath}
 \left\lbrace (\tau_1, \dots , \tau_{2n-1} ): \tau_i \ge 0, \tau_1 + \dots + 
\tau_{2n-1} \le \frac{t}{\hbar}
\right\rbrace 
\end{displaymath}
which has the size $\frac{(t/\hbar)^{2n-1}}{(2n-1)!}$; the $v_{p_i}$ are signs which 
one can assign to the chosen sequence of quantum dot states, depending on their 
particle numbers. 
The $\tau_j$ are the lengths of 
time intervals between subsequent tunneling events. They are multiplied with the 
energy differences of the states with respect to which the total density matrix is 
off-diagonal during these intervals; $\varepsilon_{l {\bf k}\sigma}$ is the 
energy of the one electron level in lead $l$ and wave vector ${\bf k}$; 
$ \Delta E_j = E_{\beta_{min J(j)}}  - E_{\alpha_{max J(j)}} $.

Because the SIAM has only the quantum dot states $0, \uparrow, \downarrow,$ and 
$2$, the expressions containing the matrix elements of the annihilation operators of 
the quantum dot can, as far as they are not zero, always be written as
\begin{displaymath}
 f_{l_{i}}^{(v_{p_i})}(\varepsilon_{l_i \sigma_i {\bf k_i}})
    < \alpha_{p_{i-1}} | D_{l_i \sigma_i { \bf k_i  }}^{(v_{p_i})} |\alpha_{p_i} >  
< \alpha_{q_{i-1}} | 
D_{l_i \sigma_i { \bf k_i  }}^{(-v_{p_i})} |\alpha_{q_i} >  =
\end{displaymath}
\begin{displaymath} 
 \pm f_{l_{i}}^{(v_{p_i})}(\varepsilon_{l_i {\bf k_i}}) \left| T_{l_i {\bf k_i}} 
\right|^2 .
\end{displaymath}
For simplicity $ \varepsilon_{l {\bf k} \sigma} = \varepsilon_{l {\bf k} } $ 
is assumed.
Next, the integration with respect to the wave vector is replaced by 
an integration with respect to the electron energy, as sketched in Ref. 
\cite{Ashcroft}. 
This yields coupling functions $\alpha_l (\varepsilon)$ which describe the energy 
dependence of the tunnel coupling. 
Upon introducing $\alpha_l^\pm
(\varepsilon) := f_l^{(\pm)} (\varepsilon) \alpha_l (\varepsilon) $, 
the part of the diagram beginning with the integration over the Brillouin zones
can be rewritten as

\begin{eqnarray*}
&&    \pm \int d \varepsilon_1 \dots 
\int d \varepsilon_n  
\end{eqnarray*}
\begin{eqnarray*}
&&  \prod_{i: p_i < q_i \le k}   \alpha_{l_i}^{(v_{p_i})}  (\varepsilon_i)   	
  \prod_{i: k+1 \le p_i < q_i} \alpha_{l_i}^{(v_{p_i})}   (\varepsilon_i)                 	
  \prod_{i: p_i \le k, k+1 \le q_i} \alpha_{l_i}^{(-v_{p_i})}  (\varepsilon_i)       	
\end{eqnarray*}

\begin{eqnarray*}
 && {\int \int \dots \int }_{\tau_1 + \tau_2 + \dots + \tau_{2n-1} \le 
\frac{t}{\hbar}} 
\quad  \prod_{j=1}^{2n-1}
\end{eqnarray*}
\begin{equation} \label{exponent_1}
 \exp \left\lbrace    - i \tau_j  \left( \Delta E_j  
    + \sum_{i: p_i  \in \tilde{J} (j), q_i \not\in \tilde{J} (j)}^{}   v_{p_i} 
    \varepsilon_i - \sum_{i: p_i  \not\in \tilde{J} (j), q_i \in \tilde{J} (j)}   
v_{p_i} 
    \varepsilon_i                 \right) \right\rbrace  .
\end{equation}

In the exponent, the sums with respect to $j$ and $i$ can be swapped; the integration 
with respect to the times can be swapped with the energy integrals, one obtains then 
Fourier transforms.  To realize this plan I subdivide the 
set of the tunneling lines into four subsets of $\left\lbrace 1, ..., n 
\right\rbrace$:
\begin{eqnarray*}
 I_1 &:=& \left\lbrace i:   p_i \ge k+1\right\rbrace,   \\
I_2 &:=& \left\lbrace i: \exists j \mbox{ such that } p_i  \in \tilde{J} (j), 
q_i \not\in \tilde{J} (j)  
\mbox{ and: } p_i \le k, q_i \ge k+1\right\rbrace ,  \\  
I_3 &:=& \left\lbrace i:  q_i \le k \right\rbrace,   \\ 
I_4 &:=& \left\lbrace i: \exists j \mbox{ such that } p_i  \not\in \tilde{J} 
(j), q_i \in \tilde{J} (j)  
\mbox{ and: } p_i \le k, q_i \ge k+1\right\rbrace .  
\end{eqnarray*}

$\left\lbrace 1, \dots , n \right\rbrace$ is the disjoint union of $I_1, I_2, I_3, 
$ and $I_4$. The expression (\ref{exponent_1}) reads in terms of Fourier transforms:
\begin{equation} \label{time integrals}
 \pm {\int \int \dots \int }_{\tau_1 + \tau_2 + \dots + \tau_{2n-1} \le 
\frac{t}{\hbar}} \quad
\exp \left( -i \sum_{j=1}^{2n-1} \tau_j \Delta E_j   \right)
\end{equation}
\begin{eqnarray*}
&& \prod_{i:i\in I_1}  {\cal F} \left( \alpha_{l_i}^{(v_{p_i})} \right)  
\left( v_{p_i}
 \sum_{j: p_i  \in \tilde{J} (j), q_i \not\in \tilde{J} (j)} \tau_j \right)  \\ 
&& \prod_{i:i\in I_2}  {\cal F} \left( \alpha_{l_i}^{(-v_{p_i})} \right)  
\left( v_{p_i}
 \sum_{j: p_i  \in \tilde{J} (j), q_i \not\in \tilde{J} (j)} \tau_j \right)  \\
 && \prod_{i:i\in I_3}  {\cal F} \left( \alpha_{l_i}^{(v_{p_i})} \right)  
\left( -v_{p_i} \sum_{j: p_i  \not\in \tilde{J} (j), q_i \in \tilde{J} (j)} 
\tau_j \right)  \\ && \prod_{i:i\in I_4}  {\cal F} \left( \alpha_{l_i}^{(-v_{p_i})} 
\right)  \left( -v_{p_i}
 \sum_{j: p_i  \not\in \tilde{J} (j), q_i \in \tilde{J} (j)} \tau_j \right) .
\end{eqnarray*}

\subsection{Expansion of the transform $K(\lambda = 0)$}
In the time space, the kernels have the diagrammatic expansion 
\begin{displaymath}
 K(t, w) = \sum_{n=1}^\infty w^{2n} K^{(2n)}_{w=1} (t),
\end{displaymath}
where $w$ is the coupling parameter and $K^{(2n)}$ is the contribution of the 
diagrams of order $2n$. The task is to prove that the expansion survives the 
transformation to the Laplace space (\ref{transformation}), even in the limit
$\lambda \to 0$. To this end, I must 
show that for sufficiently small values of $w$,i.e., for $w<w_0$, 
where $w_0 > 0$, the series of integrals
\begin{displaymath}
  \sum_{n=1}^\infty w^{2n}    \int_0^\infty dt \left| K^{(2n)}_{w=1} (t) \right|  
\end{displaymath}
is finite.

Hence, back to the time integral, expression (\ref{time integrals}): 
The integral with respect to the $\tau$-s can be calculated by fixing a time 
$\tau'$ between zero and the upper bound, $t/\hbar$, performing the integration over 
the set where the sum of the $\tau_j$-s equals this time,  
and then integrating with respect to  
$\tau'$. In this way, the integrated kernel is expressed by its time derivative.
Moreover, the appearing Fourier transforms are bounded by an exponential 
decay: $ \left|{\cal F}(\dots) (\tau) \right| \le a(w) \exp(-c |\tau|) = 
w^{2} a_{w=1} \exp(-c |\tau|) $; this is the 
case at least if one makes a simple assumption about the coupling function, for 
example Lorentzian shape. In an appendix I shall show why this is the case.  
Making use of this inequality and with the definition
\begin{displaymath}
 N_j := \mbox{ number of elements of }\left\lbrace i:   p_i  \in \tilde{J} (j), 
q_i \not\in \tilde{J} (j) 
\mbox{ or vice versa }
 \right\rbrace 
\end{displaymath}
one obtains the estimate

\begin{eqnarray*}
 w^{2n} \left| K^{(2n)}_{w=1}(t) \right|  \le  
&&  1/\hbar \quad w^{2n} \sum_{k=0}^{2n} \quad \sum_{\beta_0, \dots, \beta_k, 
\alpha_k, \dots \alpha_{2n}} 
     \quad \sum_{l_1 \sigma_1, \dots , l_{n} \sigma_{n} } \quad    \\
&& \sum_{(p,q)} \quad \sum_{\mbox{J: consistent with (p,q)}}  \quad a^n  \\
&& \frac{1}{\hbar} \frac{1}{\sqrt{2n-1}}\int_{ \tau_1 + \dots + \tau_{2n-1} = 
t/\hbar } 
\exp \left(  -c \sum_j \tau_j N_j  \right).    
\end{eqnarray*}
The notation $a_{w=1} =:a $ was used. So far, I chose first 
an arbitrary pair formation $(p,q)$ and then summed over all time orderings 
$ J$ which are 
consistent with the pair formation in the sense of irreducibility. These two sums 
can be swapped. Taking the integral of the right hand side over the positive
time axis, one obtains:
\begin{eqnarray} \label{time integral 2}
 w^{2n} \int_0^\infty dt \left| K^{(2n)}_{w=1}(t) \right|  \le  
&&  1/\hbar \quad \frac{w^{2n}  a^n}{c^{2n-1}} \quad  \sum_{k=0}^{2n} \quad 
		\sum_{\beta_0, \dots, \beta_k, \alpha_k, \dots \alpha_{2n}} 
     \quad \sum_{l_1 \sigma_1, \dots , l_{n} \sigma_{n} } \quad \nonumber   \\
&& \sum_{\mbox{J}} \quad \sum_{\mbox{(p,q): irreducible in J}} \nonumber \\
&&   \prod_{j=1}^{2n-1} \frac{1}{N_j}.  
\end{eqnarray}

\subsection{Combinatorial intermezzo}
Now, an estimate for the summand 
\begin{equation} \label{combinatorics}
\sum_{\mbox{(p,q): irreducible in J}} \quad \prod_{j=1}^{2n-1} \frac{1}{N_j}.
\end{equation}
is required.

The objects which define a single contribution to this sum are an increasing 
sequence of intervals of integers 
$\tilde{J}$, which may alternatively be viewed as an ordering of the numbers 
$1, 2, ... 2n,$ and 
a pair formation which is irreducible in $\tilde{J}$. For any $j \in 
\left\lbrace 1, \dots, 2n-1 \right\rbrace$, 
$N_j$ is the number of pairs which are separated by the set $\tilde{J} (j)$, 
i.e., $N_j$ is the number of $i$-s in $\left\lbrace 1, \dots , n \right\rbrace $ 
with the property that $p_i \in \tilde{J}(j), q_i \notin \tilde{J}(j)$,
or $p_i \notin \tilde{J}(j), q_i \in \tilde{J}(j)$.
Without loss of generality I can assume that the ordering $\tilde{J}$ is just given 
by the natural way of counting. The sum (\ref{combinatorics}) does not depend on 
$\tilde{J}$. One can view $N_j= N(p,q)_j$ as a map
\begin{displaymath}
 N(p,q): \quad \left\lbrace 1, \dots , 2n-1 \right\rbrace \rightarrow  
\left\lbrace 1,2, \dots \right\rbrace 
\end{displaymath}
 with the property that, with every step, the value of the function changes 
by $\pm1$ and that $N(p,q)(2n-1) =  N(p,q)(1) = 1$. The sum (\ref{combinatorics})
can be rewritten as 
\begin{displaymath}
 \sum_{\mbox{(p,q): irreducible}} function (N(p,q)) = \sum_N \sum_{(p,q): 
N(p,q) = N} function (N). 
\end{displaymath}

 With this I express that there are in general different irreducible pair 
formations $(p,q), (p',q')$ with  $N(p,q) = N(p',q')$. One can take the sum over all 
irreducible pair formations by first fixing a map $N$, summing over all $(p,q)$ with 
$N(p,q) = N$, and finally taking the sum with respect to $N$. All information about 
$N$ is contained in the subset 
\begin{displaymath}
 S^+ := \left\lbrace 1 \right\rbrace  \cup \left\lbrace j: N(j) > N(j-1) 
\right\rbrace .
\end{displaymath}
There are $2^{2n}$ subsets of $\left\lbrace 1, \dots , 2n \right\rbrace$, and, 
as a consequence, less than $2^{2n}$  ways of choosing the map $N$.

Now, I want to fix the map $N$ and think about
\begin{displaymath}
 \sum_{(p,q): N(p,q) = N} function (N) = \prod_{j= 1}^{2n-1} \frac{1}{N(j)}  
\sum_{(p,q): N(p,q) = N} 1.
\end{displaymath}
Define $S^- := \left\lbrace 1, \dots , 2n \right\rbrace \setminus S^+$. For sure, 
\begin{displaymath}
 \left\lbrace 1, \dots, 2n \right\rbrace = S^+ \cup S^-.
\end{displaymath}
I imagine that the numbers $1, \dots, 2n$ are listed in the 
natural order from the left to the right.
If the pair formation $(p,q)$ has the property $N(p,q) = N$, then the partner of a 
number $l$ in $S^+$ 
can always be found to the {\em right} of $l$, the partner of a number $l'$ in $S^-$ 
can always be found to the {\em left} of $l'$.  Thus, the number of possibilities 
to choose a partner for $l' \in S^-$ is always bounded by the number  of 
those elements of $S^+$ which can be found to the left of $l'$.
I refer to this number as $Z^+(l')$. Going through the elements of $S^-$ 
from the left to the right, one notices that, for the first element $l_1$, one has 
really $Z^+(l_1)$ possibilities to choose a partner, for the second element, however, 
only $Z^+ (l_2) - 1$, for the third $Z^+ (l_3)-2$, and so on. One finds that
the number of different pair formations with $N(p,q) = N$ is
\begin{displaymath}
 \prod_{i=1}^{n}  \left( Z^+ (l_i) - (i-1) \right) = \prod_{i=1}^{n-1}  
\left(  N(l_i) + 1 \right). 
\end{displaymath}

For the last equality, I used the characterization $N(j) = \left| S^+ \cap 
\left\lbrace 
1, \dots , j \right\rbrace \right| - \left| S^- \cap \left\lbrace 1, \dots , 
j \right\rbrace \right|$ of the map $N$.                                                                             
One obtains: 
\begin{eqnarray*}
 \sum_{(p,q): N(p,q) = N} function (N)   &=& \prod_{j= 1}^{2n-1} \frac{1}{N(j)}   
\cdot 
\prod_{i=1}^{n-1}  
 \left(  N(l_i) + 1 \right) \\
&\le&  \prod_{i=1}^{n-1} \frac{N(l_i) + 1}{N(l_i)} \le 2^{n-1} .
\end{eqnarray*}
The result of the combinatorial intermezzo can be summarized by the estimate 
\begin{displaymath}
 \sum_{\mbox{(p,q): irreducible in J}} \quad \prod_{j=1}^{2n-1} \frac{1}{N_j} 
\le 2^{3n}.
\end{displaymath}

\subsection{Back to the expansion of the transform $K(\lambda = 0)$}
This intermediate result can now be inserted into the inequality 
(\ref{time integral 2}). For a fixed value of 
$k$, the number of different time orderings $J$ is bounded by $2^{2n-1}$, since, 
every time the interval is enlarged, there are at most two possibilities for this 
(towards the left or, alternatively, towards the right). The sums with respect to 
the spins and the lead indices as well as the sum with respect to the quantum dot 
states can be estimated in an elementary way, just by the number of possible choices. 
Finally, the number of possibilities 
to choose $k$ is $2n+1$. One arrives at an inequality of the 
form
\begin{displaymath}
  w^{2n} \int_0^\infty dt \left| K^{(2n)}_{w=1} (t) \right|  \le  const' 
\quad const^n  w^{2n},
\end{displaymath}
 with $const, const' > 0$, independent of $n$. One can conclude that 
$K(\lambda = 0, w)$ 
is given by a power series in $w^2$ with positive radius of convergence. The 
coefficients are indeed $K^{(2n)} (\lambda = 0)$. The treatment of the 
current kernel is analogous. The combinatorial intermezzo was necessary, since
a crude estimate by the number of irreducible pair formations would have induced the 
presence of factorials. This causes problems, while the emergence of powers of         
some constant does not.

\section{From the kernels to the current}

\subsection{Representation of the current in terms of the kernels}
It was shown that for $w<w_0$:
\begin{displaymath}
 K_{(c)} (w) = w^2 K_{(c)}^{(2)}  + w^4 K_{(c)}^{(4)} + \dots .
\end{displaymath}
$K_c$ denotes the current kernel, while $K$ is the density matrix kernel.
From the recursive definitions of the kernels \cite{Kern11}, it can be concluded on a 
general level (not only for the SIAM): 
\begin{itemize}
 \item $x$ a hermitian operator on the quantum dot space $\Rightarrow$ $K^{(2n)}x$ 
hermitian; 
\item for any $x$: $ Tr \left( K^{(2n)}x \right) = 0$;
\item for any hermitian $x$: $ Tr \left( K_c^{(2n)} (x) \right)$ real. 
\end{itemize}
For the perfectly conventional SIAM considered in this text, the fact that in every
single diagram the overall amount of spin added or subtracted on each of the two 
contours is equal on the two contours implies that any
matrix of the form $|a><a|$ (with $a = 0, \uparrow, \downarrow, 2$) is 
mapped by every single order of the kernels to a linear combination of these 
four operators. Hence, the density matrix kernel and all of its 
orders can be viewed as  maps
\begin{displaymath}
 K^{(2n)}: V \rightarrow V, 
\end{displaymath}
where $V$ is defined as the four dimensional real vector space spanned by the 
operators $|a><a|$. Their image is always contained in the three dimensional 
linear subspace $U \subset V$ of the operators with vanishing trace. It is  
technically useful to introduce the operator 
\begin{displaymath}
 L(w) := \frac{1}{w^2} K (w).
\end{displaymath}
The quantum master equation in the stationary limit can be written as 
 \begin{displaymath}
 L(w) \rho (w) = 0. 
\end{displaymath}
The stationary reduced density matrix of the quantum dot, ''$\rho (w)$'', 
is obtained from this equation.

Since the image of $L(w)$ has dimension three or less, the equation has a solution. 
$L(0)$ is given exclusively by the second order 
diagrams and, in the basis $v_1:= |0><0|, v_2:= |\uparrow><\uparrow|, 
v_3:= |\downarrow><\downarrow|, v_4 := |2><2| $, it can be written as the matrix: 
\begin{displaymath}
 L(0) =  \frac{2\pi}{\hbar} \left( \begin{array}{cccc}  	
\dots            	& \alpha^-(E_{\uparrow 0})    	& \alpha^-(E_{\downarrow 0})	&    0    		\\ 
\alpha^+(E_{\uparrow 0})    	& \dots            	&     0   		&  \alpha^-(E_{2\uparrow})    	\\ 
\alpha^+(E_{\downarrow 0})    	&    0          	&  \dots	    	&   \alpha^-(E_{2\downarrow})  	\\
0                   	& \alpha^+(E_{2\uparrow})	&  \alpha^+(E_{2\downarrow})    	&    \dots    		\\
\end{array} \right).
\end{displaymath}
I used the notation $E_{ab} = E_a - E_b$ for arbitrary quantum dot states $a,b = 0, 
\uparrow, \downarrow, 2 $, where the energies $E_a$ are the eigenenergies of the 
Hamiltonian of the isolated quantum dot, Eq. \ref{dot Hamiltonian}.

$L(0)v_1, L(0)v_2 , L(0)v_3 $ and $v_4$ are linearly independent. Only the 
assumption that $\alpha^+(E_{\uparrow 0})$ etc. are strictly positive is needed for this. 
Let $End(V)$ be the vector space of the endomorphisms of $V$
(the linear maps $V \rightarrow V$). Use the map 
\begin{displaymath}
 \phi : End(V) \rightarrow End (V),
\end{displaymath}
\begin{displaymath}
 \phi (E) \left( \sum_{i=1}^{4} \lambda_i v_i \right) = \sum_{i=1}^{3} 
\lambda_i E(v_i) + \lambda_4 v_4 . 
\end{displaymath}
$\phi (L(0))$ is invertible, $\phi$ is smooth. The set of the invertible endomorphisms, 
$\left\lbrace det \neq 0 \right\rbrace$, is topologically open in $End(V)$. Hence, there is $r>0$ 
such, that for all 
\begin{displaymath}
 E \in B_r(L(0)) := \left\lbrace E' \in End(V): \left| E' - L(0) \right| < r 
\right\rbrace 
\end{displaymath}
still $\phi (E) \in \left\lbrace det \neq 0 \right\rbrace$. ( All norms on $End(V)$ 
are equivalent, since it is a space of finite dimension.) This allows me to define
\begin{eqnarray*}
 F: \quad  B_r(L(0)) &\rightarrow& V, \\
 F(E)         &:=&      \left[ inverse ( \phi (E))\quad E v_4 \right]   -v_4.  
\end{eqnarray*}

For sufficiently small values of $w$, one finds $L(w) \in B_r(L(0))$. Then, 
necessarily, 
the three operators $L(w)v_i, i = 1,2,3,$ are linearly independent and $L(w)v_4$ is a 
linear combination of those three:
\begin{displaymath}
 L(w) v_4 = \sum_{i=1}^{3} x_i L(w) v_i.
\end{displaymath}
 This implies that 
\begin{displaymath}
 F(L(w)) = \sum_{i=1}^{3} x_i v_i - v_4 , \quad \Rightarrow  L(w) \left[ F(L(w)) 
\right] = 0.
\end{displaymath}
The result obtained so far is that for, say, $w<w_0$, the space of the solutions 
to the quantum master equation in the stationary limit, 
\begin{displaymath}
 L(w) \rho (w) = 0,
\end{displaymath}
is spanned by $F(L(w))$, the dimension is one. The equation $L(0) \rho(0) = 0$ is 
solvable with an operator $\rho(0)$ with trace one. This implies that 
$Tr \left\lbrace  F(L(0)) \right\rbrace \neq 0$. The composition of maps
\begin{displaymath}
 Tr \circ F
\end{displaymath}
 is smooth, and so one can conclude that there is $r' \le r$ such, that 
\begin{displaymath}
 \mbox{ for all } E \in B_{r'} (L(0)) : \quad Tr (F (E)) \neq 0. 
\end{displaymath}
As a consequence, the map 
\begin{eqnarray}
 f:  B_{r'}(L(0))   &\rightarrow&  V, \nonumber \\
     E               &\mapsto&    \frac{F(E)}{Tr(F(E))} 
\label{the map f}
\end{eqnarray}
is still smooth. For $w < w_0'$, with  fixed $w_0' \le w_0$, one obtains that 
$L(w) \in B_{r'}(L(0)) $ and 
\begin{displaymath}
 \rho (w) = f (L(w)). 
\end{displaymath}
The equation for the current $I(w)$ reads: 
\begin{equation} \label{current_in_terms_of_kernels}
 I (w) =   Tr \quad K_c (w) \left[  f (L(w)) \right]. 
\end{equation}

\subsection{Taylor series of $I(w)$ in $w=0$}
The above representation of the current in terms of the kernels shows that it is a 
smooth function of $w$. It has a well-defined Taylor series in $w=0$; the 
coefficients $c_k$ of this power series are given by the derivatives of the current 
with respect to $w$ in $w =0$:
\begin{displaymath}
 c_k = \frac{1}{k!} I^{(k)} ( w = 0).
\end{displaymath}
(All odd derivatives vanish, since the current is also a smooth function of $w^2$; 
this is not relevant for the following arguments.) The $c_k$ can alternatively be 
defined recursively  without taking derivatives:  
\begin{eqnarray*}
 c_0 &:=& I (w=0),\\
p_0 (w) &:=& c_0,  \\
c_{k+1}  &:=& \lim_{w \to 0} \frac{(I-p_k) (w)}{w^{k+1}} , \\
p_{k+1}(w) &:=& p_k (w) +  c_{k+1} w^{k+1} .    
\end{eqnarray*}
This representation is equivalent to the definition by derivatives of $I(w)$.

\subsection{Replacing the kernels}
Perturbation theory is applied \cite{Koller_Diss} by truncating the expansions of the 
kernels at a finite order. Upon defining
\begin{displaymath}
 \bar {K}_{(c)} (w) :=  w^2 K_{(c)}^{(2)}  + \dots + w^{2n} K_{(c)}^{(2n)}, 
\end{displaymath}
the density matrix  ``$\bar{\rho} (w)$'' is determined by the equation
\begin{displaymath}
 \bar {K}(w) \bar{\rho} (w) = 0,
\end{displaymath}
and the current $\bar {I} (w)$ by 
\begin{displaymath}
 \bar{I} (w) = Tr \left( \bar{K}_c (w) \bar {\rho} (w) \right) . 
\end{displaymath}
 In the same way as done for the exact current (\ref{current_in_terms_of_kernels}), 
this can be expressed in terms of the kernels as 
\begin{displaymath}
 \bar{I} (w) =   Tr \quad \bar{K}_c (w) \left[  f (\bar{L}(w)) \right] \quad 
\mbox{for  } w \le w_0'', 
\end{displaymath}
where I used $\bar {L} (w) := \frac{1}{w^2} \bar{K} (w)$.

For sufficiently small values of $w$, both $L(w)$ and $\bar {L} 
(w)$ can be assumed to be contained in $B_{\frac{r'}{2}}(L(0))$, and hence: 
\begin{displaymath}
f\left[L(w)\right] = f \left[\bar{L} (w) + w^{2n} b(w)\right] = f \left[\bar{L}(w)
\right] + w^{2n} B (w),
\end{displaymath}  
where $b(w), B(w)$ denote bounded functions: 
$f$ is smooth on $B_{r'} (L(0))$ and thus the derivative of 
$f$ is bounded on $B_{\frac{r'}{2}} (L(0))$. Moreover, the estimate
\begin{displaymath}
 |f(x) - f(y)| \le |x- y| \sup \left\lbrace |f'(z)|: z \in B_{\frac{r'}{2}} (L(0)) 
\right\rbrace \quad 
\mbox{for  }
x, y \in B_{\frac{r'}{2}} (L(0))
\end{displaymath}
can be used.

Analogously, the current kernel can be written as 
\begin{displaymath}
 K_c (w) = \bar {K}_c (w) + w^{2n+2} b_c (w) \quad \mbox {with bounded  } b_c (w).
\end{displaymath}
In summary: 
\begin{displaymath}
 I (w) = \bar{I} (w) + w^{2n+2} b_I (w) \quad \mbox {with bounded  }  b_I (w). 
\end{displaymath}
Inserting this expression for $I(w) $ into the recursive definition of the 
coefficients of the Taylor series of $I(w)$, it is seen that the coefficients 
$c_0, \dots , c_{2n}$ equal the corresponding coefficients of the 
Taylor series of $\bar{I} (w)$.

\section{Direct calculation of diagrams} \label{theory for diagrams}
Results for the kernels up to sixth order in the tunneling coupling, 
neglecting double occupancy, are presented in the final section of this text.  
Hence, the purpose of the present section is to provide a background 
in principle sufficient for the calculation of the appearing diagrams.

\subsection*{Notation:}
\begin{itemize}
\item For any function $f$ with domain $\mathbb{R}$ define the mirrored map $S f$ of $f$
      by
      \begin{displaymath}
      (Sf) (x) := f (-x).
      \end{displaymath}       
\item For any function $f$ with domain $\mathbb{R}$ and $x_0 \in \mathbb{R}$ define the 
     the translation $T_{x_0} f$ of $f$ by $x_0$ as the map
      \begin{displaymath}
      ( T_{x_0} f ) (x) \quad := \quad f( x - x_0 ).
      \end{displaymath}
\item For a smooth and quadratically integrable function $f: \mathbb{R} \rightarrow 
       \mathbb{C}$ denote the Hilbert transform of $f$ by
       \begin{displaymath}
        H f (x) \quad := \quad \frac{1}{\pi} \quad \int_0^\infty d \omega \quad 
        \frac{f ( x + \omega) - f ( x - \omega ) }{\omega} . 
       \end{displaymath}
\item For $x_0 \in \mathbb{R}$ and a smooth function $f: \mathbb{R} \rightarrow \mathbb{C} $
     define the map $ \delta_{x_0} f : \mathbb{R} \rightarrow \mathbb{C} $ as the continuous 
     continuation to $\mathbb{R}$ of the function
     \begin{displaymath}
     (\delta_{x_0} f) (x) \quad := \quad \frac{f(x) - f(x_0)}{x - x_0}.
     \end{displaymath}
     In particular, let $\delta := \delta_0$. Note that 
     \begin{displaymath}
     \delta_{x_0} \quad = \quad T_{x_0} \delta T_{-x_0}.
     \end{displaymath}
\item For (measurable and) quadratically integrable functions $f, g: \mathbb{R} 
      \rightarrow \mathbb{C} $ define the convolution $f*g$ of $f$ with $g$ by     
      \begin{displaymath}
      ( f * g ) (x) \quad := \quad \int_\mathbb{R} dy \quad f( y ) g (y - x) .
       \end{displaymath}         
\item For $ \eta > 0 $ and, for example, quadratically integrable $f: \mathbb{R} 
      \rightarrow \mathbb {C} $
      let the transform $ t_\eta f $ of $f$ be defined as 
      \begin{displaymath}
      t_\eta f \quad := \quad f * l_\eta
      \end{displaymath}
      with $l_\eta$ the lorentzian 
      \begin{displaymath}
      l_\eta (x) \quad := \quad \frac{1}{\pi} \quad \frac{\eta}{\eta ^2 + x ^2}. 
      \end{displaymath}       
\end{itemize}

\subsection*{Remark:}
The convolution of $f$ with $g$ in $x$ can also be written as the scalar product
of quadratically integrable functions of  $f^* $
(let the asterisk denote the complex conjugate) and $T_x g$. Hence, it is well-defined.
The Fourier transform conserves the scalar product of quadratically integrable 
functions (Plancherel).
Morover, the relation
\begin{displaymath}
{ \cal F} T_x g \quad = \quad \mu_x ( {\cal F} g ) 
\end{displaymath} 
with $\mu_x$ the factor
\begin{displaymath}
\mu_x (y ) \quad := \quad e^ {-i x y}
\end{displaymath}
holds. As a consequence, the convolution of two quadratically integrable functions is 
always continuous.

\subsection*{Lemma (dualism of spaces):}
Define the space of functions ${\cal L}$ as the set of all measurable functions 
$f: \mathbb{R} \rightarrow \mathbb{C} $ which decay exponentially. The latter property
be defined by the condition that there are $a, b > 0$ such, that for all $x \in 
\mathbb{R}$:
\begin{displaymath}
|f (x)| \quad \le \quad a e^{-b|x|}.
\end{displaymath}
On the other hand, define the space of functions ${\cal R}$ as the set of all Fourier
transforms of functions in ${\cal L}$, where the terminology 
\begin{displaymath}
( {\cal F} f ) (x) \quad := \quad \int_\mathbb{R} dy \quad f (y) e^{-ixy}
\end{displaymath}
is used. Examples of elements in ${\cal L}$ are, for arbitrary $\eta > 0$, 
\begin{displaymath}
d_\eta (t) \quad := \quad e^{-\eta |t| } \quad \mbox{and} \quad r_\eta (t) \quad :=
                   \quad d_\eta (t) sign (t),
\end{displaymath}
corresponding examples of elements in ${\cal R}$ are then
\begin{displaymath}
{\cal F} d_\eta  \quad = \quad  2 \pi l_\eta \quad \mbox{and} \quad ({\cal F} r_\eta) (x)
      \quad = \quad -2i \frac{x}{\eta ^2 + x^2} \quad = \quad 
       2 \pi i  ( H l_\eta) (x). 
\end{displaymath}
In this particular example, the action of the Hilbert transform $H$ on a function in 
${\cal R}$ has, up to a factor, the same effect as the multiplication with the sign
of the corresponding exponentially decaying function. This is true {\em in general}, 
e.g. Ref. \cite{Stein}. Nevertheless, it makes sense to verify the equality of 
maps in the present situation.

All functions in ${\cal R}$ are analytic and quadratically integrable. 
Moreover, it is clear from the definition of the spaces that they are copies of each 
other, as far as the zero-set equivalence is applied in ${\cal L}$, the isomorphism 
is the Fourier transform. I shall refer to the inverse of the Fourier transform by 
${\cal F}^{-1}$. It can be quickly verified that product and convolution 
of two functions in ${\cal L}$ are functions in ${\cal L}$. The same holds true  
for functions in ${\cal R}$:

\textbf{Statement 1:} If $f, g \in {\cal R}$, then 
\begin{eqnarray*}
   (1a) \quad f g \quad &=& \quad {\cal F} \left[ ({\cal F}^{-1} f ) * 
                            ( S {\cal F}^{-1} g ) \right] \quad \in \quad {\cal R}, \\
   (1b) \quad f * g \quad &=& \quad 2 \pi \quad {\cal F } \left[ ( {\cal F}^{-1} f )
                           ( S {\cal F}^{-1} g)     \right]   \quad \in \quad {\cal R}.                       
\end{eqnarray*}
As a consequence, every function  $ f \in {\cal R}$ has the form $f = t_\eta \tilde{f}$ 
with $\eta > 0, \tilde{f} \in {\cal R}$.

\textbf{Statement 2:}  The Hilbert transform $H$ is an endomorphism of ${\cal R}$, 
        and the corresponding endomorphism of ${\cal L}$ is given by the multiplication
        with the sign and $-i$,  
    \begin{displaymath}
      \left( {\cal F}^{-1} H {\cal F} \right) (f) (t) \quad = 
                   \quad -i \quad  sign(t) f(t). 
     \end{displaymath}             
     One consequence of this equality together with the isometry of the Fourier transform
     is that the Hilbert transform, too, is isometric with respect to the scalar product
     of quadratically integrable functions.         

\textbf{Statement 3:} The maps $\delta_{x_0} = T_{x_0} \delta T_{-x_0}, x_0 \in 
     \mathbb{R}$, are endomorphisms of ${\cal R}$. In particular, the endomorphism 
     ${\cal F}^{-1}  \delta  {\cal F}$ of exponentially decaying functions
     is given by   
      \begin{eqnarray*}
      ( {\cal F}^{-1}  \delta  {\cal F} ) ( \alpha ) (x_0) &=& 
             -i \int_{x_0}^{\infty} dx \alpha (x) \quad \mbox{for } x_0 > 0,\nonumber \\  
      ( {\cal F}^{-1}  \delta  {\cal F} ) ( \alpha ) (x_0) &=& 
             i \int_{-\infty}^{x_0} dx \alpha (x) \quad \mbox{for } x_0 < 0. \nonumber \\
      \end{eqnarray*}

\subsection*{\textbf{Proof of statement 1:}}
Statement (1a) is verified easily by the convolution theorem for Fourier transforms. 
In order to show statement (1b), let
\begin{eqnarray*}
\alpha   \quad &:=& \quad {\cal F}^{-1} f, \\ 
\beta   \quad &:=& \quad {\cal F}^{-1} g. \\
\end{eqnarray*} 
Define 
\begin{displaymath}
G_\varepsilon (x) \quad := \quad \frac{1}{\varepsilon \sqrt{\pi} } \quad 
                    e^{-(x/\varepsilon)^2}, 
\end{displaymath}
choose a sequence $\varepsilon_n \to 0 \quad (n \to \infty)$, let
\begin{eqnarray*}
\alpha_n   \quad &:=& \quad \alpha * G_{\varepsilon_n}, \\ 
\beta_n   \quad &:=& \quad \beta * G_{\varepsilon_n}, \\
\end{eqnarray*}
and 
\begin{eqnarray*}
f_n   \quad &:=& \quad {\cal F} \alpha_n \quad = \quad f \gamma_n, \\ 
g_n   \quad &:=& \quad {\cal F} \beta_n \quad = \quad g \gamma_n 
\end{eqnarray*}
with $\gamma_n (x) = \exp \left( -\frac{1}{4} (\varepsilon_n x)^2 \right) $.
For every single $x \in \mathbb{R}$, the equation
\begin{equation}
f * g (x) = \lim_{n \to \infty} f_n * g_n (x)
\label{label one of dualism-lemma}
\end{equation}
holds (Lebesgue). On the other hand, $f_n * g_n$ is integrable and 
quadratically integrable, and hence
\begin{displaymath}
f_n * g_n \quad = \quad  2 \pi \quad   {\cal F}   
                      ( \alpha_n S \beta_n ) \quad \in {\cal R}. 
\end{displaymath}
For any $x\in \mathbb{R}$ the following equation holds: 
\begin{equation}
f_n * g_n (x) \quad = 2 \pi  \int_\mathbb{R} dy \quad e^{-ixy} 
       \alpha_n (y)  \beta_n (-y).  
\label{label 2 of dualism-lemma}
\end{equation}
Now, note that
\begin{eqnarray*}
f_n \quad = \quad f \gamma_n   \quad &\to & \quad f, \\
g_n \quad = \quad g \gamma_n   \quad &\to & \quad g 
\end{eqnarray*}
in $|| \quad ||_2$,  the norm of the quadratically integrable functions. 
The isometry of the Fourier transformation implies that also 
\begin{eqnarray*}
\alpha_n   \quad &\to & \quad \alpha, \\
\beta_n    \quad &\to & \quad \beta 
\end{eqnarray*}
in this norm. As a consequence, the product 
\begin{displaymath}
\alpha_n S \beta_n \quad \to \quad \alpha S \beta \quad (n \to \infty) 
\end{displaymath}
in the norm $|| \quad ||_1$ of the integrable functions. This implies the 
following convergence of the right hand side of Eq. 
(\ref{label 2 of dualism-lemma}):
\begin{displaymath}
2 \pi  \int_\mathbb{R} dy \quad e^{-ixy}  \alpha_n (y)  \beta_n (-y)
\quad \to \quad 2 \pi \quad {\cal F} (\alpha S \beta) (x) 
\quad (n \to \infty).
\end{displaymath} 
With Eq. (\ref{label one of dualism-lemma}) and the definition of $\alpha, 
\beta$ follows
\begin{displaymath}
f * g (x) \quad = \quad 2 \pi \quad {\cal F} 
                ( ({\cal F}^{-1} f) (S {\cal F}^{-1} g )) (x) ,
\end{displaymath}
and thus the statement (1b).

\subsection*{\textbf{Proof of statements 2,3:}}
Let $g = {\cal F} \beta$ be an arbitrary function in ${\cal R}$.
Consider the image of $g$ under the Hilbert transform $H$ as well
as under $\delta$, and consider also ${\cal F}^{-1} Hg = 
{\cal F}^{-1} H {\cal F} \beta$ as well as  ${\cal F}^{-1} \delta g = 
{\cal F}^{-1} \delta {\cal F} \beta$.

To this end, write 
\begin{displaymath}
g \quad = \quad t_\eta f \quad = \quad f * l_\eta
\end{displaymath}
 with $\eta > 0, f \in {\cal R}$. Then,
 \begin{eqnarray*}
 H g \quad &=&  \quad -  f * (H l_\eta)  \quad \in {\cal R}, \\
 \delta g \quad &=&  \quad \frac{1}{-2i} \left\lbrace  (f l_\eta ) * 
 {\cal F} r_\eta \quad + \quad (f ( {\cal F} r_\eta )) * l_\eta 
 \right\rbrace \quad \in {\cal R}.
 \end{eqnarray*}
Moreover, according to statement (1b): 
 \begin{eqnarray*}
 \beta :=\quad  {\cal F}^{-1} g \quad &=&  \quad d_\eta \quad 
 {\cal F}^{-1} f, \\ 
 {\cal F}^{-1} H {\cal F} \quad \beta \quad &=&  \quad -i \quad r_\eta \quad  
 {\cal F}^{-1} f  \quad =  \quad -i \quad sign \quad  \beta .
 \end{eqnarray*}
Finally, with statements (1a) and (1b):
\begin{displaymath}
({\cal F}^{-1} \delta {\cal F}) \beta  = \quad \frac{-i d_\eta}{2} \left\lbrace
\quad sign \left( ({\cal F}^{-1} f) * d_\eta \right) \quad + \quad 
({\cal F}^{-1} f) * (d_\eta  sign) \quad \right\rbrace . 
\end{displaymath}
For $x_0 > 0$ one obtains:
\begin{eqnarray*}
\left\lbrace ({\cal F}^{-1} \delta {\cal F}) \beta \right\rbrace (x_0) 
\quad &=& \quad - i e^{-\eta x_0} \quad \int_{x_0}^\infty dx \quad 
({\cal F}^{-1} f) (x) e^{-\eta ( x - x_0) }  \\ 
&= & \quad -i \int_{x_0}^\infty dx \beta (x), 
\end{eqnarray*}
the case $x_0 < 0 $ is analogous.

\subsection*{Lemma (extension of the Hilbert transform):}
Let $g \in {\cal R}, \varepsilon_1, \dots , \varepsilon_N \in \mathbb{R}$, 
$ g = t_\eta f$ with $\eta > 0, f \in {\cal R}$.
Then the limit
\begin{displaymath}
(H_{ext} g) \quad (\varepsilon_1, \dots, \varepsilon_N ) \quad := \quad 
\lim_{\lambda_N \to 0^+} \dots \lim_{\lambda_1 \to 0^+} 
\end{displaymath}
\begin{displaymath}
\quad \quad \int_{\mathbb{R}}
d \omega \quad g(\omega) \quad \frac{1}{\lambda_1 + i (\omega - \varepsilon_1)}
\dots \frac{1}{\lambda_N + i (\omega - \varepsilon_N)}
\end{displaymath}
exists, and it is given by 
\begin{eqnarray*}
(H_{ext} g)(\varepsilon_1, \dots, \varepsilon_N ) &=& \quad 
(-i)^{N-1} \pi ( 1 - iH )  ( \delta_{\varepsilon_{N-1}} \dots 
\delta_{\varepsilon_1}    g) (\varepsilon_N) \\ &= & \quad 
\int_0^\infty  \dots \int_0^\infty dt_1 \dots dt_N \\ && \quad 
\quad \quad  {\cal F} g (t_1 + \dots + t_N ) \exp i 
(\varepsilon_1 t_1 + \dots + \varepsilon_N t_N ) \\
&=& \quad \int_{\mathbb{R}}
d \omega \quad f(\omega) \quad \frac{1}{\eta + i (\omega - \varepsilon_1)}
\dots \frac{1}{\eta + i (\omega - \varepsilon_N)}.
\end{eqnarray*}

\subsection*{Proof:}
Indeed, 
\begin{displaymath}
\lim_{\lambda \to 0^+} \int_{\mathbb{R}} d\omega \frac{g(\omega)}
        {\lambda + i( \omega - \varepsilon)} \quad = \quad 
        \pi ( 1 - iH ) (g) (\varepsilon). 
\end{displaymath}
In case $N \ge 2$, write 
\begin{displaymath}
 g(\omega) \quad = \quad   (\delta_{\varepsilon_1} g)(\omega) 
 (\omega - \varepsilon_1) + g(\varepsilon_1)
\end{displaymath}
and apply the convergence theorem and the residue calculus. Repeated use 
of these manipulations yields
\begin{displaymath}
(H_{ext} g) (\varepsilon_1, \dots, \varepsilon_N ) \quad = \quad
 \lim_{\lambda_N \to 0^+} \dots \lim_{\lambda_1 \to 0^+} 
\end{displaymath}
\begin{displaymath}
\int d \omega \quad \delta_{\varepsilon_1} g (\omega ) \frac{\omega - \varepsilon_1}
{\lambda_1 + i ( \omega - \varepsilon_1) } 
\frac{1}{\lambda_2 + i (\omega - \varepsilon_2)}
\dots \frac{1}{\lambda_N + i (\omega - \varepsilon_N)} \quad = \dots =
\end{displaymath}
\begin{displaymath}
\frac{1}{i^{N-1}} \lim_{\lambda_N \to 0^+} \int d\omega ( \delta_{\varepsilon_{N-1}}
\dots \delta_{\varepsilon_1} g) (\omega) \frac{1}
{\lambda_N + i ( \omega - \varepsilon_N)}
\end{displaymath}
\begin{displaymath}
= \quad (-i) ^{N-1} \pi ( 1 - iH ) ( \delta_{\varepsilon_{N-1}}
\dots \delta_{\varepsilon_1} g ) ( \varepsilon_N).
\end{displaymath}

For the proof of the second possible representation of of $H_{ext}g$ 
with arbitrary $g \in {\cal R} $, define $ \gamma_n (x) := \exp 
\left( -1/4 (x/n)^2 \right) $ and 
\begin{displaymath}
g_n \quad := \quad \gamma_n g. 
\end{displaymath}
Note that 
\begin{displaymath}
H_{ext} g ( \varepsilon_1, \dots, \varepsilon_N ) \quad = \quad 
\lim_{\lambda_N \to 0^+} \dots \lim_{\lambda_1 \to 0^+} \quad
\lim_{n \to \infty}
\end{displaymath}
\begin{displaymath}
\quad \quad \int d \omega g_n (\omega ) \quad  \frac{1}
{\lambda_1 + i ( \omega - \varepsilon_1)  } \dots  \frac{1}
{\lambda_N + i ( \omega - \varepsilon_N)  }
\end{displaymath}
\begin{displaymath}
= \mbox{[limits]} \quad \int_0^\infty \dots \int_0^\infty dt_1 \dots dt_N 
\quad {\cal F}g_n ( t_1 + \dots + t_N )  
\end{displaymath}
\begin{displaymath}
\quad \quad e^{-\lambda_1 t_1} \dots    e^{-\lambda_N t_N}
\exp i( \varepsilon_1 t_1 + \dots + \varepsilon_N t_N ) \quad \mbox{(with Fubini)} 
\end{displaymath}
\begin{displaymath}
= \mbox{[limits]} \quad \int_0^\infty dt \quad {\cal F} g_n (t) 
 \int_{ \left\lbrace 
(t_1, \dots, t_N) \in \mathbb{R}^N: t_i \ge 0, \sum_i t_i = t \right\rbrace }  
\end{displaymath}
\begin{displaymath}
\quad \quad e^{-\lambda_1 t_1} \dots    e^{-\lambda_N t_N}
\exp i( \varepsilon_1 t_1 + \dots + \varepsilon_N t_N ) 
\end{displaymath}
\begin{displaymath}
=: \mbox{[limits]} \quad \int_0^\infty dt \quad ({\cal F} g_n) (t) \mu (t)
\end{displaymath}
\begin{displaymath}
=   \quad \lim_{\lambda_N \to 0^+} \dots \lim_{\lambda_1 \to 0^+}
\quad \int_0^\infty dt \quad ({\cal F} g) (t) \mu (t) \quad \mbox{(continuity in }
||\quad ||_2\mbox{)} 
\end{displaymath}
\begin{displaymath}
= \quad \lim_{\lambda_N \to 0^+} \dots \lim_{\lambda_1 \to 0^+} \quad 
\int_0^\infty  \dots \int_0^\infty dt_1 \dots dt_N
\quad  {\cal F} g (t_1 + \dots + t_N )
\end{displaymath}
\begin{displaymath}
\quad \quad \exp i (\varepsilon_1 t_1 + \dots + \varepsilon_N t_N )
\quad e^{-\lambda_1 t_1} \dots e^{-\lambda_N t_N}. 
\end{displaymath}
\begin{displaymath}
= \quad \int_0^\infty  \dots \int_0^\infty dt_1 \dots dt_N 
\quad  {\cal F} g (t_1 + \dots + t_N ) \exp i 
(\varepsilon_1 t_1 + \dots + \varepsilon_N t_N )  
\end{displaymath}
with Lebesgue.

The lemma's third representation of $H_{ext} g $ can be obtained with 
Tonelli/Fubini and the residue calculus, starting directly from the definition. 
\begin{displaymath}
\quad 
\end{displaymath}

\section{Application: Kernels up to sixth order, neglecting double occupancy}

\begin{displaymath}
\quad
\end{displaymath}
I shall assume that the situation is symmetric with respect to the spins, 
$E_\sigma = E_{\bar{\sigma}}$.
An rigorous application of the theory would imply that the stationary reduced 
density matrix $\rho$ is obtained from the quantum master equation in the stationary limit:
\begin{displaymath}
\quad 
\end{displaymath}
\begin{displaymath}
   \left( \begin{array}{cccc}  	
\dots            	&   L_{\sigma \sigma}^{00} 	&  L_{\bar {\sigma} \bar{\sigma}}^{00}  
& L_{22}^{00}       	\\
   L_{00}^{\sigma \sigma}  	&        \dots     	&    L_{\bar{\sigma} \bar{\sigma}}^{\sigma \sigma}  		
  &      L_{22}^{\sigma \sigma} 	\\
    L_{00}^{\bar{\sigma}\bar{ \sigma}}  	&     L_{\sigma \sigma}^{\bar{\sigma} \bar{\sigma}} 		
  &     \dots     &      L_{22}^{\bar{\sigma} \bar{\sigma}}   	\\  
    L_{00}^{22}         	&   L_{\sigma \sigma}^{22} 	&  L_{\bar{\sigma} \bar{\sigma}}^{22}  
&  \dots       	\\  
\end{array} \right)
  \left( \begin{array}{c}  	
\rho_{00}      \\  \rho_{\sigma \sigma}  \\   \rho_{  \bar{\sigma} \bar{\sigma}   }  \\  
    \rho_{22}
\end{array} \right)  \quad = \quad 0, 
\end{displaymath} 
\begin{displaymath}
\quad 
\end{displaymath}
where $L (w) = \frac{1}{w^2} K (w)$, $w$ the coupling parameter, $K = K (\lambda = 0)$ 
the density matrix kernel, and 
\begin{displaymath}
\quad 
\end{displaymath}
\begin{displaymath}
L_{aa}^{bb} \quad := \quad <b, L ( |a><a|) b>
\end{displaymath}
\begin{displaymath}
\quad 
\end{displaymath} 
for quantum dot states $a,b \in \left\lbrace 0, \sigma, \bar{\sigma}, 2 \right\rbrace$.
The normalized solution is 
\begin{displaymath}
\quad 
\end{displaymath}
\begin{displaymath}
 \left( \begin{array}{c}  	
\rho_{00}      \\  \rho_{\sigma \sigma}  \\   \rho_{  \bar{\sigma} \bar{\sigma}   }  \\  
    \rho_{22}
\end{array} \right)  \quad = \quad   \left( \begin{array}{c}  	
\lambda_N     \\  \mu_Z  \\   \mu_Z  \\  \lambda_Z
\end{array} \right) \frac{1}{N}
\end{displaymath}
with 
\begin{displaymath}
N \quad := \quad \lambda_N + \mu_Z + \mu_Z + \lambda_Z
\end{displaymath}
and
\begin{displaymath}
\quad 
\end{displaymath}
\begin{eqnarray*}
\lambda_Z \quad &:=& \quad  L_{\sigma \sigma}^{00} L_{00}^{22} \quad + \quad
                            L_{\sigma \sigma}^{22} L_{00}^{22} \quad + \quad
                          2 L_{\sigma \sigma}^{22} L_{00}^{\sigma \sigma}, \\
\mu_Z \quad &:=& \quad      L_{00}^{22} L_{22}^{\sigma \sigma} \quad + \quad
                            L_{22}^{00} L_{00}^{\sigma \sigma} \quad + \quad
                          2 L_{00}^{\sigma \sigma} L_{22}^{\sigma \sigma}, \\      
\lambda_N \quad &:=& \quad  L_{22}^{00} L_{\sigma \sigma}^{22} \quad + \quad
                            L_{22}^{00} L_{\sigma \sigma}^{00} \quad + \quad
                          2 L_{\sigma \sigma}^{00} L_{22}^{\sigma \sigma} .                                              
\end{eqnarray*}
\begin{displaymath}
\quad 
\end{displaymath}

The particle current onto the lead $l$ is  
\begin{displaymath}
\quad 
\end{displaymath}
\begin{eqnarray*}
I_l \quad &=& \quad \rho_{00} \quad Tr\left\lbrace K_c (|0><0|) \right\rbrace  
\quad +  \quad  \sum_\sigma \rho_{\sigma \sigma} \quad Tr\left\lbrace K_c (|\sigma><\sigma|)
 \right\rbrace  \\ 
\quad &+& \quad \rho_{22} \quad Tr\left\lbrace K_c (|2><2|) \right\rbrace , 
\end{eqnarray*}
\begin{displaymath}
\quad 
\end{displaymath}
where $K_c = K_{c,l} $ is the current kernel of lead $l$. It would now be consistent
with the theoretical part of this text to replace the complete kernels by their 
Taylor expansions up to sixth order in the coupling parameter. 
\begin{displaymath}
\quad 
\end{displaymath}

However, I did not or not yet calculate any diagrams of sixth order which include both 
the states $0$ and $2$. The purpose of this section is to motivate a further 
and more rigorous study of the sixth order, so I shall use the following approximation
scheme: Replace the value of all diagrams in which the state $2$ appears by zero 
and assume that both the density matrix kernel and the current kernel are given only by 
the remaining diagrams. Moreover, assume that the probability of double occupancy 
is zero, $\rho_{22} = 0$. The quantum master equation in the stationary limit turns 
then into
\begin{displaymath}
\quad 
\end{displaymath} 
\begin{displaymath}
   \left( \begin{array}{cccc}  	
\dots            	&   L_{\sigma \sigma}^{00} 	&  L_{\bar {\sigma} \bar{\sigma}}^{00}  
& 0       	\\
   L_{00}^{\sigma \sigma}  	&        \dots     	&    L_{\bar{\sigma} \bar{\sigma}}^{\sigma \sigma}  		
  &      0 	\\
    L_{00}^{\bar{\sigma}\bar{ \sigma}}  	&     L_{\sigma \sigma}^{\bar{\sigma} \bar{\sigma}} 		
  &     \dots     &    0   	\\  
    0        	&   0 	&  0  &  \dots       	\\  
\end{array} \right)  
  \left( \begin{array}{c}  	
\rho_{00}      \\  \rho_{\sigma \sigma}  \\   \rho_{  \bar{\sigma} \bar{\sigma}   }  \\  0
\end{array} \right)  \quad = \quad 0, 
\end{displaymath}
\begin{displaymath}
\quad 
\end{displaymath}
the normalized solution is 
\begin{displaymath}
\quad 
\end{displaymath}
\begin{displaymath}
 \left( \begin{array}{c}  	
\rho_{00}      \\  \rho_{\sigma \sigma}  \\   \rho_{  \bar{\sigma} \bar{\sigma}   } 
\end{array} \right)  \quad = \quad   \left( \begin{array}{c}  	
  L_{\sigma \sigma}^{00}     \\  L_{00}^{\sigma \sigma}  \\   L_{00}^{\sigma \sigma}  
\end{array} \right) \frac{1}{L_{\sigma \sigma}^{00}  +  2 L_{00}^{\sigma \sigma} }.
\end{displaymath}
\begin{displaymath}
\quad 
\end{displaymath}

Moreover, if all diagrams which contain the state $2$ are neglected, then the following 
relations between the current kernels $K_{c,l}$ and the density matrix kernel $K$
hold: If the notation
\begin{displaymath}
\quad 
\end{displaymath}
\begin{eqnarray*}
\Gamma_l^+ \quad &:=& \quad  -\frac{1}{2} Tr \left\lbrace K_{c,l} (|0><0|) \right\rbrace, \\
\Gamma_l^- \quad &:=& \quad   Tr \left\lbrace K_{c,l} (|\sigma >< \sigma |) \right\rbrace
\end{eqnarray*} 
and 
\begin{eqnarray*}
\Gamma^+ \quad &:=& \quad \sum_l \Gamma_l^+,  \\
\Gamma^- \quad &:=&  \quad  \sum_l \Gamma_l^-
\end{eqnarray*}
\begin{displaymath}
\quad 
\end{displaymath}
is used, then:
\begin{displaymath}
\quad 
\end{displaymath} 
\begin{eqnarray*}
\Gamma^+ \quad &=& \quad K_{00}^{\sigma \sigma}, \\
\Gamma^- \quad &=&  \quad  K_{\sigma \sigma}^{00}.
\end{eqnarray*}
\begin{displaymath}
\quad 
\end{displaymath}

The density matrix and the current read in terms of the rates $\Gamma_l^\pm $:
\begin{displaymath}
\quad 
\end{displaymath}
\begin{displaymath}
 \left( \begin{array}{c}  	
\rho_{00}      \\  \rho_{\sigma \sigma}  \\   \rho_{  \bar{\sigma} \bar{\sigma}   } 
\end{array} \right)  \quad = \quad   \left( \begin{array}{c}  	
  \Gamma^-     \\  \Gamma^+  \\   \Gamma^+  
\end{array} \right) \frac{1}{ \Gamma^-  +  2 \Gamma^+ },
\end{displaymath}
\begin{displaymath}
\quad 
\end{displaymath}
and 
\begin{displaymath}
\quad 
\end{displaymath}
\begin{equation}
I_l \quad = \quad  \rho_{00} \left( \Gamma_{\bar {l}}^+ -  \Gamma_l^+  \right)
          \quad + \quad \rho_{\sigma \sigma}
           \left( \Gamma_l^- -  \Gamma_{\bar {l}}^-  \right).
\label{current in terms of rates}
\end{equation}
\begin{displaymath}
\quad 
\end{displaymath}

The exact values of the rates $\Gamma_l^\pm $ have not been calculated, but the sums of the 
contributions of all diagrams within certain diagram selections could be determined. 
One diagram selection is the dressed second order (DSO) \cite{Kern epjb}. The value of the rates 
obtained within this selection are the following: 
\begin{displaymath}
\quad
\end{displaymath}
\begin{displaymath}
\Gamma_l^+ (DSO) \quad = \quad \frac{2 \pi}{\hbar} \int d\varepsilon \quad 
\frac{\left((\alpha + \alpha ^+) \alpha_l^+ \right)(\varepsilon) }
{\left( \pi (\alpha + \alpha^+) \right)^2 (\varepsilon) + ( \varepsilon - E_{10}
+ p_{\alpha + \alpha^+} (\varepsilon) )^2 }
\end{displaymath}
\begin{displaymath}
\quad 
\end{displaymath}   
\begin{displaymath} 
\quad \quad \quad \quad \quad \quad = \quad  \frac{2 \pi}{\hbar} \int 
    d\varepsilon \quad \left\lbrace |\Pi |^2 
     (\alpha + \alpha ^+) \alpha_l^+  \right\rbrace       (\varepsilon)
\end{displaymath}
\begin{displaymath}
\quad 
\end{displaymath}
with
\begin{displaymath}
\Pi (\varepsilon) \quad := \quad \frac{1}{ \pi (\alpha + \alpha^+)  (\varepsilon) \quad + 
\quad i( \varepsilon - E_{10}  + p_{\alpha + \alpha^+} (\varepsilon) )} ,
\end{displaymath}
\begin{displaymath}
\quad 
\end{displaymath}
\begin{displaymath}
p_{f} (\varepsilon) \quad := \quad \int_0^\infty d\omega 
\quad \frac{f (\varepsilon + \omega ) - f (\varepsilon - \omega) }{\omega} \quad 
= \quad \pi (Hf)( \varepsilon ), 
\end{displaymath}
\begin{displaymath}
\quad 
\end{displaymath} 
$H$ the Hilbert transform; $E_{10} := E_\sigma - E_0$. 
The rate $\Gamma_l^-(DSO)$ is
\begin{displaymath}
\quad 
\end{displaymath}
\begin{displaymath}
\Gamma_l^- (DSO) \quad = 
 \quad  \frac{2 \pi}{\hbar} \int   \left\lbrace |\Pi |^2 
     (\alpha + \alpha ^+) \alpha_l^-  \right\rbrace, 
\end{displaymath}
\begin{displaymath}
\quad 
\end{displaymath}
where the integration variable and the measure have been omitted for simplicity. 
\begin{displaymath}
\quad 
\end{displaymath}

A diagram selection which has been considered earlier and which contains the DSO 
selection is the resonant tunneling approximation (RTA) \cite{König96, Schoeller97}.  
The rates of the RTA selection were obtained by solving an integral equation, 
they are 
\begin{displaymath}
\quad 
\end{displaymath}
\begin{eqnarray*}
\Gamma_l^+ (RTA) \quad &=& \quad \Gamma_l^+ (DSO) \\ && + \quad \frac{2 \pi}{\hbar} 
\quad \frac{\int   \left\lbrace |\Pi |^2 (\alpha_l + \alpha_l^+)  
   \right\rbrace \quad   \int   \left\lbrace |\Pi |^2 \alpha ^+  \right\rbrace}
             {\int   \left\lbrace |\Pi |^2  \right\rbrace } \\ && - \quad 
    \frac{2 \pi}{\hbar} \quad \int   \left\lbrace |\Pi |^2 
    (\alpha_l + \alpha_l^+) \alpha^+    \right\rbrace ,  
\end{eqnarray*}
\begin{displaymath}
\quad 
\end{displaymath}
\begin{eqnarray*}
\Gamma_l^- (RTA) \quad &=& \quad \Gamma_l^- (DSO) \\ && + \quad \frac{2 \pi}{\hbar} 
\quad \frac{\int   \left\lbrace |\Pi |^2 (\alpha_l + \alpha_l^+)  
   \right\rbrace  \quad \int   \left\lbrace |\Pi |^2 \alpha ^-  \right\rbrace}
             {\int   \left\lbrace |\Pi |^2  \right\rbrace } \\ && - \quad 
    \frac{2 \pi}{\hbar} \quad \int   \left\lbrace |\Pi |^2 
    (\alpha_l + \alpha_l^+) \alpha^-    \right\rbrace . 
\end{eqnarray*}
\begin{displaymath}
\quad 
\end{displaymath}

In this section, the Taylor expansions of the exact rates
$\Gamma_l^\pm$ up to sixth order in the coupling are 
presented and discussed. Let the coefficients 
be $\Gamma_l^\pm (2n)$. At first, note the Taylor expansions 
of the rates of the DSO  and of the RTA. For $SEL = RTA, DSO$, let
$\Gamma_l^\pm (SEL)(2n)$ be the expansion coefficients of 
 $\Gamma_l^\pm (SEL)$, and define

\begin{displaymath}
\quad 
\end{displaymath}
\begin{displaymath}
\Gamma_l^\pm (SEL,III) \quad := \quad \Gamma_l^\pm (SEL) (2)  +  
\Gamma_l^\pm (SEL) (4)  +  \Gamma_l^\pm (SEL) (6).  
\end{displaymath}
\begin{displaymath}
\quad 
\end{displaymath}

\subsection{Sixth order DSO rates and discussion}

\begin{displaymath}
\quad 
\end{displaymath}
Take into account the DSO rates up to sixth order and discuss
the existence of the zero temperature limit of current and density matrix on the one 
hand, and the temperature dependence of the linear conductance for $T \to 0$
on the other hand.

The expansion coefficients $\Gamma_l^\pm (DSO) (2n)$ can be obtained by App. B or, 
alternatively, by direct calculation of the corresponding diagrams. The results 
for $n = 1, 2, 3$ are:
\begin{displaymath}
\quad 
\end{displaymath} 
\begin{displaymath}
\Gamma_l^\pm (DSO)(2) \quad = \quad \frac{2  }{\hbar} \quad  
                                    \pi  \alpha_l^\pm (E_{10}),
\end{displaymath}
\begin{displaymath}
\quad 
\end{displaymath}
\begin{displaymath}
\Gamma_l^\pm (DSO)(4) \quad = \quad \frac{2 }{\hbar} \quad \pi^2 \left\lbrace
               H ( \alpha_l^\pm ( \alpha + \alpha^+ ) ) \quad - \quad 
               \alpha_l^\pm  H ( \alpha + \alpha^+ ) \right\rbrace ' (E_{10})  
\end{displaymath}               
\begin{displaymath}
\quad 
\end{displaymath}
\begin{displaymath}
 \quad \quad  \quad \quad \quad \quad \quad = \quad \frac{2 }{\hbar} \quad 
               Bil (DSO) ( \alpha_l^\pm, \alpha + \alpha^+ ),  
\end{displaymath}
\begin{displaymath}
\quad 
\end{displaymath}
where the bilinear map $Bil (DSO): {\cal R} \times {\cal R} \rightarrow \mathbb{C}$
is given by the [real part of the] fourth order DSO diagram, 
\begin{displaymath}
\quad 
\end{displaymath}
\begin{displaymath}
Bil (DSO) (f, g) \quad = \quad \pi^2 \left\lbrace
    H ( f  g ) \quad - \quad   f  (H  g ) \right\rbrace ' (E_{10}).
\end{displaymath}
\begin{displaymath}
\quad 
\end{displaymath}

Finally, the sixth order contribution is
\begin{displaymath}
\quad
\end{displaymath} 
\begin{eqnarray*}
\Gamma_l^\pm (DSO)(6) \quad &=& \quad \frac{2 }{\hbar} \quad  \pi^3 
\begin{LARGE} \textbf{\{} \end{LARGE} \quad \quad  
     \alpha_l^\pm   H \begin{LARGE}
     \textbf{[}\end{LARGE}  ( \alpha + \alpha^+ ) 
      H( \alpha + \alpha^+ ) 
       \begin{LARGE}\textbf{]} \end{LARGE}   
       \\ &&   \quad \quad \quad \quad \quad  - \quad H 
       \begin{LARGE} \textbf{[} \end{LARGE}   
        \alpha_l^\pm       
       ( \alpha + \alpha^+ )   
      H( \alpha + \alpha^+ )   \begin{LARGE} \textbf{]} \end{LARGE} \quad    
     \begin{LARGE} \textbf{\}} \end{LARGE}  '' (E_{10} ) 
\end{eqnarray*}     
\begin{displaymath}
\quad 
\end{displaymath}     
\begin{displaymath}    
 \quad \quad \quad \quad \quad \quad \quad     = \quad \frac{2 }{\hbar} \quad 
      Tril (DSO) ( \alpha_l^\pm, \alpha + \alpha^+, \alpha + \alpha^+ ), 
\end{displaymath}
\begin{displaymath}
\quad 
\end{displaymath}
where $Tril (DSO)$ is the trilinear map ${\cal R} \times {\cal R} 
\times {\cal R} \rightarrow \mathbb{C} $ given by the sixth order DSO diagram
\begin{displaymath}
\quad 
\end{displaymath}
\begin{eqnarray*}
Tril(DSO) (f,g,h) \quad &=& \quad \frac{\pi^3}{2} \quad \begin{LARGE} 
      \textbf{\{} \end{LARGE} \quad f (Hg) (Hh) \quad - \quad f g h \\
      && \quad \quad \quad \quad - \quad H   \begin{LARGE} 
      \textbf{[} \end{LARGE} f g (Hh) + f (Hg) h \begin{LARGE}   \textbf{]} 
      \end{LARGE} \quad \begin{LARGE}   \textbf{\}} \end{LARGE}''(E_{10}).
\end{eqnarray*}  
\begin{displaymath}
\quad 
\end{displaymath}

\subsection*{\textbf{Existence of $\lim_{T \to 0} \Gamma_l^\pm (DSO, III)$} }
\begin{displaymath}
\quad 
\end{displaymath}
Assume that 
\begin{displaymath}
\mu_l, \mu_{\bar{l}} \quad \neq \quad E_{10}.
\end{displaymath}

For the discussion of the zero temperature limit make
several general notes:
\begin{displaymath}
\quad 
\end{displaymath}
\begin{itemize}

\item If $g \in {\cal R}, n \in \mathbb{N} , \varepsilon \in \mathbb{R}$, then
\begin{displaymath}
\quad 
\end{displaymath}
\begin{displaymath}
( \delta_\varepsilon ^n g ) (\varepsilon ) \quad = \quad \frac{1}{n!} 
  g^{(n)} (\varepsilon ) , 
\end{displaymath}
\begin{displaymath}
\quad 
\end{displaymath}
and for general $x \in \mathbb{R}$: 
\begin{displaymath}
\quad 
\end{displaymath}
\begin{displaymath}
( \delta_\varepsilon ^n g ) (x) \quad = \quad \frac{g(x) - p(g,\varepsilon )_{n-1} (x)}
   {(x - \varepsilon) ^n}, 
\end{displaymath}
\begin{displaymath}
\quad 
\end{displaymath}
where $p(g, \varepsilon )_{n-1}$ is the Taylor polynomial of $g$ around $\varepsilon$ 
of degree $n-1$, in particular $p(g, \varepsilon )_{-1} := 0$.  
\begin{displaymath}
\quad 
\end{displaymath}

\item As a consequence, the $n$-th derivative of the Hilbert transform of $g$ has 
the representation
\begin{displaymath}
\quad 
\end{displaymath}
\begin{eqnarray*}
(H g)^{(n)} (\varepsilon ) \quad &=& \quad \frac{n!}{\pi} \quad \int_0^\infty 
\frac {d \omega}{\omega^{n+1}}  \begin{LARGE} \textbf{\{ } \end{LARGE} \quad
 ( g - p (g,\varepsilon )_{n-1} ) (\varepsilon + \omega )  \\  && \quad \quad \quad \quad \quad 
 + \quad (-1)^{n+1}  ( g - p (g,\varepsilon )_{n-1} ) (\varepsilon - \omega ) \quad
 \begin{LARGE} \textbf{\}} \end{LARGE}   \\ &=:& \quad H^{(n)} g (\varepsilon). 
\end{eqnarray*}
\begin{displaymath}
\quad 
\end{displaymath}

\item In particular 
\begin{eqnarray*}
H ^{(0)} g (\varepsilon ) \quad &=& \quad \frac{1}{\pi} \quad \int_0^\infty 
\frac {d \omega}{\omega}  \begin{LARGE} \textbf{\{ } \end{LARGE} 
  g   (\varepsilon + \omega ) - g ( \varepsilon - \omega )  
 \begin{LARGE} \textbf{\}} \end{LARGE} ,   
\end{eqnarray*}
\begin{eqnarray*}
H^{(1)} g (\varepsilon ) \quad &=& \quad \frac{1}{\pi} \quad \int_0^\infty 
\frac {d \omega}{\omega^2}  \begin{LARGE} \textbf{\{ } \end{LARGE} 
  g  (\varepsilon + \omega )  +  g (\varepsilon - \omega )  - 
   2 g (\varepsilon )    
  \begin{LARGE} \textbf{\}} \end{LARGE},    
\end{eqnarray*}
\begin{eqnarray*}
H^{(2)} g (\varepsilon ) \quad &=& \quad \frac{2}{\pi} \quad \int_0^\infty 
\frac {d \omega}{\omega^3}  \begin{LARGE} \textbf{\{ } \end{LARGE} 
  g  (\varepsilon + \omega )  -  g (\varepsilon - \omega )  - 
   2\omega g' (\varepsilon )    
  \begin{LARGE} \textbf{\}} \end{LARGE}.    
\end{eqnarray*}
\begin{displaymath}
\quad 
\end{displaymath}

\item The integrand appearing in $H^{(n)} g ( \varepsilon )$ can be represented like
\begin{displaymath}
\quad 
\end{displaymath}
\begin{displaymath}
\frac {1}{\omega^{n+1}}  \begin{LARGE} \textbf{\{ } \end{LARGE} \quad
 ( g - p (g,\varepsilon )_{n-1} ) (\varepsilon + \omega )   
\end{displaymath}
\begin{displaymath}
 \quad \quad \quad + \quad (-1)^{n+1}  ( g - p (g,\varepsilon )_{n-1} ) 
 (\varepsilon - \omega ) \quad \begin{LARGE} \textbf{\}} \end{LARGE}  
\end{displaymath}
\begin{displaymath}
\quad 
\end{displaymath}
\begin{displaymath}
= \quad \frac{1}{\omega^{n+1}} \int  \dots \int_{0 \le t_{n+1} \le 
\dots \le t_1 \le \omega} \quad dt_1 \dots dt_{n+1} 
\end{displaymath}
\begin{displaymath}
\quad \quad \quad 
\begin{LARGE} \textbf{(} \end{LARGE} \quad 
g^{(n+1)}  ( \varepsilon + t_{n+1} ) \quad   + \quad  g^{(n+1)} ( \varepsilon - t_{n+1} )  
\quad \begin{LARGE} \textbf{)} \end{LARGE}.
\end{displaymath}

\begin{displaymath}
\quad 
\end{displaymath}
\item All of the derivatives of the normalized Fermi function 
$f (x) = 1/ (1 + e^x)$ decay exponentially: For all $n \ge 1$ there is 
$K_n > 0$ such, that for all $x \in \mathbb{R}$:
\begin{displaymath}
\quad 
\end{displaymath}
\begin{displaymath}
| f^{(n)} (x) | \quad \le \quad K_n e^{-|x|} . 
\end{displaymath}
\begin{displaymath}
\quad 
\end{displaymath}
As a consequence, the following statement about the Fermi function at chemical
potential $\mu_l$ and temperature $T$, 
\begin{displaymath}
\quad 
\end{displaymath}
\begin{displaymath}
f_l (\varepsilon ) \quad := \quad f \left( \frac{\varepsilon - \mu_l}{k_B T } \right)
\end{displaymath}
\begin{displaymath}
\quad 
\end{displaymath}
holds: For arbtitrary $n \ge 0$ and $r > 0$ there is a constant $const (n, r)$ such, 
that for all $\varepsilon \in \mathbb{R}$ with $|\varepsilon - \mu_l| \ge r$, and 
 {\em independently} of $T > 0$ : 
\begin{displaymath}
\quad 
\end{displaymath} 
 \begin{displaymath}
 |f_l^{(n)} (\varepsilon )| \quad \le \quad const (n, r).  
\end{displaymath}  
\begin{displaymath}
\quad 
\end{displaymath}
In particular, the limit $\lim_{T \to 0}f_l^{(n)} (\varepsilon)$ exitsts for all $n \ge 0$
and $\varepsilon \neq \mu_l$.  
\end{itemize}
\begin{displaymath}
\quad 
\end{displaymath}

 Finally, note the following 
 \subsubsection*{\textbf{Lemma} (discussion of $\Gamma_l^\pm (DSO, III)$ and 
                                                       $G (DSO, III )$):}
                                                       
\begin{displaymath}
\quad 
\end{displaymath}                                                       
Let $\gamma \in  {\cal R}$, $m \in \{1, 2, \dots \}$, 
$\mu_1, \dots , \mu_m \in \mathbb{R}.$ For $T > 0$ 
and $j = 1, 2, \dots , m$ let 
\begin{displaymath}
f_j (\varepsilon ) \quad := \quad f \left( \frac{\varepsilon - \mu_j}{k_B T}   \right),
\end{displaymath}
and assume $n \ge 0$. Then the following statements hold:
\begin{displaymath}
\quad 
\end{displaymath}

\begin{enumerate}
\item For all $\varepsilon \in \mathbb{R} \setminus \{ \mu_1, \dots, \mu_m \}$ the limit
\begin{displaymath}
\lim_{T \to 0} \quad H^{(n)} ( \gamma f_1 \dots  f_m ) (\varepsilon ) 
\end{displaymath}
exists. Moreover, for arbitrary $r > 0$ there is a constant $c > 0$ 
{\em independent} of $T$ such, that for all $\varepsilon \in \mathbb{R}$ with
\begin{displaymath}
\min \{ | \varepsilon - \mu_j |: j = 1, \dots, m \} \quad \ge \quad r 
\end{displaymath}
the inequality 
\begin{displaymath}
|H^{(n)} ( \gamma f_1 \dots  f_m ) (\varepsilon ) | \quad \le \quad c
\end{displaymath} 
holds. 
\begin{displaymath}
\quad 
\end{displaymath} 
 
\item Let $\kappa \in \{ 0, 1, \dots \}$. 
Then for all $\varepsilon \in \mathbb{R} \setminus  \mu_1 $ the limit
\begin{displaymath}
\lim_{T \to 0} \quad H^{(n)} ( \gamma f_1' f_1^\kappa ) (\varepsilon ) 
\end{displaymath}
exists.

\item Let $\alpha \in {\cal R} $ possess the properties of the function 
$\alpha_l$ in App. A, let $\mu_0 \in \mathbb{R}$, 
\begin{displaymath}
f_0 (\varepsilon ) \quad := \quad f \left( \frac{\varepsilon - \mu_0}{k_B T}  
 \right),
\end{displaymath} 
and $ \kappa_1, \dots ,  \kappa_m \in \{ 0, 1, \dots \} $. Then the limit
\begin{displaymath}
\quad 
\end{displaymath}
\begin{displaymath}
\lim_{T \to 0} \quad    H^{(n)} \left[ \gamma f_1^{\kappa_1} \dots f_m^{\kappa_m} 
          \quad H (\alpha f_0 ) \right] (\varepsilon)
\end{displaymath}
\begin{displaymath}
\quad 
\end{displaymath}
exists for all $\varepsilon \in \mathbb{R} \setminus \{ \mu_0, \mu_1,        
\dots , \mu_m \}$. 
\begin{displaymath}
\quad 
\end{displaymath}

\item Let additionally $\kappa \in \{ 0, 1, \dots \}$ and choose $T_0 > 0$ 
arbitrarily. Then for all $\varepsilon \in \mathbb{R} \setminus \mu_0$: 
\begin{displaymath}
\quad
\end{displaymath}
\begin{displaymath}
 H^{(n)} \left[ \gamma (-f_0') f_0^\kappa   
\quad H (\alpha f_0 ) \right] \quad (\varepsilon ) 
\end{displaymath}
\begin{displaymath}
\quad 
\end{displaymath}
\begin{displaymath}
\quad \quad \equiv \quad \quad \frac{n!}{\pi^2} \quad 
\frac{(\alpha \gamma) (\mu_0)}{\kappa + 1} 
\quad 
\frac{1}{ ( \mu_0 - \varepsilon )^{n+1}}
\quad \log \left( \frac{T}{T_0} \right)  
\end{displaymath}
\begin{displaymath}
\quad 
\end{displaymath}
in the sense that the difference between these two functions of the temperature
converges as $T \to 0$; $\log := \exp^{-1}$.

\end{enumerate} 
\begin{displaymath}
\quad 
\end{displaymath}

\subsubsection*{\textbf{Proof of statement (i):}}
Let $r > 0$ and $\varepsilon \in \mathbb{R}$ with 
\begin{displaymath}
\min \{ |\varepsilon - \mu_j |: j = 1, \dots , m  \}  \quad \ge \quad r
\end{displaymath}
be given. Decompose

\begin{displaymath}
\quad 
\end{displaymath}
\begin{equation} 
H^{(n)} g (\varepsilon ) \quad =
\label{decomposition of H^n g (eps)}
\end{equation}
\begin{displaymath}
\quad 
\end{displaymath}    
\begin{displaymath}
\quad \quad  \frac{n!}{\pi} \quad 
\int_{0}^{r/2} \frac{d \omega}{\omega^{n+1}} \quad \int  \dots \int_{0 \le t_{n+1} \le 
\dots \le t_1 \le \omega}   
\end{displaymath}
\begin{displaymath}
 \quad \quad  \quad \quad \quad \quad \quad \quad \quad \quad 
\begin{LARGE} \textbf{(} \end{LARGE}  
g^{(n+1)}  ( \varepsilon + t_{n+1} ) \quad   + \quad  g^{(n+1)} ( \varepsilon - t_{n+1} )  
 \begin{LARGE} \textbf{)} \end{LARGE} 
 \end{displaymath}          
\begin{displaymath}
\quad 
\end{displaymath}      
\begin{displaymath}
 + \quad \frac{n!}{\pi} \quad \int_{r/2}^\infty 
\frac {d \omega}{\omega^{n+1}}  \begin{LARGE} \textbf{\{ } \end{LARGE} 
 - p(g, \varepsilon )_{n-1}  (\varepsilon + \omega )  
  \quad   + \quad (-1)^{n}   p(g,\varepsilon )_{n-1}  (\varepsilon - \omega ) 
 \begin{LARGE} \textbf{\}} \end{LARGE} 
\end{displaymath}  
\begin{displaymath}
\quad 
\end{displaymath}
\begin{displaymath}
 + \quad \frac{n!}{\pi} \quad \int_{r/2}^\infty 
\frac {d \omega}{\omega^{n+1}}  \begin{LARGE} \textbf{\{ } \end{LARGE} \quad
 g  (\varepsilon + \omega )  
  \quad \quad  + \quad (-1)^{n+1}   g  (\varepsilon - \omega ) \quad
 \begin{LARGE} \textbf{\}} \end{LARGE} 
\end{displaymath}      
with 
\begin{displaymath}
g \quad := \quad \gamma f_1 \dots f_m.
\end{displaymath}

The integrand of the integral over $[0, r/2]$ is pointwise convergent, which can be 
seen from the alternative representation of this integrand. Moreover, for every 
$l \in \{0, 1, \dots \}$ there is 
$c_l > 0$ such, that for all $x \in \mathbb{R}$ with 
\begin{displaymath}
\min \{|x - \mu_j|: j = 1, \dots, m \} \quad \ge \quad r/2
\end{displaymath}
the inequality 
\begin{displaymath}
|g^{(l)} (x)| \quad \le \quad c_l  
\end{displaymath}
holds independently of the value of the temperature $T$. In particular, the integrand 
of the integral over $[0, r/2]$ is bounded by 
\begin{displaymath}
\frac{2 c_{n+1}}{(n+1)!},
\end{displaymath}
so the integral is convergent for $T \to 0$ with Lebesgue. 
Moreover, note that the upper bound for its value, 
\begin{displaymath}
\frac{c_{n+1}}{n+1} \frac{r}{\pi},
\end{displaymath}
does depend on $r$, but that it is independent of the temperature $T$ and of 
the value of the initially given $\varepsilon$.  
\begin{displaymath}
\quad 
\end{displaymath}

The numerator of the second contributing integral, the first integral over
$[ r/2, \infty [$, is a polynomial in the integration variable $\omega$ 
of degree smaller or equal $n - 1$. Its coefficients are obtained from the 
derivatives of $g$ in $\varepsilon$. Hence, they are convergent with $T \to 0$, 
and upper  bounds for their absolute value are obtained from the above 
chosen $c_l, l = 0, \dots, n-1$. 
\begin{displaymath}
\quad 
\end{displaymath}

What remains to be shown is the convergence of the rest contribution
\begin{displaymath}
\quad \frac{n!}{\pi} \quad \int_{r/2}^\infty \quad 
\frac {d \omega}{\omega^{n+1}}  \quad \textbf{\{ }  
  g  (\varepsilon + \omega ) \quad  + \quad (-1)^{n+1}   g  
 (\varepsilon - \omega ) \textbf{\}}
\end{displaymath}
\begin{displaymath}
= \quad \frac{n!}{\pi} \quad \int_\mathbb{R} \quad 
d \omega  \quad   g  (\varepsilon + \omega ) \quad 
\frac{\textbf{1}_{\mathbb{R}\setminus B_{r/2}(0)} (\omega) }
  {\omega ^{n+1}}
\end{displaymath}
with $T \to 0$ and the existence of an upper bound which is independent 
of the temperature $T$ and of $\varepsilon$. However, the integrand of the latter
integral is pointwise convergent, an integrable upper bound of the integrand is
given by  
\begin{displaymath}
   |\gamma  (\varepsilon + \omega )| \quad 
\frac{\textbf{1}_{\mathbb{R}\setminus B_{r/2}(0)} (\omega) }
  {|\omega | ^{n+1}},
\end{displaymath}
and so an upper bound for the integral independent of temperature and the 
value of $\varepsilon$ is obtained by the Cauchy-Schwarz inequality, 
\begin{displaymath}
 \quad \left| 
 \frac{n!}{\pi} \quad \int_\mathbb{R} \quad 
d \omega  \quad   g  (\varepsilon + \omega ) \quad 
\frac{\textbf{1}_{\mathbb{R}\setminus B_{r/2}(0)} (\omega) }
  {\omega ^{n+1}} \right|
\end{displaymath}
\begin{displaymath}
\le \quad \frac{n!}{\pi} \quad || \gamma ||_2 \quad || \omega \mapsto 
\frac{\textbf{1}_{\mathbb{R}\setminus B_{r/2}(0)} (\omega) }
  {\omega ^{n+1}}  ||_2 .
\end{displaymath}

\subsubsection*{\textbf{Proof of statement (ii):}}
Let $\varepsilon \in \mathbb{R} \setminus \mu_1 $ be given. Let
\begin{displaymath}
r \quad := \quad  |\mu_1  - \varepsilon |
\end{displaymath}
and decompose $H^{(n)} g (\varepsilon)$ as in the proof of statement (i), 
where now
\begin{displaymath}
g \quad := \quad \gamma f_1' f_1^\kappa .
\end{displaymath} 
The convergence of the integral over $[0,r/2]$ as well as of the first integral over
$[r/2, \infty [$ is seen in the same way as in part (i).

Finally, the contribution 
\begin{displaymath}
\quad \quad \frac{n!}{\pi} \quad \int_\mathbb{R} \quad 
d \omega  \quad   
\gamma  (\varepsilon + \omega ) \quad 
\frac{\textbf{1}_{\mathbb{R}\setminus B_{r/2}(0)} (\omega) }
  {\omega ^{n+1}} \quad (f_1' f_1^\kappa ) (\varepsilon + \omega )
\end{displaymath}
\begin{displaymath}
= \quad \frac{n!}{\pi} \quad \int_\mathbb{R} \quad 
d x  \quad  h(x) 
 \quad (f_1' f_1^\kappa ) (x )
\end{displaymath}
with 
\begin{displaymath}
h(x) \quad := \quad \gamma  (x) \quad 
\frac{\textbf{1}_{\mathbb{R}\setminus B_{r/2}(0)} (x - \varepsilon) }
  {( x- \varepsilon )^{n+1}}
\end{displaymath}
needs to be considered. Note that the function $h$ is bounded and smooth 
on $\mathbb{R} \setminus \{ \varepsilon - r/2, \varepsilon + r/2 \}$. 
Upon applying an integral transformation the integral turns into 
\begin{displaymath}
\frac{n!}{\pi} \quad \int_\mathbb{R}dz \quad  f'(z) f^\kappa (z) 
\quad h ( \mu_1 + k_B T z) .
\end{displaymath}  
The integrand of this integral is pointwise convergent for $T \to 0$, and an 
integrable upper bound is given by $||h||_\infty |f'(z)|$, so the convergence 
follows with Lebesgue.
\begin{displaymath}
\quad 
\end{displaymath}

\subsubsection*{\textbf{Proof of statement (iii):}}
Let $\varepsilon \in \mathbb{R} \setminus \{ \mu_0, \dots, \mu_m \}$ be given. 
Let 
\begin{displaymath}
r \quad := \quad \min \left\lbrace | \varepsilon - \mu_j | : \quad j = 0, \dots , 
                                                           m   \right\rbrace,
\end{displaymath} 
 \begin{displaymath}
 g \quad := \quad \gamma f_1^{\kappa_1} \dots f_m^{\kappa_m} 
          \quad H (\alpha f_0 ),
 \end{displaymath}
and decompose $H^{(n)} g (\varepsilon )$ as in the proof of statement (i), Eq. 
(\ref{decomposition of H^n g (eps)}). 
Note that $ \alpha f_0  \in {\cal R}$, hence for arbitrary $l \ge 0$: 
$ H ( \alpha f_0 ) ^{( l)} = H^{(l)} ( \alpha f_0 ) $. [I did not investigate the
question if and how the condition of being an element in $\cal R$ could be weakened
without loss of the equality.] Part (i) of this lemma yields thus statements 
about the derivatives of $H( \alpha f_0 ) $. It follows that the first and the second
integral contributing to $H^{(n)} g (\varepsilon)$ are convergent for $T \to 0$.

The third integral is 
\begin{displaymath}
\quad \frac{n!}{\pi} \quad \int_{r/2}^\infty \quad 
\frac {d \omega}{\omega^{n+1}}  \quad \textbf{\{ }  
  g  (\varepsilon + \omega ) \quad  + \quad (-1)^{n+1}   g  
 (\varepsilon - \omega ) \textbf{\}}
\end{displaymath}
\begin{displaymath}
= \quad \frac{n!}{\pi} \quad \int_\mathbb{R} \quad 
d \omega  \quad  \left( \gamma f_1^{\kappa_1} \dots f_m^{\kappa_m} \right)   ( \omega ) \quad 
\frac{\textbf{1}_{\mathbb{R}\setminus B_{r/2}(\varepsilon)} (\omega) }
  {( \omega - \varepsilon ) ^{n+1}}  \quad H (\alpha f_0 ) (\omega) 
\end{displaymath}
\begin{displaymath}
= \quad \frac{n!}{\pi} \quad \int_\mathbb{R} \quad 
d \omega \quad h (\omega) \quad H (\alpha f_0 ) (\omega)
\end{displaymath}
with a corresponding definition of $h$. Note that $h$ as well as 
$\alpha f_0$ are pointwise convergent for $T \to 0$, and because this pointwise 
convergence is bounded by a square integrable function, the convergence is 
satisfied in the $|| \quad ||_2 $ norm as well. The Hilbert transform is isometric 
with respect to this norm, so $H ( \alpha f_0 ) $ is convergent in $ || \quad ||_2$. 
With Cauchy-Schwarz follows the convergence of the integral of the product. 
\begin{displaymath}
\quad 
\end{displaymath}

\subsubsection*{\textbf{Proof of statement (iv):}}
Let $\varepsilon \in \mathbb{R} \setminus \mu_0$, 
\begin{displaymath}
r \quad := \quad | \varepsilon - \mu_0 |,
\end{displaymath}
\begin{displaymath}
g \quad := \quad \gamma (-f_0') f_0^\kappa   \quad H (\alpha f_0 ),
\end{displaymath}
and decompose $H^{(n)}g (\varepsilon) $ as in part (i). Note that the first two 
integrals in the decomposition are convergent for $T \to 0$.  
The third integral is
\begin{displaymath}
\quad 
\end{displaymath} 
\begin{displaymath}
\frac{n!}{\pi} \quad \int_\mathbb{R} d \omega \quad 
\left[   (-f_0') f_0^\kappa   \quad H (\alpha f_0 )  \right] (\omega)  
\quad h (\omega)
\end{displaymath}
with 
\begin{displaymath}
h (\omega ) \quad := \quad \gamma (\omega ) \quad  
\frac{\textbf{1}_{\mathbb{R}\setminus B_{r/2}(\varepsilon)} (\omega) }
{( \omega - \varepsilon ) ^{n+1}}. 
\end{displaymath}
After a transformation the integral reads
\begin{displaymath}
\quad 
\end{displaymath}
\begin{displaymath}
\frac{n!}{\pi} \quad \int_\mathbb{R} dx \quad (-f'(x)) f^\kappa (x) \quad  
H (\alpha f_0) (\mu_0 + x k_B T) \quad h (\mu_0 + x k_B T).
\end{displaymath}
\begin{displaymath}
\quad 
\end{displaymath}
For a further analysis apply now a method related to the approximation in Ref. 
\cite{Kern epjb}: The difference 
\begin{displaymath}
\gamma_T (x) \quad := \quad H (\alpha f_0) (\mu_0 + x k_B T)  \quad - \quad 
                            H (\alpha f_0) (\mu_0)
\end{displaymath}
\begin{displaymath}
\quad 
\end{displaymath}
is pointwise convergent for $T \to 0$, and there is $c > 0$ such, that for 
all $ x \in \mathbb{R}, T \in ]0, T_0]$:
\begin{displaymath}
\quad 
\end{displaymath}
\begin{equation}
|\gamma_T (x) | \quad \le \quad c |x|.
\label{estimate for gamma_T}
\end{equation}
\begin{displaymath}
\quad 
\end{displaymath}

For the latter statement, represent 
\begin{displaymath}
\gamma_T (x_0) \quad = \quad \int_0^{x_0} dx \quad \gamma_T' (x) ,
\end{displaymath}
\begin{displaymath}
\gamma_T' (x) \quad = \quad \frac{1}{\pi} \int_0^\infty dy \quad G_{x,T}(y),
\end{displaymath}
where
\begin{displaymath}
G_{x,T} (y) \quad := \quad \frac{1}{y^2} \left\lbrace F_{x,T} (y) + 
                     F_{x, T} (-y) - 2 F_{x,T} (0) \right\rbrace , 
\end{displaymath}
\begin{displaymath}
F_{x,T} (z) \quad := \quad \alpha (\mu_0 + k_B T (x + z)) \quad f (x + z).
\end{displaymath}
\begin{displaymath}
\quad 
\end{displaymath}
The integrand $G_{x,T}$ is pointwise convergent for $T \to 0$, and there is
an integrable map 
\begin{displaymath}
G_{T_0} : \quad ]0, \infty[ \rightarrow \mathbb{R}
\end{displaymath}
such, that for all $x \in \mathbb{R}$,  $T \in ]0, T_0]$:
\begin{displaymath}
|G_{x,T}| \quad \le \quad G_{T_0}
\end{displaymath}
pointwise. 
To obtain such an upper bound for $|G_{x,T}|(y)$, distinguish between the cases 
$y \le 1$ and $y > 1$. Represent 
\begin{displaymath}
F_{x,T} (y) + F_{x,T} (-y) - 2 F_{x,T} (0) \quad = 
\end{displaymath}
\begin{displaymath}
\quad \int  \int_{0 \le t_2 \le t_1 \le y} \quad 
\left( F_{x,T}'' (t_2) + F_{x,T}'' (-t_2) \right) 
\end{displaymath} 
in case $y \le 1$. 
\begin{displaymath}
\quad 
\end{displaymath}

It follows from the pointwise convergence of $\gamma_T$ and from the estimate 
(\ref{estimate for gamma_T}), that the integral 
\begin{displaymath}
\frac{n!}{\pi} \quad \int_\mathbb{R} dx \quad (-f'(x)) f^\kappa (x) \quad  
\gamma_T (x)  \quad h (\mu_0 + x k_B T).
\end{displaymath}
\begin{displaymath}
\quad 
\end{displaymath}
is convergent for $T \to 0$. Hence, 
\begin{displaymath}
\quad 
\end{displaymath}
\begin{displaymath}
\quad \quad \frac{n!}{\pi} \quad \int_\mathbb{R} dx \quad (-f'(x)) f^\kappa (x) 
\quad  H (\alpha f_0) (\mu_0 + x k_B T) \quad h (\mu_0 + x k_B T)
\end{displaymath}
\begin{displaymath}
\equiv \quad \frac{n!}{\pi} \quad H (\alpha f_0) (\mu_0) \quad \int_\mathbb{R} 
dx \quad (-f'(x)) f^\kappa (x) 
\quad  \quad h (\mu_0 + x k_B T)
\end{displaymath}
\begin{displaymath}
\quad 
\end{displaymath}

Show that 
\begin{displaymath}
 H (\alpha f_0 ) (\mu_0) \quad \equiv \quad \frac{\alpha (\mu_0)}{\pi}
 \log \left( \frac{T}{T_0} \right): 
\end{displaymath}
Represent 
\begin{displaymath}
\frac{d}{dT} H (\alpha f_0 ) (\mu_0) \quad = \frac{\pi^{-1}}{ T}
\int_\mathbb{R} dx \quad (-f'(x) ) \alpha ( \mu_0 + k_B T x), 
\end{displaymath}
\begin{displaymath}
\quad 
\end{displaymath}
and for $ T' \in ]0, T_0]$: 
\begin{displaymath}
\quad \quad H (\alpha f_0 ) (\mu_0) (T = T')  \quad - \quad 
      \frac{\alpha (\mu_0) }{ \pi} 
       \log \left(\frac{T'}{T_0}  \right)  
\end{displaymath}
\begin{displaymath}
\quad 
\end{displaymath}
\begin{displaymath}
= \quad H (\alpha f_0 ) (\mu_0) (T = T_0) 
\end{displaymath}
\begin{displaymath}
\quad \quad - \quad \frac{k_B}{\pi} \quad \int_{T'}^{T_0} dT \quad 
\int_\mathbb{R} dx \quad (-f'(x)) x ( \delta_{\mu_0} \alpha ) 
(\mu_0 + k_B T x),   
\end{displaymath}
\begin{displaymath}
\quad 
\end{displaymath}
and this is convergent for $T' \to 0$. 
\begin{displaymath}
\quad 
\end{displaymath}

Finally, 
\begin{displaymath}
\quad \quad \frac{n!}{\pi} \quad \frac{\alpha (\mu_0)}{\pi}
 \log \left( \frac{T}{T_0} \right) \quad \int_\mathbb{R} 
dx \quad (-f'(x)) f^\kappa (x) 
\quad  \quad h (\mu_0 + x k_B T)
\end{displaymath}
\begin{displaymath}
\equiv \quad \frac{n!}{\pi} \quad \frac{\alpha (\mu_0)}{\pi}
 \log \left( \frac{T}{T_0} \right) \quad \int_\mathbb{R} 
dx \quad (-f'(x)) f^\kappa (x) 
\quad  \quad h (\mu_0 ),
\end{displaymath}
\begin{displaymath}
\quad 
\end{displaymath}
since
\begin{displaymath}
\quad \quad \left| \log  \frac{T}{T_0}  \right| \quad \int_\mathbb{R} 
dx \quad (-f'(x)) f^\kappa (x) 
\quad  \quad | h (\mu_0 + x k_B T) - h (\mu_0) | 
\end{displaymath}
\begin{displaymath}
\le \quad \left| \log  \frac{T}{T_0}  \right| \quad k_B T \quad 
||\delta_{\mu_0} h||_\infty \quad 
\int_\mathbb{R} 
dx \quad (-f'(x)) f^\kappa (x) |x| 
\end{displaymath}
\begin{displaymath}
\to \quad 0 \quad (T \to 0)
\end{displaymath}
with de l'Hospital.
\begin{displaymath}
\quad
\end{displaymath}

In summary, 
\begin{displaymath}
\quad \quad H^{(n)} \left[ \gamma (-f_0') f_0^\kappa   
\quad H (\alpha f_0 ) \right] \quad (\varepsilon ) 
\end{displaymath}
\begin{displaymath}
\equiv \quad \frac{n!}{\pi^2} \quad \frac{(\alpha \gamma) (\mu_0)}{\kappa + 1} 
\quad 
\frac{1}{ ( \mu_0 - \varepsilon )^{n+1}}
\quad \log \left( \frac{T}{T_0} \right) . 
\end{displaymath}

\begin{displaymath}
\quad 
\end{displaymath}
The lemma implies the existence
of the zero temperature limit of $\Gamma_l^\pm (DSO, III)$ in case 
\begin{displaymath}
E_{10} \quad \neq \quad \mu_l, \mu_{\bar{l}}.
\end{displaymath}
\begin{displaymath}
\quad 
\end{displaymath}

\subsection*{Linear conductance within the sixth order DSO}
\begin{displaymath}
\quad 
\end{displaymath}
Within the present approximation scheme the stationary density matrix is given by
\begin{displaymath}
\quad 
\end{displaymath}
\begin{displaymath}
 \left( \begin{array}{c}  	
\rho_{00}      \\  \rho_{\sigma \sigma}   \\   
\rho_{  \bar{\sigma} \bar{\sigma}   } 
\end{array} \right)  \quad = \quad   \left( \begin{array}{c}  	
  \Gamma^-     \\  \Gamma^+   \\   \Gamma^+    
\end{array} \right) \frac{1}{ \Gamma^-   +  2 \Gamma^+ },
\end{displaymath}
\begin{displaymath}
\quad 
\end{displaymath}
and the particle current onto lead $l$ is 
\begin{displaymath}
\quad 
\end{displaymath}
\begin{displaymath}
I_l \quad = \quad  \rho_{00} \left( \Gamma_{\bar {l}}^+ -  \Gamma_l^+  \right)
          \quad + \quad \rho_{\sigma \sigma}
           \left( \Gamma_l^- -  \Gamma_{\bar {l}}^-  \right).
\end{displaymath}
\begin{displaymath}
\quad 
\end{displaymath}
Replace now the exact rates $\Gamma_{l'}^\pm $  by the sixth order DSO rates 
$ \Gamma_{l'}^\pm (DSO, III)$. Then the particle current
\begin{displaymath}
\quad
\end{displaymath}
\begin{displaymath}
I_l (DSO, III) \quad =  
\end{displaymath}
\begin{displaymath}
\quad \frac{2}{\hbar} \begin{LARGE} \textbf{\{} \end{LARGE} \quad
\pi  ( \alpha_{\bar{l}}^+ - \alpha_l^+ ) (E_{10}) \quad + \quad Bil (DSO)     
( \alpha_{\bar{l}}^+ - \alpha_l^+ , \alpha + \alpha ^+ ) 
\end{displaymath}
\begin{displaymath}
\quad \quad \quad  + \quad \quad Tril (DSO) ( \alpha_{\bar{l}}^+ - \alpha_l^+ , 
\alpha + \alpha^+, \alpha + \alpha^+) \quad \quad 
\begin{LARGE} \textbf{\}} \end{LARGE} \quad \rho_{00} \quad +
\end{displaymath}

\begin{displaymath}
\quad \frac{2}{\hbar} \begin{LARGE} \textbf{\{} \end{LARGE} \quad
\pi  ( \alpha_l^- - \alpha_{\bar {l}}^- ) (E_{10}) \quad + \quad Bil (DSO)
(   \alpha_l^- - \alpha_{\bar {l}}^-   , \alpha + \alpha ^+ ) 
\end{displaymath}
\begin{displaymath}
\quad \quad \quad  + \quad \quad Tril (DSO) ( \alpha_l^- - \alpha_{\bar {l}}^- , 
\alpha + \alpha^+, \alpha + \alpha^+) \quad \quad 
\begin{LARGE} \textbf{\}} \end{LARGE} \quad \rho_{\sigma \sigma} 
\end{displaymath}
\begin{displaymath}
\quad 
\end{displaymath}
is obtained.
\begin{displaymath}
\quad
\end{displaymath}

Let the chemical potentials be a function of the bias voltage according to 
\begin{displaymath}
\quad 
\end{displaymath}
\begin{displaymath}
\mu_l \quad = \quad E_F + e V_b,
\end{displaymath}
\begin{displaymath}
\mu_{\bar{l}} \quad = \quad const \quad = \quad E_F .
\end{displaymath}
\begin{displaymath}
\quad 
\end{displaymath}
The linear conductance $G$ is the differential conductance at zero bias, the derivative
of the electric current with respect to the bias, evaluated at zero bias, 
\begin{displaymath}
\quad 
\end{displaymath}
\begin{displaymath}
G \quad = \quad \frac{d (-e)I_l}{d V_b} (V_b = 0). 
\end{displaymath}
\begin{displaymath}
\quad 
\end{displaymath}

Assume symmetric coupling, $ \alpha_l = \alpha_{\bar{l}}$.  
The linear conductance obtained within the sixth order DSO is
\begin{displaymath}
\quad 
\end{displaymath}
\begin{displaymath}
G (DSO, III) \quad =  
\end{displaymath}
\begin{displaymath}
\quad \frac{4 \pi e^2}{h} \begin{LARGE} \textbf{\{} \end{LARGE} \quad
\pi  ( \alpha_l f_T ) (E_{10}) \quad + \quad Bil (DSO)
( \alpha_l f_T , \alpha + \alpha ^+ ) 
\end{displaymath}
\begin{displaymath}
\quad \quad \quad  + \quad \quad Tril (DSO) ( \alpha_l f_T , 
\alpha + \alpha^+, \alpha + \alpha^+) \quad \quad 
\begin{LARGE} \textbf{\}} \end{LARGE} \quad \rho_{00} \quad +
\end{displaymath}

\begin{displaymath}
\quad \frac{4 \pi e^2}{h} \begin{LARGE} \textbf{\{} \end{LARGE} \quad
\pi  ( \alpha_l f_T ) (E_{10}) \quad + \quad Bil (DSO)
(   \alpha_l f_T   , \alpha + \alpha ^+ ) 
\end{displaymath}
\begin{displaymath}
\quad \quad \quad  + \quad \quad Tril (DSO) ( \alpha_l f_T , 
\alpha + \alpha^+, \alpha + \alpha^+) \quad \quad 
\begin{LARGE} \textbf{\}} \end{LARGE} \quad \rho_{\sigma \sigma} 
\end{displaymath}
\begin{displaymath}
\quad 
\end{displaymath}

with
\begin{displaymath}
f_T (\varepsilon) \quad := \quad \frac{-1}{k_B T} f'\left( 
\frac{\varepsilon - E_F}{ k_B T }\right) \quad = \quad 
  \frac{\partial}{\partial \varepsilon} (1-f) \left( 
  \frac{\varepsilon - E_F}{ k_B T } \right).
\end{displaymath}
\begin{displaymath}
\quad 
\end{displaymath}
Note that the function $f_T$ is positive, has total weight one, and
that its weight is distributed over a region around the Fermi level
$E_F$ whose size is proportional to the temperature. 
\begin{displaymath}
\quad 
\end{displaymath}

\subsection*{Divergence of $G(DSO, III)$ with $T \to 0$}
\begin{displaymath}
\quad 
\end{displaymath}
Investigate the dependence of $G(DSO, III)$ on the temperature 
as $T \to 0$. Assume for this, that 
\begin{displaymath}
 E_{10} \quad \neq \quad  E_F .
\end{displaymath}

\begin{displaymath}
\quad 
\end{displaymath}
Consider the contribution 
\begin{displaymath}
\quad \quad ( \rho_{00} + \rho_{\sigma \sigma} ) \quad Tril (DSO) 
( \alpha_l f_T , 
\alpha + \alpha^+, \alpha + \alpha^+ ) 
\end{displaymath}
\begin{displaymath}
\equiv \quad (\rho_{00} + \rho_{\sigma \sigma} ) (-\pi^3) \quad 
H \left[ \alpha_l f_T (\alpha + \alpha^+ ) \quad H (\alpha + \alpha^+ ) 
\right]'' (E_{10}) .
\end{displaymath}

\begin{displaymath}
\quad 
\end{displaymath}
With the statements (ii) and (iv) of the lemma follows 
\begin{displaymath}
\quad 
\end{displaymath}
\begin{displaymath}
G (DSO, III) \quad \equiv \quad \frac{e^2 \pi^2}{h} (\rho_{00} + \rho_{\sigma \sigma}) 
   \quad  \frac{ \alpha^2 (E_F)}{ (E_{10} - E_F)^3} \quad  
   \log \left( \frac{T}{T_0} \right) \quad 
\end{displaymath}   
\begin{displaymath}   
 \quad \quad \quad \quad \quad \quad \quad \quad \quad \quad   \left\lbrace 6 \alpha (E_F) 
   \right\rbrace .     
\end{displaymath}
\begin{displaymath}
\quad 
\end{displaymath}
The linear conductance obtained by the sixth order DSO diverges logarithmically to 
infinity with $T \to 0$ in case $ E_{10} < E_F$, and it diverges logarithmically to 
minus infinity if $E_{10} > E_F$. 
\begin{displaymath}
\quad 
\end{displaymath}

\subsection{Sixth order RTA rates and discussion}
\begin{displaymath}
\quad 
\end{displaymath}
The sixth order RTA rates can be determined either indirectly from the complete RTA rates
by deriving these three times with respect to the square of the coupling parameter
(App. B) or else by direct calculation of the corresponding diagrams. They are
\begin{displaymath}
\quad 
\end{displaymath}
\begin{displaymath}
\Gamma_l^\pm (RTA, III) \quad = \quad \Gamma_l^\pm (RTA)(2) + \Gamma_l^\pm (RTA)(4) + 
           \Gamma_l^\pm (RTA)(6)
\end{displaymath}
with 
\begin{displaymath}
\Gamma_l^\pm (RTA) (2) \quad = \quad \Gamma_l^\pm (DSO) (2),
\end{displaymath}
\begin{displaymath}
\Gamma_l^+ (RTA) (4) \quad = \quad \Gamma_l^+ (DSO) (4) 
\end{displaymath}
\begin{displaymath}
\quad \quad \quad \quad \quad \quad \quad \quad  + \quad \frac{2}{\hbar} 
\quad \pi^2  \begin{LARGE} \textbf{\{} \end{LARGE}
 (\alpha_l + \alpha_l^+ ) (H \alpha^+)' \quad + \quad \alpha^+ 
 H (\alpha_l + \alpha_l^+ )' 
\end{displaymath} 
  \begin{displaymath}
 \quad \quad \quad \quad \quad \quad \quad \quad \quad \quad \quad \quad 
 \quad \quad \quad - 
 \quad H \left[ (\alpha_l + \alpha_l^+ )  \alpha^+  \right]' 
 \begin{LARGE}  \textbf{\}} \end{LARGE} (E_{10}), 
\end{displaymath}
\begin{displaymath}
\quad\quad \quad \quad \quad \quad  \quad = \quad \Gamma_l^+ (DSO) (4)
\end{displaymath}
\begin{displaymath}
 \quad \quad \quad \quad \quad \quad \quad \quad\quad 
+\quad \frac{2}{\hbar} \quad Bil (RTA \setminus DSO ) 
( \alpha_l + \alpha_l^+, \alpha^+ ),
\end{displaymath}

analogously

\begin{displaymath}
\Gamma_l^- (RTA) (4) \quad = \quad \Gamma_l^- (DSO) (4) 
\end{displaymath}
\begin{displaymath}
\quad \quad \quad \quad \quad \quad \quad \quad  + \quad \frac{2}{\hbar} 
\quad \pi^2  \begin{LARGE} \textbf{\{} \end{LARGE}
 (\alpha_l + \alpha_l^+ ) (H \alpha^-)' \quad + \quad \alpha^- 
 H (\alpha_l + \alpha_l^+ )' 
\end{displaymath} 
  \begin{displaymath}
 \quad \quad \quad \quad \quad \quad \quad \quad \quad \quad \quad \quad 
 \quad \quad \quad - 
 \quad H \left[ (\alpha_l + \alpha_l^+ )  \alpha^-  \right]' 
 \begin{LARGE}  \textbf{\}} \end{LARGE} (E_{10}), 
\end{displaymath}
\begin{displaymath}
\quad\quad \quad \quad \quad \quad  \quad = \quad \Gamma_l^- (DSO) (4)
\end{displaymath}
\begin{displaymath}
 \quad \quad \quad \quad \quad \quad \quad \quad\quad 
+\quad \frac{2}{\hbar} \quad Bil (RTA \setminus DSO ) 
( \alpha_l + \alpha_l^+, \alpha^- ).
\end{displaymath}
\begin{displaymath}
\quad 
\end{displaymath}
The bilinear map $Bil (RTA \setminus DSO ) : {\cal R} \times {\cal R} \rightarrow
\mathbb{C}$ is given by the fourth order RTA$\setminus$DSO diagram, 
\begin{displaymath}
Bil (RTA \setminus DSO ) (f, g) \quad = \quad \pi^2 \left\lbrace 
       f H g'  +  g  Hf'  -   H ( fg )' \right\rbrace  (E_{10}).  
\end{displaymath}
\begin{displaymath}
\quad 
\end{displaymath}

Finally, 
\begin{displaymath}
\Gamma_l^+ (RTA) (6) \quad = \quad \Gamma_l^+ (DSO) (6) \quad 
\end{displaymath}
\begin{displaymath}
\quad \quad \quad \quad \quad \quad \quad \quad + \quad 
\frac{2}{\hbar} \quad  
Tril (RTA \setminus DSO ) \quad ( \alpha_l + \alpha_l^+, \alpha + \alpha^+, \alpha^+), 
\end{displaymath}
\begin{displaymath}
\Gamma_l^- (RTA) (6) \quad = \quad \Gamma_l^- (DSO) (6) \quad 
\end{displaymath}
\begin{displaymath}
\quad \quad \quad \quad \quad \quad \quad \quad + \quad 
\frac{2}{\hbar} \quad  
Tril (RTA \setminus DSO ) \quad ( \alpha_l + \alpha_l^+, \alpha + \alpha^+, \alpha^-), 
\end{displaymath}
\begin{displaymath}
\quad 
\end{displaymath}
where $Tril (RTA \setminus DSO ) $ is a trilinear map 
${\cal R} \times {\cal R} \times {\cal R} \rightarrow \mathbb{C}$ given by the sum 
of RTA$\setminus$DSO diagrams, 
\begin{displaymath}
\quad 
\end{displaymath}
\begin{displaymath}
Tril (RTA \setminus DSO ) (f,g,h) \quad = 
\end{displaymath}

\begin{displaymath}
 \pi^3 \quad 
\begin{LARGE} \textbf{\{} \end{LARGE} \quad 
H \left( f h Hg \right) ''
\end{displaymath}
\begin{displaymath}
\quad \quad \quad - \quad  f \quad H ( h Hg)'' \quad    -   \quad    
  h \quad H ( f Hg )'' 
\end{displaymath}
\begin{displaymath}
 \quad \quad \quad + \quad g f'h' \quad - \quad f h g'' \quad + \quad g (H f') (H h')
\end{displaymath}
\begin{displaymath}
\quad \quad \quad - \quad H g \left( \quad f' H h' \quad +\quad  h' H f' \quad 
\right) \quad  
\begin{LARGE} \textbf{\}} \quad  (E_{10}) \end{LARGE} .
\end{displaymath}
\begin{displaymath}
\quad 
\end{displaymath}

\subsection*{\textbf{Existence of $\lim_{T \to 0} \Gamma_l^\pm (RTA, III)$} }
The existence of this limit is seen in the same way as the existence of the zero 
temperature limit of the sixth order DSO rates, still assuming 
\begin{displaymath}
\mu_l, \mu_{\bar{l}} \quad \neq \quad E_{10}.
\end{displaymath}
\begin{displaymath}
\quad 
\end{displaymath}

\subsection*{Linear conductance within the sixth order RTA}
\begin{displaymath}
\quad 
\end{displaymath}
The stationary reduced density matrix of the sixth order RTA is given by
\begin{displaymath}
\quad 
\end{displaymath}
\begin{displaymath}
 \left( \begin{array}{c}  	
\rho_{00}      \\  \rho_{\sigma \sigma}   \\   
\rho_{  \bar{\sigma} \bar{\sigma}   } 
\end{array} \right)  \quad = \quad   \left( \begin{array}{c}  	
  \Gamma^-     \\  \Gamma^+   \\   \Gamma^+    
\end{array} \right) \frac{1}{ \Gamma^-   +  2 \Gamma^+ },
\end{displaymath}
\begin{displaymath}
\quad 
\end{displaymath}
and the particle current onto lead $l$ is 
\begin{displaymath}
\quad           
\end{displaymath}
\begin{displaymath}
I_l \quad = \quad  \rho_{00} \left( \Gamma_{\bar {l}}^+ -  \Gamma_l^+  \right)
          \quad + \quad \rho_{\sigma \sigma}
           \left( \Gamma_l^- -  \Gamma_{\bar {l}}^-  \right),
\end{displaymath}
\begin{displaymath}
\quad 
\end{displaymath}
where the exact rates $\Gamma_{l'}^\pm $  are to be replaced by the sixth 
order RTA rates $ \Gamma_{l'}^\pm (RTA, III)$. Assuming symmetric coupling, 
$\alpha_l = \alpha_{\bar{l}}$, the particle current
\begin{displaymath}
\quad 
\end{displaymath}
\begin{displaymath}
I_l (RTA, III) \quad =  
\end{displaymath}
\begin{displaymath}
 \frac{2}{\hbar} \begin{LARGE} \textbf{\{} \end{LARGE} \quad
\pi  ( \alpha_{\bar{l}}^+ - \alpha_l^+ ) (E_{10}) \quad + \quad Bil (DSO)  \quad \quad
( \alpha_{\bar{l}}^+ - \alpha_l^+ \quad ,\quad \alpha + \alpha ^+ ) 
\end{displaymath}
\begin{displaymath}
 \quad \quad  \quad \quad \quad \quad \quad \quad  \quad \quad \quad 
 + \quad  Bil (RTA\setminus DSO)
\quad ( \alpha_{\bar{l}}^+ - \alpha_l^+ 
\quad , \quad \alpha^+ ) 
\end{displaymath}
\begin{displaymath}
 \quad  + \quad \quad  \quad  Tril (DSO)\quad \quad  
( \alpha_{\bar{l}}^+ - \alpha_l^+ \quad , 
\quad \alpha + \alpha^+ \quad ,\quad \alpha + \alpha^+)  
\end{displaymath}
\begin{displaymath}
 \quad   + \quad  Tril (RTA\setminus DSO) \quad ( \alpha_{\bar{l}}^+ - \alpha_l^+ 
\quad ,\quad  \alpha + \alpha^+ \quad ,\quad  \alpha^+) \quad  
\begin{LARGE} \textbf{\}} \end{LARGE}  \rho_{00} \quad +
\end{displaymath}
\begin{displaymath}
\quad 
\end{displaymath}

\begin{displaymath}
 \frac{2}{\hbar} \begin{LARGE} \textbf{\{} \end{LARGE} \quad
\pi  ( \alpha_l^- - \alpha_{\bar{l}}^- ) (E_{10}) \quad + \quad Bil (DSO)  \quad \quad
( \alpha_l^- - \alpha_{\bar{l}}^- \quad ,\quad \alpha + \alpha ^+ ) 
\end{displaymath}
\begin{displaymath}
 \quad \quad  \quad \quad \quad \quad \quad \quad  \quad \quad \quad 
 + \quad  Bil (RTA\setminus DSO)
\quad ( \alpha_l^+ - \alpha_{\bar{l}}^+ 
\quad , \quad \alpha^- ) 
\end{displaymath}
\begin{displaymath}
 \quad  + \quad \quad  \quad  Tril (DSO)\quad \quad  
( \alpha_l^- - \alpha_{\bar{l}}^- \quad , 
\quad \alpha + \alpha^+ \quad ,\quad \alpha + \alpha^+)  
\end{displaymath}
\begin{displaymath}
 \quad   + \quad  Tril (RTA\setminus DSO) \quad ( \alpha_l^+ - \alpha_{\bar{l}}^+ 
\quad ,\quad  \alpha + \alpha^+ \quad ,\quad  \alpha^-) \quad  
\begin{LARGE} \textbf{\}} \end{LARGE}  \rho_{\sigma \sigma} 
\end{displaymath}
is obtained.
\begin{displaymath}
\quad 
\end{displaymath}

Assume that a bias voltage is applied and determine the linear conductance.    
Including the electron charge, the linear conductance within the sixth order RTA is
\begin{displaymath}
\quad 
\end{displaymath}
\begin{displaymath}
G (RTA, III) \quad =  
\end{displaymath}
\begin{displaymath}
 \frac{4 \pi e^2}{h} \begin{LARGE} \textbf{\{} \end{LARGE} \quad
\pi  (  \alpha_l f_T ) (E_{10}) \quad + \quad Bil (DSO)  \quad \quad
(  \alpha_l f_T \quad ,\quad \alpha + \alpha ^+ ) 
\end{displaymath}
\begin{displaymath}
 \quad \quad  \quad \quad \quad \quad \quad \quad  \quad \quad \quad 
 + \quad  Bil (RTA\setminus DSO)
\quad (  \alpha_l f_T 
\quad , \quad \alpha^+ ) 
\end{displaymath}
\begin{displaymath}
 \quad  + \quad \quad  \quad  Tril (DSO)\quad \quad  
(  \alpha_l f_T \quad , 
\quad \alpha + \alpha^+ \quad ,\quad \alpha + \alpha^+)  
\end{displaymath}
\begin{displaymath}
 \quad   + \quad  Tril (RTA\setminus DSO) \quad (  \alpha_l f_T 
\quad ,\quad  \alpha + \alpha^+ \quad ,\quad  \alpha^+) \quad  
\begin{LARGE} \textbf{\}} \end{LARGE}  \rho_{00} \quad +
\end{displaymath}
\begin{displaymath}
\quad 
\end{displaymath}

\begin{displaymath}
 \frac{4 \pi e^2}{h} \begin{LARGE} \textbf{\{} \end{LARGE} \quad
\pi  (  \alpha_l f_T ) (E_{10}) \quad + \quad Bil (DSO)  \quad \quad
(  \alpha_l f_T \quad ,\quad \alpha + \alpha ^+ ) 
\end{displaymath}
\begin{displaymath}
 \quad \quad  \quad \quad \quad \quad \quad \quad  \quad \quad \quad 
 + \quad  Bil (RTA\setminus DSO)
\quad ( -  \alpha_l f_T
\quad , \quad \alpha^- ) 
\end{displaymath}
\begin{displaymath}
 \quad  + \quad \quad  \quad  Tril (DSO)\quad \quad  
(  \alpha_l f_T \quad , 
\quad \alpha + \alpha^+ \quad ,\quad \alpha + \alpha^+)  
\end{displaymath}
\begin{displaymath}
 \quad   + \quad  Tril (RTA\setminus DSO) \quad ( -  \alpha_l f_T 
\quad ,\quad  \alpha + \alpha^+ \quad ,\quad  \alpha^-) \quad  
\begin{LARGE} \textbf{\}} \end{LARGE}  \rho_{\sigma \sigma} .
\end{displaymath}  
\begin{displaymath}
\quad 
\end{displaymath}

\subsection*{Divergence of $G (RTA, III) $ with $T \to 0$}
\begin{displaymath}
\quad 
\end{displaymath}
Assume 
\begin{displaymath}
E_{10} \quad \neq \quad E_F.
\end{displaymath}
\begin{displaymath}
\quad 
\end{displaymath}
The only divergent terms contributing to the linear conductance within the 
sixth order RTA  which have not yet been investigated are
\begin{displaymath}
\quad 
\end{displaymath}
\begin{displaymath}
\frac{4 \pi^4 e^2}{h} \quad \rho_{00} \quad 
\begin{LARGE} \textbf{\{} \end{LARGE} \quad 
H \left( \alpha_l f_T \alpha^+ \quad H \alpha^+ \right)'' (E_{10})
\end{displaymath}
\begin{displaymath}
\quad \quad \quad \quad \quad \quad - \quad \alpha^+ (E_{10}) \quad 
H \left( \alpha_l f_T \quad H \alpha^+   \right)'' (E_{10}) \quad 
\begin{LARGE} \textbf{\}} \end{LARGE} \quad +  
\end{displaymath}
\begin{displaymath}
\frac{4 \pi^4 e^2}{h} \quad \rho_{\sigma \sigma} \quad 
\begin{LARGE} \textbf{\{} \end{LARGE} \quad 
- \quad H \left( \alpha_l f_T \alpha^- \quad H \alpha^+ \right)'' (E_{10})
\end{displaymath}
\begin{displaymath}
\quad \quad \quad \quad \quad \quad + \quad \alpha^- (E_{10}) \quad 
H \left( \alpha_l f_T \quad H \alpha^+   \right)'' (E_{10}) \quad 
\begin{LARGE} \textbf{\}} \end{LARGE}
\end{displaymath}
\begin{displaymath}
\quad 
\end{displaymath}

\begin{displaymath}
\equiv \quad \frac{\pi ^2 e^2 }{h } \quad \rho_{00} \quad  
\frac{ \alpha^2  (E_F) }{(E_{10} - E_F )^3} \quad \log \left( 
\frac{T}{T_0} \right) \quad
\left\lbrace -2 \alpha (E_F) 
+ 4 \alpha^+ (E_{10}) \right\rbrace \quad +
\end{displaymath}
\begin{displaymath}
\quad \quad \frac{\pi ^2 e^2 }{h } \quad \rho_{\sigma \sigma} \quad  
\frac{ \alpha^2  (E_F) }{(E_{10} - E_F )^3} \quad \log \left( 
\frac{T}{T_0} \right) \quad
\left\lbrace 2 \alpha (E_F) 
- 4 \alpha^- (E_{10}) \right\rbrace .
\end{displaymath}
\begin{displaymath}
\quad 
\end{displaymath}

In summary, 
\begin{displaymath}
G (RTA, III ) \quad \equiv \quad 
\end{displaymath}
\begin{displaymath}
\quad 
\end{displaymath}
\begin{displaymath}
 \frac{\pi ^2 e^2 }{h } \quad \rho_{00}  \quad 
\frac{ \alpha^2 (E_F) }{(E_{10} - E_F )^3} \quad \log \left( \frac{T}{T_0} \right) 
\quad \left\lbrace 4 \alpha (E_F) 
+ 4 \alpha^+ (E_{10}) \right\rbrace \quad +
\end{displaymath}
\begin{displaymath}
 \frac{\pi ^2 e^2 }{h } \quad \rho_{\sigma \sigma}  \quad 
\frac{ \alpha^2 (E_F) }{(E_{10} - E_F )^3} 
\quad \log \left( \frac{T}{T_0}  \right) 
\quad \left\lbrace 8 \alpha (E_F) 
- 4 \alpha^- (E_{10}) \right\rbrace .
\end{displaymath}
\begin{displaymath}
\quad 
\end{displaymath}

In particular in the case $E_{10}  > E_F$:
\begin{displaymath}
\quad 
\end{displaymath}
\begin{displaymath}
G (RTA, III) \quad \equiv \quad \frac{ \pi^2 e^2  }{h} \quad 
 \frac{ \alpha^2  (E_F) } {(E_{10} - E_F )^3} \quad
 \log \left( \frac{T}{T_0}  \right)
\end{displaymath}
\begin{displaymath}
\quad \quad \quad \quad \quad \quad \quad \quad 
\begin{LARGE} \textbf{\{} \end{LARGE} 
4 \alpha (E_F ) \rho_{00} \quad + \quad 
8 \alpha (E_F) \rho_{\sigma \sigma} \quad -  \quad 4 \alpha (E_{10}) 
\rho_{\sigma \sigma} 
\begin{LARGE} \textbf{\}} \end{LARGE},
\end{displaymath}
\begin{displaymath}
\quad 
\end{displaymath}
and in case $ E_{10} < E_F$: 
\begin{displaymath}
\quad 
\end{displaymath}
\begin{displaymath}
G (RTA, III) \quad \equiv \quad \frac{ \pi^2 e^2  }{h} \quad 
 \frac{ \alpha^2  (E_F) } {(E_{10} - E_F )^3} \quad 
  \log \left( \frac{T}{T_0}  \right)
\end{displaymath}
\begin{displaymath}
\quad \quad \quad \quad \quad \quad \quad \quad 
\begin{LARGE} \textbf{\{} \end{LARGE} 
4 \alpha (E_F ) \rho_{00} \quad + \quad 
8 \alpha (E_F) \rho_{\sigma \sigma} \quad + \quad 4 \alpha (E_{10}) 
\rho_{00} 
\begin{LARGE} \textbf{\}} \end{LARGE}.
\end{displaymath}
\begin{displaymath}
\quad 
\end{displaymath}

The linear conductance obtained by the sixth order RTA diverges logarithmically 
to infinity with $T \to 0$ in case $E_{10} < E_F$. 
In case $E_{10} > E_F$  the situation is not clear, since the sign of the sum in 
the curly bracket might be negative. 
Assuming that the term containing the probability $\rho_{00}$ is dominant in the 
regime $E_{10} > E_F$, the linear conductance diverges logarithmically to minus 
infinity.

\subsection{Exact sixth order rates and discussion}

The components of the exact sixth order rates are 
\begin{displaymath}
\Gamma_l^\pm (2) \quad = \quad \Gamma_l^\pm (RTA)(2), 
\end{displaymath}
\begin{displaymath}
\Gamma_l^\pm (4) \quad = \quad \Gamma_l^\pm (RTA)(4), 
\end{displaymath}
\begin{displaymath}
\Gamma_l^+ (6) \quad = \quad \Gamma_l^+ (RTA)(6) 
\end{displaymath}
\begin{displaymath}
 \quad \quad \quad \quad \quad + \quad \frac{2}{\hbar} \quad  
       \begin{LARGE} \textbf{\{} \end{LARGE} \quad 
   Tril(a) \quad (\alpha_l^+, \alpha^-, \alpha^+ ) \quad + \quad 
\end{displaymath}
\begin{displaymath}
\quad \quad \quad \quad \quad \quad \quad \quad \quad \quad \quad 
 Tril(b) \quad ( \alpha_l^-, \alpha^+,
       \alpha^+ ) \quad \begin{LARGE}   \textbf{\}}   \end{LARGE} ,
\end{displaymath}

\begin{displaymath}
\Gamma_l^- (6) \quad = \quad \Gamma_l^- (RTA)(6) 
\end{displaymath}
\begin{displaymath}
 \quad \quad \quad \quad \quad + \quad \frac{2}{\hbar} \quad  
       \begin{LARGE} \textbf{\{} \end{LARGE} \quad 
   Tril(a) \quad (\alpha_l^-, \alpha^+, \alpha^- ) \quad + \quad 
\end{displaymath}
\begin{displaymath}
\quad \quad \quad \quad \quad \quad \quad \quad \quad \quad \quad 
 Tril(b) \quad ( \alpha_l^+, \alpha^-,
       \alpha^- ) \quad \begin{LARGE}   \textbf{\}}   \end{LARGE} ,
\end{displaymath}
where $Tril(a), Tril(b)$ are trilinear forms ${\cal R} \times {\cal R} 
\times {\cal R} \rightarrow \mathbb{C} $ given by sums of non-RTA diagrams. 
They have decompositions
\begin{displaymath}
Tril(a) \quad = \quad Tril (a') \quad - \quad Tril(RTA \setminus DSO ) 
\quad - \quad 2  Tril (DSO), 
\end{displaymath}
\begin{displaymath}
Tril(b) \quad = \quad Tril (b') \quad - \quad Tril (RTA \setminus DSO ) , 
\end{displaymath}

where I arrived at                               
\begin{displaymath}
Tril (a') (f,g,h) \quad = \quad 2\pi \quad I^{conv} (\quad  T_{-E_{10}} \delta_{E_{10}}^2 
  f \quad ,\quad  g\quad ,\quad  \delta_{E_{10}}^2 h \quad  )  
\end{displaymath}
\begin{displaymath}
\quad 
\end{displaymath}
\begin{displaymath}
 \quad \quad \quad \quad + \quad \pi^3 \quad 
\begin{LARGE} \textbf{\{} \end{LARGE}  \quad \quad 
2h' \quad  H [ f Hg ]' \quad 
- \quad 2 Hh' \quad H[fg]'   
\end{displaymath}
\begin{displaymath}
 \quad \quad \quad \quad \quad \quad \quad \quad \quad 
+ \quad f' \quad H[h Hg]' \quad - \quad f' \quad H [g Hh]'
\end{displaymath}
\begin{displaymath}
\quad \quad \quad \quad \quad \quad \quad \quad \quad 
+ \quad 2h \quad H[f  (Hg)']' \quad - \quad h \quad H [f Hg ]''
\end{displaymath}
\begin{displaymath}
\quad \quad \quad \quad \quad \quad \quad \quad \quad \quad \quad 
\quad \quad \quad \quad 
- \quad Hh \quad H[fg]'' \quad \quad \quad \quad \quad \begin{LARGE}
\textbf{\}} (E_{10}) 
\end{LARGE} 
\end{displaymath}

\begin{displaymath}
\quad 
\end{displaymath}
\begin{displaymath}
 \quad \quad \quad \quad + \quad \pi^3 \quad 
\begin{LARGE} \textbf{\{} \end{LARGE}  \quad \quad 
 (Hf)' g (Hh)'    
 \quad + \quad 3 f' (Hg)' (Hh)      
\end{displaymath}
\begin{displaymath}
 \quad \quad \quad \quad \quad \quad \quad \quad \quad 
- \quad  (Hf)' (Hg) h'  
\end{displaymath}
\begin{displaymath}
\quad \quad \quad \quad \quad \quad \quad \quad \quad 
- \quad f'' g h  
    \quad + \quad 2 f g' h' \quad + \quad 2 f' g h' 
\end{displaymath}
\begin{displaymath}
\quad \quad \quad \quad \quad \quad \quad \quad \quad 
+ \quad f g'' h
    \quad + \quad f' g' h 
\end{displaymath}
\begin{displaymath}
\quad \quad \quad \quad \quad \quad \quad \quad \quad 
+ \quad  2 f' (Hg) (Hh)' 
    \quad + \quad 2f (Hg)' (Hh)'  
\end{displaymath}
\begin{displaymath}
  \quad \quad \quad \quad \quad \quad \quad \quad \quad 
  + \quad  f (Hg)'' (Hh) 
    \quad + \quad f'' (Hg) (Hh) \quad \quad \quad 
    \begin{LARGE}  \textbf{\}}  \end{LARGE} (E_{10}),    
\end{displaymath}
\begin{displaymath}
\quad 
\end{displaymath}

\begin{displaymath}
Tril (b') (f,g,g) \quad = \quad - 2\pi \quad I^{conv} (\quad  T_{-E_{10}} \delta_{E_{10}}^2 
  g \quad ,\quad  \delta_{E_{10}} f \quad ,\quad  \delta_{E_{10}} g \quad  )  
\end{displaymath}
\begin{displaymath}
\quad 
\end{displaymath}
\begin{displaymath}
 \quad \quad \quad \quad + \quad \pi^3 \quad 
\begin{LARGE} \textbf{\{} \end{LARGE}  \quad \quad 
 - 2 g \quad H[f (Hg)']'      \quad 
+ \quad g \quad H[f Hg]''       
\end{displaymath}
\begin{displaymath}
 \quad \quad \quad \quad \quad \quad \quad \quad \quad \quad \quad 
+ \quad  Hg \quad H[fg]''   \quad + \quad 2 g' \quad 
H[f Hg]' 
\end{displaymath}
\begin{displaymath}
\quad \quad \quad \quad \quad \quad \quad \quad \quad \quad \quad 
\quad \quad \quad \quad 
+ \quad 2Hg' \quad H[fg]' \quad \quad \quad \quad \quad \begin{LARGE}
\textbf{\}} (E_{10}) 
\end{LARGE} 
\end{displaymath}
\begin{displaymath}
\quad 
\end{displaymath}

\begin{displaymath}
 \quad \quad \quad \quad  + \quad \pi^3 \quad 
\begin{LARGE} \textbf{\{} \end{LARGE}  \quad \quad 
 -2(Hf)'' g (Hg)    
 \quad - \quad 3 (Hf)' g' (Hg)      
\end{displaymath}
\begin{displaymath}
 \quad \quad \quad \quad  \quad \quad  \quad \quad \quad \quad \quad 
- \quad  (Hf)' g (Hg)'  
\end{displaymath}
\begin{displaymath}
\quad \quad \quad \quad \quad \quad \quad \quad \quad \quad \quad 
+ \quad 2 f g' g'  
    \quad + \quad f' g g' \quad - \quad  f g g'' 
\end{displaymath}
\begin{displaymath}
\quad \quad \quad \quad \quad \quad \quad \quad \quad \quad \quad 
- \quad f' (Hg) (Hg)'
    \quad - \quad 2f (Hg)' (Hg)' 
\end{displaymath}
\begin{displaymath}
\quad  \quad  \quad \quad \quad \quad \quad \quad \quad \quad \quad \quad \quad 
 \quad \quad \quad  - \quad  f (Hg) (Hg)''  \quad  \quad \quad \quad 
    \begin{LARGE}  \textbf{\}}  \end{LARGE} (E_{10}).    
\end{displaymath}
\begin{displaymath}
\quad 
\end{displaymath}

The map 
\begin{displaymath}
I^{conv} : \quad {\cal R} \times {\cal R} \times {\cal R}  \quad \rightarrow 
\mathbb{C},
\end{displaymath}
\begin{displaymath}
\quad \quad \quad \quad (f,g,h) \quad \mapsto \quad  \int_\mathbb{R} dx \quad f(x) \quad
(g * h) (x) 
\end{displaymath}
appears regularly in the calculation of the non-RTA diagrams. Note that the 
representation is not unique, since for real $f,g,h \in {\cal R}$:
\begin{displaymath}
\quad 
\end{displaymath}
\begin{displaymath}
I^{conv} ( \delta f, \delta g, \delta \delta h ) \quad + \quad 
I^{conv} ( \delta \delta f, \delta g, \delta h )
\end{displaymath}
\begin{displaymath}
\quad \quad \quad \quad \quad \quad = \quad I^{conv} 
( \delta \delta f,  g, \delta \delta h ) \quad  + \quad  \pi^2 
\quad Re \left\lbrace  {f^a} ' g^s {h^a} ' 
\right\rbrace (0),
\end{displaymath}
where for $\alpha \in {\cal R}$: 
\begin{displaymath}
\alpha^s \quad := \quad \alpha \quad - \quad  i H \alpha,
\end{displaymath}                                     
\begin{displaymath}                                
\alpha^a \quad := \quad \alpha \quad + \quad  i H \alpha .
\end{displaymath}
[Represent $I^{conv} (a,b,c) $ by use of the Fourier back transforms of
$a,b,c$. Integrate by parts.] Moreover, there is the relation 
\begin{displaymath}
H \left[ f Hg \quad + \quad g Hf \right] \quad = \quad (Hf) (Hg) \quad - \quad fg.
\end{displaymath}

\begin{displaymath}
\quad 
\end{displaymath}

\subsection*{\textbf{Existence of $\lim_{T \to 0} \Gamma_l^\pm (III) $}}

Verify now the existence of the zero temperature limit of the exact rates
$\Gamma_l^\pm (III)$, once more assuming 
\begin{displaymath}
E_{10} \quad \neq \quad \mu_l, \mu_{\bar l}.
\end{displaymath}

For the treatment of those terms which contain the map
$I^{conv}$, note the following remark and lemma:
\begin{displaymath}
\quad 
\end{displaymath}

\emph{Remark (representations of $\delta_\varepsilon ^n f$):}
\begin{displaymath}
\quad 
\end{displaymath}
For smooth $f: \mathbb{R} \rightarrow \mathbb{C}$ , $ \varepsilon \in \mathbb{R}$,
$n \ge 1$:
\begin{displaymath}
(\delta_\varepsilon ^n f) (\varepsilon) \quad = \quad \frac{f^{(n)}(\varepsilon)}{n!}, 
\end{displaymath}
\begin{displaymath}
(\delta_\varepsilon ^n f) (x) \quad = \quad \frac{f (x) - p(f, \varepsilon)_{n-1} (x)}
{ ( x- \varepsilon )^n} 
\end{displaymath}
for $x \neq \varepsilon$, and for all $x \in \mathbb{R}$:
\begin{displaymath}
(\delta_\varepsilon ^n f) (x) \quad = \quad \int \dots \int_{0 \le t_n \le \dots \le t_1 
\le 1} f^{(n)} ( \varepsilon + t_n (x - \varepsilon)). 
\end{displaymath}
\begin{displaymath}
\quad
\end{displaymath}

\textbf{Lemma (decay of $\delta_{E_{10}}^n \alpha_l^+$; consequences):} 
\begin{displaymath}
\quad
\end{displaymath}
Let the function $\alpha_l$ be given as in App. A. Moreover, let 
$n \in \{ 0, 1, \dots \}$, 
\begin{displaymath}
f_l (\omega) \quad := \quad f \left( \frac{\omega - \mu_l}{ k_B T } \right), 
\end{displaymath}
$\varepsilon \in \mathbb{R} \setminus \mu_l$, and $\alpha_l^+ := \alpha_l f_l$. 
\begin{displaymath}
\quad 
\end{displaymath}

\underline{Statement 1:} There are $c, x_0 > 0$ such, that for all $x \in 
\mathbb{R}$: 
\begin{displaymath}
|(\delta_\varepsilon ^n \alpha_l ) (x)| \quad \le \quad \frac{c}{|x| + x_0} 
\quad =: \quad c \quad \iota_{x_0} (x).
\end{displaymath}
\begin{displaymath}
\quad 
\end{displaymath}

\underline{Statement 2:} There are $c', x_0' > 0$ such, that for all $x \in 
\mathbb{R}$ and $T > 0$: 
\begin{displaymath}
|(\delta_\varepsilon ^n \alpha_l^+ ) (x)| \quad \le \quad c' \quad \iota_{x_0'} (x).
\end{displaymath}
Moreover, the pointwise limit $\lim_{T \to 0} \delta_\varepsilon^n \alpha_l^+ $
exists. The convergence is thus fulfilled in the $||\quad ||_2$ norm too.  
\begin{displaymath}
\quad
\end{displaymath}

\underline{Statement 3:} Let $(f_T)_{T > 0}, (g_T)_{T > 0}$ be families of functions 
in ${\cal R}$. Let the pointwise limits $h_0 := \lim_{T \to 0} h_T $ exist for $h = f,g$. 
Moreover, assume there are $x_0, c > 0$ such, that for all $T > 0$: 
\begin{displaymath}
|f_T|, |g_T| \quad \le \quad c \quad \iota_{x_0}.
\end{displaymath}  
Then the convolution $f_T * g_T$ converges pointwise and even uniformly to 
$f_0 * g_0$, and
\begin{displaymath}
|f_T * g_T| (x) \quad \le \quad c^2 \quad \iota_{x_0} * \iota_{x_0} \quad (x) .
\end{displaymath}
The latter function is bounded by a constant and quadratically integrable. In particular,
the convolution $f_T * g_T$ converges in $|| \quad ||_2$. 
\begin{displaymath}
\quad
\end{displaymath}

\underline{Statement 4:} Let $(h_T)_{T > 0}$ be a familiy of functions in 
${\cal R}$ with the properties that it converges pointwise and uniformly 
to a function $h_0$ as $T \to 0$ and that there is a bounded and quadratically 
integrable function $h_\infty : \mathbb{R} \rightarrow [0,\infty[$ such, that for all 
$ T > 0$: $ |h_T| \le h_\infty$ everywhere. Then the integral 
\begin{displaymath}
\quad 
\end{displaymath} 
\begin{displaymath}
\int_\mathbb{R} dy \quad \delta^n_\varepsilon \left( \alpha_l f_l'\right) 
(y) 
 \quad h_T (y)
\end{displaymath}
\begin{displaymath}
\quad 
\end{displaymath}
is convergent with $T \to 0$.

\subsubsection*{Proof of statement 1:}
According to the additional statement in App. A consider only the case $n \ge 1$. 
Choose $\rho > 0$ arbitrary and estimate for $x \in \mathbb{R}$ with $
| x - \varepsilon |  <  \rho $:
\begin{displaymath}           
|( \delta_\varepsilon ^n \alpha_l ) (x)| \quad =  \quad \left| \int \dots 
\int_{0 \le t_n \le \dots \le t_1 \le 1} 
\alpha_l^{(n)} ( \varepsilon + t_n (x - \varepsilon)) \right|
\end{displaymath}
\begin{displaymath}
\quad \quad \quad \quad \quad \quad \le \quad \frac{1}{n!} S_0,
\end{displaymath}
where
\begin{displaymath}
S_0 \quad := \quad  \sup \left\lbrace  |\alpha_l^{(n)}(\omega)|: \quad \omega 
\in B_\rho (\varepsilon) \right\rbrace , 
\end{displaymath}
\begin{displaymath}
\quad 
\end{displaymath}
choose $x_0 > 0 $ arbitrary, and estimate for $x \in \mathbb{R}$ with 
$ |x - \varepsilon | \ge \rho$:
\begin{displaymath}
\quad 
\end{displaymath}
\begin{displaymath}
|( \delta_\varepsilon ^n \alpha_l ) (x)| \quad \le  \quad 
\frac{|\alpha_l (x)| + |p (\alpha_l, \varepsilon)_{n-1} (x)| } 
{ | x- \varepsilon |^n}
\end{displaymath} 
\begin{displaymath}
 \quad \quad \quad  \quad \quad  \le \quad \left\lbrace 
 \frac{|\alpha_l (x)|\quad (x_0 + |x|) }
 {|x - \varepsilon |^n} 
 \quad  + \quad   \frac{|p (\alpha_l, \varepsilon)_{n-1} (x)| \quad 
( x_0 + |x|)}{ | x - \varepsilon |^n } \right\rbrace
\end{displaymath}
\begin{displaymath}
\quad \quad \quad  \quad \quad \quad \quad \quad \frac{1}{ x_0 + |x|}
\end{displaymath}
\begin{displaymath}
\quad \quad \quad  \quad \quad  \le \quad \frac{S_1 + S_2 }{  x_0 + |x| }
\end{displaymath}
with 
\begin{displaymath}
S_1 \quad := \quad \sup \left\lbrace \frac{ |\alpha_l (\omega)|\quad (x_0 + |\omega |) }
{| \omega - \varepsilon |^n}: \quad \omega \in \mathbb{R}, |\omega - \varepsilon |
\ge \rho \right\rbrace ,
\end{displaymath}
\begin{displaymath}
S_2 \quad := \quad \sup \left\lbrace \frac{ |p( \alpha_l, \varepsilon )_{n-1} 
(\omega)|\quad (x_0 + |\omega |) }
{| \omega - \varepsilon |^n}: \quad \omega \in \mathbb{R}, |\omega - \varepsilon |
\ge \rho \right\rbrace .
\end{displaymath}
\begin{displaymath}
\quad 
\end{displaymath}
The suprema $S_0, S_1, S_2$ are finite, so there is $c > 0$ such, that everywhere 
$|\delta_\varepsilon ^n \alpha_l | \le c \iota_{x_0}$.

\subsubsection*{Proof of statement 2:}
According to the additional statement of App. A consider only the case $n \ge 1$. 
Let 
\begin{displaymath}
\rho \quad := \quad \frac{1}{2} | \varepsilon - \mu_l |,
\end{displaymath}
\begin{displaymath}
\quad 
\end{displaymath}
and note that for $x \in \mathbb{R}$ with $ |x - \varepsilon | < \rho$:
\begin{displaymath}
\quad 
\end{displaymath} 
\begin{displaymath}           
|( \delta_\varepsilon ^n \alpha_l^+ ) (x)| \quad \le \quad \frac{1}{n!} S_0',
\end{displaymath}
where
\begin{displaymath}
S_0' \quad := \quad  \sup \left\lbrace  |\alpha_l^{+ (n)}(\omega)|: \quad \omega 
\in B_\rho (\varepsilon), T > 0 \right\rbrace , 
\end{displaymath}
\begin{displaymath}
\quad 
\end{displaymath}
and for $x \in \mathbb{R}$ with $ | x  - \varepsilon | \ge \rho$: 
\begin{displaymath}
\quad 
\end{displaymath}
\begin{displaymath}
|( \delta_\varepsilon ^n \alpha_l^+ ) (x)| \quad \le  \quad 
 \frac{S_1' + S_2' }{  x_0 + |x| }
\end{displaymath}
with
\begin{displaymath}
S_1' \quad := \quad \sup \left\lbrace \frac{ |\alpha_l^+ (\omega)|\quad (x_0 + |\omega |) }
{| \omega - \varepsilon |^n}: \quad \omega \in \mathbb{R} \setminus B_\rho (\varepsilon), 
T > 0 \right\rbrace ,
\end{displaymath}
\begin{displaymath}
S_2' \quad := \quad \sup \left\lbrace \frac{ |p( \alpha_l^+, \varepsilon )_{n-1} 
(\omega)|\quad (x_0 + |\omega |) }
{| \omega - \varepsilon |^n}: \quad \omega \in \mathbb{R} \setminus B_\rho (\varepsilon), 
T > 0 \right\rbrace .
\end{displaymath}
\begin{displaymath}
\quad 
\end{displaymath}
The suprema $S_0', S_1', S_2'$ are finite, so there is $c' > 0 $ such, that 
independently of the temperature and the argument $|\delta_\varepsilon ^n \alpha_l^+ | 
\le c' \iota_{x_0}$.
\begin{displaymath}
\quad 
\end{displaymath}

To verify the pointwise convergence with $T \to 0$, write 
\begin{displaymath}
(\delta_\varepsilon ^n \alpha_l^+) (\varepsilon) \quad = 
\quad \frac{\alpha_l^{+ (n)}(\varepsilon)}{n!}, 
\end{displaymath}
and for $x \neq \varepsilon$: 
\begin{displaymath}
(\delta_\varepsilon ^n \alpha_l^+) (x) \quad = \quad \frac{\alpha_l^+ (x) - 
p ( \alpha_l^+ , \varepsilon )_{n-1} (x)}    { ( x- \varepsilon )^n}. 
\end{displaymath}

\subsubsection*{Proof of statement 3:}
For arbitrary $y \in \mathbb{R}$:
\begin{displaymath}
f_T * g_T (y) \quad = \quad \int_\mathbb{R} dx \quad f_T (x) (T_y g_T) (x).
\end{displaymath}
Note that 
\begin{displaymath}
f_T \quad \to  \quad  f_0,
\end{displaymath}
\begin{displaymath}
T_y g_T \quad \to \quad T_y g_0
\end{displaymath}
pointwise and in $|| \quad ||_2$. With the Cauchy-Schwarz inequality follows the 
uniform convergence 
\begin{displaymath}
\int_\mathbb{R} dx \quad f_T (x) (T_y g_T) (x) \quad \to \quad 
\int_\mathbb{R} dx \quad f_0 (x) (T_y g_0) (x) \quad (T \to 0).
\end{displaymath}
\begin{displaymath}
\quad
\end{displaymath}

For the estimate note that 
\begin{displaymath}
\left| f_T * g_T (y) \right| \quad \le \quad \int_\mathbb{R} dx \quad |f_T| (x) 
|g_T| (x-y)
\end{displaymath}
\begin{displaymath}
\quad \quad \quad \quad \quad \quad \le \quad c^2 \quad 
\iota_{x_0} * \iota_{x_0} \quad (y)
\end{displaymath}
\begin{displaymath}
\quad \quad \quad \quad \quad \quad \le \quad 4 c^2 \quad 
F ( y/ x_0 ),
\end{displaymath}
where 
\begin{displaymath}
F(t) \quad := \quad \frac{\log \left( 1 + |t| \right) }{ |t| }.
\end{displaymath}
The latter function is bounded and quadratically integrable. With
Lebesgue follows the convergence of the convolution $f_T * g_T$ 
in $||\quad ||_2 $. 
\begin{displaymath}
\quad 
\end{displaymath}

\subsubsection*{Proof of statement 4:}
\begin{displaymath}
\quad 
\end{displaymath}
Let 
\begin{displaymath}
\rho \quad := \quad \frac{1}{2}| \varepsilon - \mu_l |
\end{displaymath}
\begin{displaymath}
\quad 
\end{displaymath}
and decompose in case $n \ge 1$  
\begin{displaymath}
\quad 
\end{displaymath}
\begin{displaymath}
\delta_\varepsilon ^n ( \alpha_l f_l' ) (y) \quad = \quad \quad
 \delta_\varepsilon ^n ( \alpha_l f_l' ) (y) \quad \quad 
 \textbf{1}_{B_\rho (\varepsilon) }(y)
\end{displaymath}
\begin{displaymath}
\quad 
\end{displaymath}
\begin{displaymath}
\quad \quad \quad \quad \quad \quad \quad + \quad 
\frac{ - p( \alpha_l f_l' , \varepsilon)_{n-1} (y)}
{ ( y - \varepsilon )^n} \quad \textbf{1}_{\mathbb{R}\setminus B_\rho (\varepsilon) }(y)
\end{displaymath}
\begin{displaymath}
\quad
\end{displaymath}
\begin{displaymath}
\quad \quad \quad \quad \quad \quad \quad + \quad \frac{ ( \alpha_l f_l' ) (y)}
{ ( y - \varepsilon )^n} \quad 
\textbf{1}_{\mathbb{R}\setminus B_\rho (\varepsilon) }(y)
\end{displaymath}
\begin{displaymath}
\quad 
\end{displaymath}
\begin{displaymath}
\quad \quad \quad \quad \quad \quad =: \quad f_1 (y) \quad +\quad  
f_2(y) \quad  +\quad  f_3 (y).
\end{displaymath}
\begin{displaymath}
\quad
\end{displaymath}
Note that for $y \in B_{\rho} (\varepsilon) $:
\begin{displaymath}
|f_1 (y)| \quad \le \quad \frac{1}{n!} \quad \sup \left\lbrace
 | (f_l' \alpha_l)^{(n)} |(x): \quad x \in B_\rho (\varepsilon), T > 0 
 \right\rbrace 
\end{displaymath}
\begin{displaymath}
\quad \quad \quad \quad =: \quad \frac{1}{n!} \quad S_1, 
\end{displaymath}
where $S_1$ is finite because $\mu_l \notin B_\rho (\varepsilon)$. 
From the alternative representation of $\delta_\varepsilon^n (\alpha_l 
f_l')$ it can be seen that $f_1$ is pointwise convergent. 
It follows that $f_1$ converges in particular in $ ||\quad ||_2$.

\begin{displaymath}
\quad 
\end{displaymath}
The function $f_2$ is a linear combination of the functions 
\begin{displaymath}
y \quad \mapsto \quad \frac{1}{(y - \varepsilon)^j} \quad 
  \textbf{1}_{\mathbb{R}\setminus B_\rho (\varepsilon) }(y), \quad j = 
  1, \dots , n,
\end{displaymath}
\begin{displaymath}
\quad  
\end{displaymath}
the coefficients are given by the derivatives of $f_l' \alpha_l$ in 
$\varepsilon$ and thus convergent with $T \to 0$. $f_2$ converges
in $||\quad ||_2$ as $T \to 0$. As a consequence, the integrals
\begin{displaymath}
\quad 
\end{displaymath}
\begin{displaymath}
\int_\mathbb{R} dy \quad f_\kappa (y)   \quad h_T (y)
\end{displaymath}
\begin{displaymath}
\quad 
\end{displaymath}
are convergent as $T \to 0$ for $\kappa = 1,2$.

\begin{displaymath}
\quad 
\end{displaymath}
Consider finally
\begin{displaymath}
\quad 
\end{displaymath}
\begin{displaymath}
\int_\mathbb{R} dy \quad f_3 (y) \quad h_T (y) \quad = \quad 
\int_\mathbb{R} dy \quad f_l' (y) \quad \varphi_T (y) 
\end{displaymath}
\begin{displaymath}
\quad 
\end{displaymath}
with
\begin{displaymath}
\varphi_T (y) \quad := \quad \alpha_l (y) \quad h_T (y) \quad 
         \frac{  \textbf{1}_{\mathbb{R}\setminus B_\rho (\varepsilon) }(y)   }
{ ( y - \varepsilon )^n} .
\end{displaymath}
\begin{displaymath}
\quad 
\end{displaymath}
$\varphi_T$ converges uniformly to its pointwise limit
\begin{displaymath}
\quad 
\end{displaymath}
\begin{displaymath}
\varphi_0 (y) \quad = \quad \alpha_l (y) \quad h_0 (y) \quad 
         \frac{  \textbf{1}_{\mathbb{R}\setminus B_\rho (\varepsilon) }(y)   }
{ ( y - \varepsilon )^n}
\end{displaymath}
\begin{displaymath}
\quad 
\end{displaymath}
as $T \to 0$. Moreover, note that $h_0$ is continuous, and so $\varphi_0$ is 
continuous in $\mu_l$. In summary,  
\begin{displaymath}
\quad 
\end{displaymath}
\begin{displaymath}
\int_\mathbb{R} dy \quad f_3 (y) \quad h_T (y) \quad \equiv \quad 
\int_\mathbb{R} dy \quad f_l' (y) \quad \varphi_0 (y) 
\end{displaymath}
\begin{displaymath}
\quad 
\end{displaymath}
\begin{displaymath}
\quad \quad \quad \quad \quad \quad \quad \quad \quad \quad \equiv 
\quad \int_\mathbb{R} dx \quad 
 f' (x) \quad \varphi_0 ( \mu_l + k_B T x)   
\end{displaymath}
\begin{displaymath}
\quad 
\end{displaymath}
\begin{displaymath}
\quad \quad \quad \quad \quad \quad \quad \quad \quad \quad \equiv \quad  
 - \varphi_0 ( \mu_l).   
\end{displaymath}

\begin{displaymath}
\quad 
\end{displaymath}
\begin{displaymath}
\quad 
\end{displaymath}

Consider now the zero temperature limits of those contributions to the rate 
$\Gamma_l^+ (6)$ in which the map $I^{conv}$ appears. They are up to prefactors
\begin{displaymath}
\quad 
\end{displaymath}
\begin{displaymath}
       I^{conv} ( \quad  T_{-E_{10}} \delta_{E_{10}}^2 
  \alpha_l^+ \quad ,\quad  \alpha^- \quad ,\quad  \delta_{E_{10}}^2 \alpha^+ \quad  ) 
\end{displaymath} 
and 
\begin{displaymath}
   I^{conv} (\quad  T_{-E_{10}} \delta_{E_{10}}^2 
  \alpha^+ \quad ,\quad  \delta_{E_{10}} \alpha_l^- \quad ,
  \quad  \delta_{E_{10}} \alpha^+ \quad  ).
\end{displaymath}
\begin{displaymath}
\quad
\end{displaymath}
In each of the two terms, the convolution of the second with the third argument 
converges in $||\quad ||_2 $ according to the last lemma, and so by another 
application of the Cauchy-Schwarz inequality the integral of the product of this 
convolution with the first argument converges as well. Those contributions to 
the rate $\Gamma_l^- (6)$ which contain the map $I^{conv}$ can be treated in the same
way.  
\begin{displaymath}
\quad
\end{displaymath}

The only remaining contributions to $\Gamma_l^+ (III) - \Gamma_l^+ (RTA, III)$ 
\begin{displaymath}
\quad 
\end{displaymath}
\begin{displaymath}
 \quad = \quad \frac{2}{\hbar} \quad \left\lbrace \quad 
 Tril (a) ( \alpha_l^+, \alpha^-, \alpha^+ ) \quad + \quad 
Tril (b) ( \alpha_l^-, \alpha^+, \alpha^+ ) \quad \right\rbrace
\end{displaymath}
\begin{displaymath}
\quad 
\end{displaymath} 
whose convergence is not clear with the lemma formulated for 
the discussion of 
the sixth order DSO are, up to a prefactor containing $\alpha^+ ( E_{10}) $,
\begin{displaymath}
\quad 
\end{displaymath}
\begin{displaymath}
  \left\lbrace   H[\alpha_l^+  (H\alpha^-)'] \quad  -   
 \quad H[\alpha_l^- (H \alpha^+)']  \right\rbrace' (E_{10}) \quad =       
\end{displaymath}
\begin{displaymath}
\quad
\end{displaymath} 
 \begin{displaymath}
  \left\lbrace   H[\alpha_l  (H\alpha^-)]' \quad  -   
 \quad H[\alpha_l' (H \alpha^-) ] 
\quad - \quad  H [ \alpha_l^- ( H \alpha )' ] 
  \right\rbrace' (E_{10})      \quad \equiv \quad 0.
\end{displaymath}
\begin{displaymath}
\quad
\end{displaymath}  
The existence of the zero temperature limit of  $\Gamma_l^- (III)$ can be
shown in the same way. 
\begin{displaymath}
\quad 
\end{displaymath}

\subsection*{Linear conductance within the sixth order} 
\begin{displaymath}
\quad 
\end{displaymath}
Within the present approximation scheme the stationary density matrix is given by
\begin{displaymath}
\quad 
\end{displaymath}
\begin{displaymath}
 \left( \begin{array}{c}  	
\rho_{00}      \\  \rho_{\sigma \sigma}   \\   
\rho_{  \bar{\sigma} \bar{\sigma}   } 
\end{array} \right)  \quad = \quad   \left( \begin{array}{c}  	
  \Gamma^-     \\  \Gamma^+   \\   \Gamma^+    
\end{array} \right) \frac{1}{ \Gamma^-   +  2 \Gamma^+ },
\end{displaymath}
\begin{displaymath}
\quad 
\end{displaymath}
and the particle current onto lead $l$ is 
\begin{displaymath}
\quad 
\end{displaymath}
\begin{displaymath}
I_l \quad = \quad  \rho_{00} \left( \Gamma_{\bar {l}}^+ -  \Gamma_l^+  \right)
          \quad + \quad \rho_{\sigma \sigma}
           \left( \Gamma_l^- -  \Gamma_{\bar {l}}^-  \right).
\end{displaymath}
\begin{displaymath}
\quad 
\end{displaymath}
Replace now the exact rates $\Gamma_{l'}^\pm $  by the exact sixth order rates 
$ \Gamma_{l'}^\pm (III)$ and assume symmetric coupling. Then the particle current onto 
lead $l$ is

\begin{displaymath}
\quad 
\end{displaymath}
\begin{displaymath}
I_l (III) \quad =  
\end{displaymath}
\begin{displaymath}
\quad 
\end{displaymath}
\begin{displaymath}
 \frac{2}{\hbar} \begin{LARGE} \textbf{\{} \end{LARGE} \quad
\pi  ( \alpha_{\bar{l}}^+ - \alpha_l^+ ) (E_{10}) \quad + \quad Bil (DSO)  \quad \quad
( \alpha_{\bar{l}}^+ - \alpha_l^+ \quad ,\quad \alpha + \alpha ^+ ) 
\end{displaymath}
\begin{displaymath}
 \quad \quad  \quad \quad \quad \quad \quad \quad  \quad \quad \quad 
 + \quad  Bil (RTA\setminus DSO)
\quad ( \alpha_{\bar{l}}^+ - \alpha_l^+ 
\quad , \quad \alpha^+ ) 
\end{displaymath}
\begin{displaymath}
 \quad  + \quad \quad  \quad  Tril (DSO)\quad \quad  
( \alpha_{\bar{l}}^+ - \alpha_l^+ \quad , 
\quad \alpha + \alpha^+ \quad ,\quad \alpha + \alpha^+)  
\end{displaymath}
\begin{displaymath}
 \quad   + \quad  Tril (RTA\setminus DSO) \quad ( \alpha_{\bar{l}}^+ - \alpha_l^+ 
\quad ,\quad  \alpha + \alpha^+ \quad ,\quad  \alpha^+) \quad  
\end{displaymath}
\begin{displaymath}
 \quad   + \quad  \quad Tril (a) \quad \quad ( \alpha_{\bar{l}}^+ - \alpha_l^+ 
\quad ,\quad  \alpha^-  \quad ,\quad  \alpha^+) \quad  
\end{displaymath}
\begin{displaymath}
 \quad   + \quad  \quad Tril (b) \quad \quad ( \alpha_{\bar{l}}^- - \alpha_l^- 
\quad ,\quad  \alpha^+  \quad ,\quad  \alpha^+) \quad  \quad \quad 
\begin{LARGE} \textbf{\}} \end{LARGE}  \rho_{00} \quad +
\end{displaymath}
\begin{displaymath}
\quad 
\end{displaymath}

\begin{displaymath}
 \frac{2}{\hbar} \begin{LARGE} \textbf{\{} \end{LARGE} \quad
\pi  ( \alpha_l^- - \alpha_{\bar{l}}^- ) (E_{10}) \quad + \quad Bil (DSO)  \quad \quad
( \alpha_l^- - \alpha_{\bar{l}}^- \quad ,\quad \alpha + \alpha ^+ ) 
\end{displaymath}
\begin{displaymath}
 \quad \quad  \quad \quad \quad \quad \quad \quad  \quad \quad \quad 
 + \quad  Bil (RTA\setminus DSO)
\quad ( \alpha_l^+ - \alpha_{\bar{l}}^+ 
\quad , \quad \alpha^- ) 
\end{displaymath}
\begin{displaymath}
 \quad  + \quad \quad  \quad  Tril (DSO)\quad \quad  
( \alpha_l^- - \alpha_{\bar{l}}^- \quad , 
\quad \alpha + \alpha^+ \quad ,\quad \alpha + \alpha^+)  
\end{displaymath}
\begin{displaymath}
 \quad   + \quad  Tril (RTA\setminus DSO) \quad ( \alpha_l^+ - \alpha_{\bar{l}}^+ 
\quad ,\quad  \alpha + \alpha^+ \quad ,\quad  \alpha^-) 
\end{displaymath}
\begin{displaymath}
 \quad   + \quad  \quad Tril (a) \quad \quad ( \alpha_l^- - \alpha_{\bar{l}}^- 
\quad ,\quad  \alpha^+  \quad ,\quad  \alpha^-) \quad  
\end{displaymath}
\begin{displaymath}
 \quad   + \quad  \quad Tril (b) \quad \quad ( \alpha_l^+ - \alpha_{\bar{l}}^+ 
\quad ,\quad  \alpha^-  \quad ,\quad  \alpha^-) \quad  \quad \quad 
\begin{LARGE} \textbf{\}} \end{LARGE}  \rho_{\sigma \sigma}.
\end{displaymath}
\begin{displaymath}
\quad 
\end{displaymath}

Including the electron charge, the linear conductance is

\begin{displaymath}
\quad 
\end{displaymath}
\begin{displaymath}
G (III) \quad =  
\end{displaymath}
\begin{displaymath}
\quad 
\end{displaymath}
\begin{displaymath}
 \frac{4 \pi e^2}{h} \begin{LARGE} \textbf{\{} \end{LARGE} \quad
\pi  ( \alpha_l f_T ) (E_{10}) \quad + \quad Bil (DSO)  \quad \quad
( \alpha_l f_T \quad ,\quad \alpha + \alpha ^+ ) 
\end{displaymath}
\begin{displaymath}
 \quad \quad  \quad \quad \quad \quad \quad \quad  \quad \quad \quad 
 + \quad  Bil (RTA \setminus DSO)
\quad ( \alpha_l f_T 
\quad , \quad \alpha^+ ) 
\end{displaymath}
\begin{displaymath}
 \quad  + \quad \quad  \quad  Tril (DSO)\quad \quad  
( \alpha_l f_T \quad , 
\quad \alpha + \alpha^+ \quad ,\quad \alpha + \alpha^+)  
\end{displaymath}
\begin{displaymath}
 \quad   + \quad  Tril (RTA\setminus DSO) \quad ( \alpha_l f_T 
\quad ,\quad  \alpha + \alpha^+ \quad ,\quad  \alpha^+) \quad  
\end{displaymath}
\begin{displaymath}
 \quad   + \quad  \quad Tril (a) \quad \quad ( \alpha_l f_T 
\quad ,\quad  \alpha^-  \quad ,\quad  \alpha^+) \quad  
\end{displaymath}
\begin{displaymath}
 \quad   + \quad  \quad Tril (b) \quad \quad ( - \alpha_l f_T 
\quad ,\quad  \alpha^+  \quad ,\quad  \alpha^+) \quad  \quad \quad 
\begin{LARGE} \textbf{\}} \end{LARGE}  \rho_{00} \quad +
\end{displaymath}
\begin{displaymath}
\quad 
\end{displaymath}

\begin{displaymath}
 \frac{ 4 \pi e^2}{ h } \begin{LARGE} \textbf{\{} \end{LARGE} \quad
\pi  ( \alpha_l f_T ) (E_{10}) \quad + \quad Bil (DSO)  \quad \quad
( \alpha_l f_T \quad ,\quad \alpha + \alpha ^+ ) 
\end{displaymath}
\begin{displaymath}
 \quad \quad  \quad \quad \quad \quad \quad \quad  \quad \quad \quad 
 + \quad  Bil (RTA\setminus DSO)
\quad ( - \alpha_l f_T 
\quad , \quad \alpha^- ) 
\end{displaymath}
\begin{displaymath}
 \quad  + \quad \quad  \quad  Tril (DSO)\quad \quad  
( \alpha_l f_T \quad , 
\quad \alpha + \alpha^+ \quad ,\quad \alpha + \alpha^+)  
\end{displaymath}
\begin{displaymath}
 \quad   + \quad  Tril (RTA\setminus DSO) \quad ( - \alpha_l f_T 
\quad ,\quad  \alpha + \alpha^+ \quad ,\quad  \alpha^-) 
\end{displaymath}
\begin{displaymath}
 \quad   + \quad  \quad Tril (a) \quad \quad ( \alpha_l f_T 
\quad ,\quad  \alpha^+  \quad ,\quad  \alpha^-) \quad  
\end{displaymath}
\begin{displaymath}
 \quad   + \quad  \quad Tril (b) \quad \quad ( - \alpha_l f_T 
\quad ,\quad  \alpha^-  \quad ,\quad  \alpha^-) \quad  \quad \quad 
\begin{LARGE} \textbf{\}} \end{LARGE}  \rho_{\sigma \sigma}.
\end{displaymath}
\begin{displaymath}
\quad 
\end{displaymath}

\subsection*{Divergence of $G(III)$ with $T \to 0$}
\begin{displaymath}
\quad 
\end{displaymath}
Assume that $E_{10} \neq E_F$ and investigate the behaviour of  $G (III)$ as 
$ T \to 0$. Consider first those terms contributing to $G (III)$ which contain 
the map $I^{conv}$ and which have the probability $\rho_{00}$ as prefactor. 
They are, up to convergent prefactors, 
\begin{displaymath}
\quad 
\end{displaymath}
\begin{displaymath}
I^{conv} \quad ( T_{-E_{10}} \delta_{E_{10}}^2 ( \alpha_l f_T ) \quad ,\quad  
\alpha^-  \quad , \quad  \delta_{E_{10}}^2 \alpha^+ )
\end{displaymath}
\begin{displaymath}
\quad 
\end{displaymath}
\begin{displaymath}
= \quad \int dy \quad \left( \delta_{E_{10}}^2 ( \alpha_l f_T ) \right) (y)
\quad \left\lbrace T_{E_{10}} \left[ \alpha^- * ( \delta_{E_{10}}^2 \alpha^+ ) 
\right] \right\rbrace (y)  
\end{displaymath}
\begin{displaymath}
\quad 
\end{displaymath}
and
\begin{displaymath}
\quad 
\end{displaymath}
\begin{displaymath}
I^{conv} \quad ( T_{-E_{10}} \delta_{E_{10}}^2 \alpha^+ \quad ,\quad  \delta_{E_{10}} 
( \alpha_l f_T ) \quad ,\quad  \delta_{E_{10}} \alpha^+ ) 
\end{displaymath}                  
\begin{displaymath}
\quad 
\end{displaymath}
\begin{displaymath}
= \quad I^{conv} \quad (  \delta_{E_{10}} ( \alpha_l f_T ) \quad , \quad 
T_{-E_{10}} \delta_{E_{10}}^2 \alpha^+ \quad  ,
\quad S \delta_{E_{10}} \alpha^+ ) 
\end{displaymath}              
\begin{displaymath}
\quad 
\end{displaymath}
\begin{displaymath}
= \quad \int dy \quad \left( \delta_{E_{10}} ( \alpha_l f_T ) \right) (y)
\quad  \left[ \left( T_{-E_{10}} \delta_{E_{10}}^2 \alpha^+ \right) * 
( S \delta_{E_{10}} \alpha^+ ) 
\right] (y).  
\end{displaymath}                                   
\begin{displaymath}
\quad
\end{displaymath}
Both of these terms are convergent according to the last lemma. The additive 
contributions to the linear conductance which have $\rho_{\sigma \sigma}$ as
prefactor and in which $I^{conv}$ appears can be treated in the same way.  
The divergence of all other contributions to the sum 
\begin{displaymath}
\quad 
\end{displaymath}      
\begin{displaymath}
Tril (a)   ( \alpha_l f_T 
\quad ,\quad  \alpha^-  \quad ,\quad  \alpha^+) \quad +
 \quad Tril (b)  ( - \alpha_l f_T 
\quad ,\quad  \alpha^+  \quad ,\quad  \alpha^+) 
\end{displaymath}      
\begin{displaymath}
\quad 
\end{displaymath}      
can be determined with the lemma formulated for the discussion of the sixth order
DSO, obtaining       
\begin{displaymath}
\quad 
\end{displaymath}      
\begin{displaymath}
\frac{4 \pi e^2}{h}\quad \rho_{00} \quad \begin{Large}
\textbf{\{} \end{Large} \quad \quad   
Tril (a)  \quad  ( \alpha_l f_T 
\quad ,\quad  \alpha^-  \quad ,\quad  \alpha^+) 
\end{displaymath}
\begin{displaymath}
 \quad \quad \quad  \quad \quad \quad +
 \quad Tril (b) \quad ( - \alpha_l f_T 
\quad ,\quad  \alpha^+  \quad ,\quad  \alpha^+)\quad  \begin{LARGE}
\textbf{\}} \end{LARGE} 
\end{displaymath}
\begin{displaymath}
\quad 
\end{displaymath}  
\begin{displaymath}
\equiv \quad \frac{\pi^2 e^2}{h} \quad \rho_{00} \quad  
\frac{\alpha^2 (E_F) }{ (E_{10} - E_F )^3} \quad 
\log \left( \frac{T }{T_0 } \right) 
\end{displaymath}
\begin{displaymath}
\quad 
\end{displaymath}  
\begin{displaymath}
\quad \quad \left\lbrace \quad 8 \alpha^+ (E_{10}) \quad + \quad
8 {\alpha^+}' (E_{10}) ( E_F - E_{10}) \quad - \quad 4 \alpha (E_F) 
\quad \right\rbrace , 
\end{displaymath}          
\begin{displaymath}
\quad 
\end{displaymath}
and analogously 
\begin{displaymath}
\quad 
\end{displaymath}  
\begin{displaymath}
\frac{4 \pi e^2}{h}\quad \rho_{\sigma \sigma} \quad \begin{Large}
\textbf{\{} \end{Large} \quad \quad   
Tril (a)  \quad  ( \alpha_l f_T 
\quad ,\quad  \alpha^+  \quad ,\quad  \alpha^-) 
\end{displaymath}
\begin{displaymath}
 \quad \quad \quad  \quad \quad \quad +
 \quad Tril (b) \quad ( - \alpha_l f_T 
\quad ,\quad  \alpha^-  \quad ,\quad  \alpha^-)\quad  \begin{LARGE}
\textbf{\}} \end{LARGE} 
\end{displaymath}
\begin{displaymath}
\quad 
\end{displaymath}  
\begin{displaymath}
\equiv \quad \frac{\pi^2 e^2}{h} \quad \rho_{\sigma \sigma} \quad  
\frac{\alpha^2 (E_F) }{ (E_{10} - E_F )^3} \quad 
\log \left( \frac{T }{T_0 } \right) 
\end{displaymath}
\begin{displaymath}
\quad 
\end{displaymath}  
\begin{displaymath}
\quad \quad \left\lbrace \quad -8 \alpha^- (E_{10}) \quad - \quad
8 {\alpha^-}' (E_{10}) ( E_F - E_{10}) \quad + \quad 4 \alpha (E_F) 
\quad \right\rbrace . 
\end{displaymath}          
\begin{displaymath}
\quad 
\end{displaymath}

In summary, 
\begin{displaymath}
\quad 
\end{displaymath}  
\begin{displaymath}
G (III)  
\end{displaymath}
\begin{displaymath}
\quad 
\end{displaymath}
\begin{displaymath}
\equiv \quad \frac{\pi^2 e^2}{h} \quad \rho_{00} \quad  
\frac{\alpha^2 (E_F) }{ (E_{10} - E_F )^3} \quad 
\log \left( \frac{T }{T_0 } \right) 
\end{displaymath}
\begin{displaymath}
\quad 
\end{displaymath}  
\begin{displaymath}
\quad \quad \left\lbrace  8 \alpha^+ (E_{10}) + 
8 {\alpha^+}' (E_{10}) ( E_F - E_{10})  +  4 \alpha^+ (E_{10}) 
 \right\rbrace  
\end{displaymath}          
\begin{displaymath}
\quad 
\end{displaymath}
\begin{displaymath}
+ \quad \frac{\pi^2 e^2}{h} \quad \rho_{\sigma \sigma} \quad  
\frac{\alpha^2 (E_F) }{ (E_{10} - E_F )^3} \quad 
\log \left( \frac{T }{T_0 } \right) 
\end{displaymath}
\begin{displaymath}
\quad 
\end{displaymath}  
\begin{displaymath}
\quad \quad \left\lbrace  -8 \alpha^- (E_{10})  - 
8 {\alpha^-}' (E_{10}) ( E_F - E_{10})  -  4 \alpha^- (E_{10}) 
\quad + \quad 12 \alpha (E_F) 
 \right\rbrace . 
\end{displaymath}          
\begin{displaymath}
\quad 
\end{displaymath}

In particular in the case $E_{10} < E_F$: 
\begin{displaymath}
\quad 
\end{displaymath}
\begin{displaymath}
G (III) 
\end{displaymath}
\begin{displaymath}
\quad 
\end{displaymath}
\begin{displaymath}
\equiv \quad \frac{4 \pi^2 e^2}{h} \quad   
\frac{\alpha^2 (E_F) }{ (E_{10} - E_F )^3} \quad 
\log \left( \frac{T }{T_0 } \right) 
\end{displaymath}
\begin{displaymath}
\quad 
\end{displaymath}  
\begin{displaymath}
\quad \quad \left\lbrace \quad  \rho_{00} \quad \left[ \alpha (E_{10}) 
+ 2 p (\alpha, E_{10} )_1 (E_F) \right] \quad + \quad 3 \rho_{\sigma \sigma }
\alpha (E_F)  \quad   \right\rbrace ,  
\end{displaymath}          
\begin{displaymath}
\quad 
\end{displaymath}                 
and in case $E_{10} > E_F$:                  
\begin{displaymath}
\quad 
\end{displaymath}
\begin{displaymath}
G (III) 
\end{displaymath}
\begin{displaymath}
\quad 
\end{displaymath}
\begin{displaymath}
\equiv \quad \frac{4 \pi^2 e^2}{h} \quad   
\frac{\alpha^2 (E_F) }{ (E_{10} - E_F )^3} \quad 
\log \left( \frac{T }{T_0 } \right) 
\end{displaymath}
\begin{displaymath}
\quad 
\end{displaymath}  
\begin{displaymath}
\quad \quad \left\lbrace \quad  \rho_{\sigma \sigma} \quad \left[ - \alpha (E_{10}) 
- 2 p (\alpha, E_{10} )_1 (E_F)  \right] \quad + \quad 3 \rho_{\sigma \sigma }
\alpha (E_F)  \quad   \right\rbrace ,  
\end{displaymath}          
\begin{displaymath}
\quad 
\end{displaymath}                 
where 
\begin{displaymath}
p(\alpha, \varepsilon_0)_1 (\varepsilon) \quad :=\quad \alpha ( \varepsilon_0 ) 
\quad + \quad \alpha' (\varepsilon_0) (\varepsilon - \varepsilon_0).
\end{displaymath}                 
\begin{displaymath}
\quad 
\end{displaymath}                 
The linear conductance can be expected to diverge logarithmically to infinity in case 
$ E_{10} < E_F $. The situation is ambiguous in case $E_{10} > E_F$, since the sign and 
magnitude of the possibly divergent term depends on the sign and magnitude of                  
\begin{displaymath}
\quad 
\end{displaymath}  
\begin{displaymath}
\rho_{\sigma \sigma} \quad \left\lbrace \quad   - \quad \alpha (E_{10}) \quad 
- \quad  2 p (\alpha, E_{10} )_1 (E_F)   \quad + \quad 3 
\alpha (E_F)  \quad   \right\rbrace .  
\end{displaymath}          
\begin{displaymath}
\quad 
\end{displaymath}                 
The sum in the curly bracket depends on the changing behaviour of the coupling function 
$\alpha$ between $E_F $ and $E_{10}$, while the factor $\rho_{\sigma \sigma }$ can be 
expected to be small in the regime $ E_{10} > E_F $.            
\begin{displaymath}
\quad 
\end{displaymath}

\subsubsection*{Differential conductance as function of the bias in sixth order 
                (expectation):}                 
\begin{displaymath}
\quad 
\end{displaymath}
The current at an arbitrary positive bias $V_b$ can be represented as 
\begin{displaymath}
\quad 
\end{displaymath}     
\begin{displaymath}
I_l \quad = \quad \int_0^{V_b} \quad d V_b' \quad \frac{d I_l}{d V_b} (V_b').
\end{displaymath}      
\begin{displaymath}
\quad 
\end{displaymath}      
On the other hand, the representation of the current in terms of the rates, 
Eq. (\ref{current in terms of rates}), 
implies the the zero temperature limit of the current exists, since the zero 
temperature limits of the rates exist. Hence, the divergence of the differential 
conductance at zero bias can be expected to be a singular behaviour at zero bias.
In particular in case $E_{10} < E_F$, the differential conductance versus the bias 
can be expected to display a maximum at zero bias which is getting more and more 
pronounced with decreasing temperature.

\begin{displaymath}
\quad 
\end{displaymath}

\section{Conclusion}
Within the real time approach the stationary reduced density of a quantum dot coupled 
to leads is obtained from the quantum master equation. The current across the 
quantum dot is obtained in a second step from the stationary reduced density matrix 
by the current kernel. The quantum master equation in the form of an equation for 
$\dot{\rho_\odot} (t) $ reads
\begin{displaymath}
\quad 
\end{displaymath}
\begin{displaymath}
\dot{\rho_\odot} (t) \quad = \quad \frac{-i}{\hbar} \left[ H_\odot , 
\rho_\odot (t) \right] \quad + \quad \int_0^t ds \quad K (t - s) 
\rho_\odot (s). 
\end{displaymath}
\begin{displaymath}
\quad 
\end{displaymath}
The kernel $K$ appearing in this equation has the structure 
\begin{displaymath}
\quad 
\end{displaymath}
\begin{displaymath}
K (t) \quad = \quad K (w,t) \quad = \quad 
      \sum_{n=1}^{\infty} w^{2n} K^{(2n)} (t), 
\end{displaymath}
\begin{displaymath}
\quad 
\end{displaymath}
where $w$ is the parameter which expresses the strength of the tunnel coupling 
between the leads and the quantum dot (coupling parameter). Assuming regularity 
conditions about those functions which describe the energy dependence of the 
tunnel coupling (coupling functions), it has been shown that for sufficiently small
values of $w$: 
\begin{displaymath}
\quad 
\end{displaymath}  
\begin{displaymath}
\sum_{n=1}^\infty \quad w^{2n} \quad \int_0^\infty dt \quad |K^{(2n)} (t)| \quad 
< \quad \infty . 
\end{displaymath}
\begin{displaymath}
\quad 
\end{displaymath}
If the initial reduced density matrix commutes with the Hamiltonian of the quantum dot, 
then this is the case at all later times (Ref. \cite{Kern11}, construction of the 
solution). It can  be concluded that for sufficiently small coupling the limit 
\begin{displaymath}
\quad 
\end{displaymath} 
\begin{displaymath}
\lim_{\lambda \to 0} \quad \lambda \quad \int_0^\infty dt \quad 
              \rho_\odot (t) e^{-\lambda t}
\end{displaymath}
\begin{displaymath}
\quad 
\end{displaymath}
exists, and that it is the normalized solution of the equation 
$K (\lambda = 0) \rho = 0$. The existence of the stationary limit of the current follows
from this. 
\begin{displaymath}
\quad 
\end{displaymath}

The stationary reduced density matrix is thus obtained from an equation of the form
\begin{displaymath}
\quad 
\end{displaymath}
\begin{displaymath}
\sum_{n=1}^{2n} w^{2n} K^{(2n)} (\lambda = 0) \quad \rho \quad = \quad 0,
\end{displaymath}
\begin{displaymath}
\quad 
\end{displaymath}
while the current is obtained from 
\begin{displaymath}
\quad 
\end{displaymath}
\begin{displaymath}
I_l \quad = \quad \sum_{n=1}^\infty w^{2n} Tr \left\lbrace
            K_c^{(2n)} (\lambda = 0) \quad \rho \right\rbrace .
\end{displaymath}
\begin{displaymath}
\quad 
\end{displaymath}
The kernels are analytic in the coupling parameter $w$ around $w = 0$. 
It follows that density matrix and current are analytic in the coupling parameter 
around $w = 0$. Indeed, the coefficients of their Taylor expansions 
up to order $2n$ are won by truncating the kernels at the corresponding 
order and calculating density matrix and current by the use of these 
approximate kernels.

\section{Why a perturbation theory for the Anderson model?}

There is a nice and straightforward interpretation of the current $I^{(2)}$ obtained from 
the second order kernels as the net effect of energy conserving 
one electron processes: If $a, b$ are quantum dot states with particle numbers
$N(b) = N(a) +1$, and if the dot is in the state $a$, then electrons can tunnel from
the leads to the dot in case their energy equals the difference $E_b - E_a$. 
On the other hand, {\em unoccupied} electron levels in the leads with this energy are 
needed for the inverse process. The rate of processes with the initial state 
$a$ and final state $b$ is proportional to the number of available occupied 
(unoccupied) electron levels in the leads with fitting energy. The probabilities 
of finding the dot in the 
possible states are obtained by the condition that in spite of the tunneling 
processes the effective change of the probabilities is zero. As an example, in the case of
the spinless quantum dot with only the states $0$ and $1$ the second order 
quantum master equation in the stationary limit up to second order reads:
 \begin{displaymath}
\quad 
\end{displaymath}
\begin{displaymath}
\rho_{00} {K^{(2)}}_{00}^{11} \quad = \quad \rho_{11} {K^{(2)}}_{11}^{00}. 
\end{displaymath}
\begin{displaymath}
\quad 
\end{displaymath}
The stationary electron current onto lead $l$ is obtained in a second step 
by balancing the 
absolute number of one electron processes during which the particle number on 
this particular lead changes.

\begin{displaymath}
\quad 
\end{displaymath}
On the other hand, the definition of the stationary current $I_l$ across the quantum dot 
within the real time approach is abstract, even the existence is non-trivial. 
The central statement of the perturbation theory connects the objects $I_l$ and 
$I_l^{(2)}$.

\begin{displaymath}
\quad 
\end{displaymath}
One unsystematic possibility to take into account simultaneous tunneling of {\em two} 
electrons is the following (Ref. \cite{Report12}): Assume a quantum dot 
with the states $\sigma, \bar{\sigma}$, $\sigma \in \{ \uparrow,\downarrow \}$ with not 
necessarily equal energies $E_\sigma$. If the quantum dot is in the state $\sigma$, then an 
electron of the opposite spin might tunnel from an occupied level $\nu_{\bar{\sigma}}$
in the leads  onto the dot while the electron on the dot 
leaves it towards an unoccupied level $\nu_\sigma$ in one of the leads.
The equation expressing the energy conservation reads: 
\begin{displaymath}
\quad 
\end{displaymath}
\begin{displaymath}
\varepsilon_{\nu_{\bar{\sigma}}} \quad - \quad \varepsilon_{\nu_\sigma}
\quad = \quad E_{\bar{\sigma} \sigma } \quad := \quad E_{\bar{\sigma}} 
- E_\sigma .
\end{displaymath}
\begin{displaymath}
\quad 
\end{displaymath}
Let ${\cal D}^\pm_{l \sigma} (\varepsilon ) d \varepsilon $ be the density of 
occupied/unoccupied electron 
levels in lead $l$ with spin $\sigma $ in an interval of width $d\varepsilon$ around 
$\varepsilon$, and let $ {\cal D}_\sigma^\pm := \sum_l {\cal D}_{l\sigma}^\pm$.

A measure for the number of pairs consisting of an 
unoccupied electron level $\nu_\sigma$ with spin $\sigma$ and an occupied 
electron level $\nu_{\bar{\sigma}}$ with spin $\bar{\sigma}$ with the property that 
the difference of their energies is smaller than some arbitrary given constant $E$ 
is 
\begin{displaymath}
\quad 
\end{displaymath} 
\begin{displaymath}
{\cal N}(E) \quad := \quad 
\int \int_{\{ ( \varepsilon , \varepsilon') : \quad \varepsilon' - \varepsilon < E \}} \quad  
d \varepsilon d \varepsilon' \quad  
{\cal D}_\sigma^- (\varepsilon ) {\cal D}_{\bar{\sigma}}^+ (\varepsilon').  
\end{displaymath}
\begin{displaymath}
\quad 
\end{displaymath}
Hence, the number of such pairs of electron levels with the property that the
difference of the two energies lies in an interval of width $dE $ around $E$ is 
measured by 
\begin{displaymath}
\quad 
\end{displaymath}
\begin{displaymath}
dE \quad \frac{d {\cal N}}{ dE } (E) \quad = \quad dE \quad \int d\varepsilon 
\quad {\cal D}_\sigma^- (\varepsilon ) {\cal D}_{\bar{\sigma}}^+ (E + \varepsilon).
\end{displaymath}
\begin{displaymath}
\quad 
\end{displaymath}
For any two lead indices $l,l'$ define 
\begin{displaymath}
\quad 
\end{displaymath}
\begin{displaymath}
\Gamma_{l'}^l (\sigma \to \bar{\sigma} ) \quad := \quad 
\int d\varepsilon \quad {\cal D}_{l \sigma}^- (\varepsilon ) 
     {\cal D}_{l'\bar{\sigma}}^+ (E_{\bar{\sigma} \sigma} + \varepsilon).
\end{displaymath}
Assume a rate of energy conserving two electron processes during which 
the quantum dot state switches from $\sigma $ to $\bar{\sigma}$, and during 
which one electron enters lead $l$ while one electron of opposite spin 
leaves lead $l'$ - assume such a rate which is proportional to 
the quantity $\Gamma_{l'}^l (\sigma \to \bar{\sigma} )$. Let 
$\Gamma (\sigma \to \bar{\sigma} ) := \sum_{l,l'}
\Gamma_{l'}^l (\sigma \to \bar{\sigma} ) $, determine the stationary 
probabilities $\rho_{\sigma \sigma}$  by 
\begin{displaymath}
\quad 
\end{displaymath}
\begin{displaymath}
\rho_{\sigma \sigma } \Gamma (\sigma \to \bar{\sigma} )  \quad = \quad 
\rho_{\bar{\sigma} \bar{ \sigma} } \Gamma (\bar{\sigma} \to \sigma )
\end{displaymath}
\begin{displaymath}
\quad 
\end{displaymath}
and the normalization, and assume an effective electron current onto lead $l$
given by 
\begin{displaymath}
\quad 
\end{displaymath}
\begin{displaymath}
I_l \quad = \quad const \quad \sum_\sigma \quad \rho_{\sigma \sigma} \quad 
    \left[   \Gamma_{\bar{l}}^l (\sigma \to \bar{\sigma} ) \quad - \quad 
        \Gamma_l^{\bar{l}} (\sigma \to \bar{\sigma} ) \right]. 
\end{displaymath}
\begin{displaymath}
\quad 
\end{displaymath}
Consider the dependence of this quantity on an applied bias at first in the case
$E_\sigma = E_{\bar{\sigma}}$. Assume symmetry in the leads with respect to the 
spin,  ${\cal D}_{l \sigma} = {\cal D}_{ l \bar{\sigma} }$.
In the case of equal energies the current turns into 
\begin{displaymath}
\quad 
\end{displaymath}
\begin{displaymath}
I_l \quad = \quad const \quad \int d \varepsilon \quad \left( 
{\cal D}_{l \sigma} {\cal D}_{\bar{l} \sigma }   \right) ( \varepsilon ) \quad 
( f_{\bar{l}} -  f_l ) (\varepsilon) 
\end{displaymath}
\begin{displaymath}
\quad 
\end{displaymath}
with $f_{l'}$  the Fermi Dirac distribution of lead $l'$. Assuming constant densities 
${\cal D}_{l \sigma} (\varepsilon) $, the current grows linearly with the bias, the differential 
conductance is constant and positive. 
\begin{displaymath}
\quad 
\end{displaymath}

Consider now the quantity $I_l$ in the case of different energies, 
$E_\sigma < E_{\bar{\sigma}}$. For simplicity assume constant and equal 
densities, ${\cal D}_{l \sigma } (\varepsilon) = D$, and consider only the 
zero temperature limit. Then 
\begin{displaymath}
\quad 
\end{displaymath}
\begin{displaymath}
\Gamma_{l'}^l (\sigma \to \bar{\sigma} ) \quad = \quad D^2 \quad 
 h ( \mu_{l'} - \mu_l - E_{\bar{\sigma} \sigma }  )
\end{displaymath}        
\begin{displaymath}
\quad 
\end{displaymath}      
where
\begin{displaymath}
h (E) \quad := \quad \frac{1}{2} (sign (E) + 1) \quad E. 
\end{displaymath}      
\begin{displaymath}
\quad 
\end{displaymath}      
As a consequence, 
\begin{displaymath}
\quad 
\end{displaymath}      
\begin{displaymath}
\Gamma ( \sigma \to \bar{\sigma} ) \quad = \quad D^2 \quad  \left[ 
2 h ( -E_{\bar{\sigma} \sigma } ) + h ( eV_b -E_{\bar{\sigma} \sigma } )
+ h ( -eV_b -E_{\bar{\sigma} \sigma } )  \right],
\end{displaymath}      
\begin{displaymath}
\quad 
\end{displaymath}      
\begin{displaymath}
\Gamma ( \bar{\sigma} \to \sigma ) \quad = \quad D^2 \quad  \left[ 
2 h ( E_{\bar{\sigma} \sigma } ) + h ( eV_b + E_{\bar{\sigma} \sigma } )
+ h ( -eV_b + E_{\bar{\sigma} \sigma } )  \right].
\end{displaymath}      
\begin{displaymath}
\quad 
\end{displaymath}  
It follows that in the assumed case $E_{\bar{\sigma} \sigma} > 0$ the probability of 
finding the dot in the energetically higher state, 
\begin{displaymath}
\quad 
\end{displaymath}      
\begin{displaymath}
\rho_{\bar{\sigma} \bar{\sigma}} \quad = \quad \frac{\Gamma ( \sigma \to \bar{\sigma} )}
{ \Gamma ( \bar{\sigma} \to \sigma )  +  \Gamma ( \sigma \to \bar{\sigma} )  } ,
\end{displaymath}      
\begin{displaymath}
\quad 
\end{displaymath}
is zero as long as the absolute value of the bias is smaller than 
$E_{\bar{\sigma} \sigma}$. The current is then 
\begin{displaymath}
\quad 
\end{displaymath}       
\begin{displaymath}
I_l \quad = \quad const \quad  \rho_{\sigma \sigma} \quad 
    \left[   \Gamma_{\bar{l}}^l (\sigma \to \bar{\sigma} ) \quad - \quad 
        \Gamma_l^{\bar{l}} (\sigma \to \bar{\sigma} ) \right]  
\end{displaymath}      
\begin{displaymath}
\quad \quad  = \quad 0.
\end{displaymath} 
\begin{displaymath}
\quad 
\end{displaymath}     
In case $|e V_b| > E_{\bar{\sigma} \sigma}$ both probabilities are strictly 
positive, and the current can be expected to become a non-constant function 
of the bias.           
\begin{displaymath}
\quad 
\end{displaymath}

In summary, the behaviour of the differential conductance as a function of the bias
as obtained within the present unsystematic approach displays the following features: 
\begin{itemize}
\item In case of equal energies $E_\sigma = E_{\bar{\sigma}}$ the assumed energy 
conserving two electron processes give rise to a finite positive value of the differential 
conductance. Not any anomaly at zero bias can be expected.    
\item In case of a nonzero difference $E_{\bar{\sigma}} - E_\sigma =: E_{\bar{\sigma} \sigma}
\neq 0$ the differential conductance as function of the bias is zero as long as 
$|e V_b | \le |E_{\bar{\sigma} \sigma}|$, while presumably positive values can be expected to 
be regained outside this interval of values of the bias.  
\end{itemize}         
\begin{displaymath}
\quad 
\end{displaymath}         
Both of these properties are in agreement with the systematic theoretical treatment of 
Ref. \cite{Koller_Diss}, which takes into account kernels of the diagrammatic real time 
approach up to fourth order. 
\begin{displaymath}
\quad 
\end{displaymath}

There is an argument which seems to indicate that the basic principle of 
energy conserving two electron processes is not contained in the fourth order of the 
diagrammatic real time approach, and hence not in the Anderson model: The equation by 
which the stationary density matrix of the quantum dot is determined is the quantum 
master equation in the stationary limit. For any quantum dot state 
$ a_0 \in \left\lbrace 0, \uparrow, \downarrow, 2 \right\rbrace $: 
\begin{displaymath}
\quad 
\end{displaymath}
\begin{displaymath}
0 \quad = \quad <a_0| ( K \rho) a_0> \quad = \quad \sum_a \rho_{aa} K_{aa}^{a_0 a_0}
\end{displaymath}
\begin{displaymath}
\quad \quad  = \quad \sum_{a \neq a_0 }\rho_{aa} K_{aa}^{a_0 a_0} \quad - \quad 
      \rho_{a_0a_0}\sum_{a \neq a_0 } K_{a_0 a_0}^{aa} .                
\end{displaymath}
The equation would allow the interpretation of the kernel elements $K_{aa}^{bb}$ as 
a rate with which the quantum dot state switches from $a$ to $b$. The kernel element 
$K_{\sigma \sigma}^{\bar{\sigma} \bar{\sigma}}$ is zero in second order. Neglecting 
the state $2$, there is only one fourth order diagram contributing to the 
kernel element (Fig. \ref{topology}). Its direct calculation yields 
\begin{displaymath}
\quad 
\end{displaymath}         
\begin{displaymath}
K_{\sigma \sigma}^{\bar{\sigma} \bar{\sigma}} \quad = \quad \frac{2}{\hbar}
  \quad I (RTA \setminus DSO ) \quad \left(  T_{-E_{\sigma 0}} \alpha^- , 
  T_{-E_{\bar{\sigma} 0 }} \alpha^+ \right)
\end{displaymath}      
\begin{displaymath}
\quad
\end{displaymath}      
with
\begin{displaymath}
I (RTA \setminus DSO )\quad (f,g)  \quad = \quad \pi \int_\mathbb{R} d \varepsilon 
\quad  \left[ ( \delta f ) (\delta g )\right] (\varepsilon )    
\end{displaymath}         
\begin{displaymath}
\quad 
\end{displaymath}
\begin{displaymath}
\quad \quad \quad \quad \quad \quad \quad \quad \quad \quad \quad = \quad \pi^2 
               \left\lbrace H^{(1)} (fg) - 
               f H^{(1)} g - g H^{(1)} f
\right\rbrace (0) .
\end{displaymath}         
The map $I(RTA \setminus DSO ) $ can take smooth and bounded functions as arguments. 
If the coupling functions $\alpha_l (\varepsilon )$ are assumed to be constant, then 
the equality 
\begin{displaymath}
\quad 
\end{displaymath}         
\begin{displaymath}
K_{\sigma \sigma}^{\bar{\sigma} \bar{\sigma}} \quad = \quad 
K_{\bar{\sigma}\bar{ \sigma}}^{\sigma \sigma}
\end{displaymath}         
\begin{displaymath}
\quad 
\end{displaymath}         
holds even if $E_{\bar{\sigma}} \neq E_\sigma$. Moreover, the matrix element is 
negative in case of constant coupling and equal energies.        
\begin{displaymath}
\quad 
\end{displaymath}

In the last section of this text I presented and discussed results for the
kernels up to sixth order in the coupling, neglecting the doubly occupied state
and in the case of equal energies $E_{\sigma 0} = E_{\bar{\sigma}0} =: E_{10}$. 
The resulting differential conductance versus the bias is expected to display a
maximum at zero bias which becomes more and more pronounced with lower and lower 
temperatures in case $E_{10}$ lies below the Fermi level of the leads.   
However, I do not or not yet find it intuitive that a process including altogether 
three electron levels in the leads should give rise to a resonance at zero bias.

\section*{Acknwoledgements}
Thanks go to Milena Grifoni, Prof., Univ. Regensburg, for remarks on a previous 
version of version one of this arXiv article. Moreover, thanks go to the DFG for 
financial support within the framework of the GRK 1570 during that part of the work 
which is covered by version one. Many thanks go to Simone Gutzwiller, Dr., for 
system administration.

Ref. \cite{Koller_Diss} (introduction) has been a reason for me to try and 
calculate sixth order diagrams. Thanks for correspondences in the context of 
Ref. \cite{Koller_Diss} and \cite{Schmid98} go to Sonja Koller, Dr., and J\"urgen Weis, 
Dr., Max-Planck-Institut f\"ur Festk\"orperforschung, Stuttgart, respectively.

Thanks go to Universit\"at Regensburg and FSU Jena for electronic mailing address and 
library services, respectively.

\newpage

\section*{Appendix A: Exponential decay of the Fourier transforms}

Let $\alpha_l$ be a non-negative, real-valued function of a real
variable $x$, which is integrable, quadratically integrable, and whose 
Fourier transform decays exponentially. More precisely, assume, that, with 
constants $c_{\alpha_l}, K_{\alpha_l} > 0$, the inequality 
\begin{displaymath}
|( {\cal F} \alpha_l ) (y)| \quad \le \quad K_{\alpha_l} e^{- c_{\alpha_l} |y|} 
\end{displaymath}
holds for all $y \in \mathbb{R}$.

The function $\alpha_l$ is then an 
element in the space of functions ${\cal R} $ defined in section
\ref{theory for diagrams}. Moreover, assume that the Fourier transform of 
$\alpha_l$ is Lipschitz continuous with constant $L_{\alpha_l} < \infty$, 
i.e., for all $y,y' \in \mathbb{R}$: 
\begin{displaymath}
|{\cal F} \alpha_l (y) - {\cal F} \alpha_l (y') | \quad \le \quad 
|y - y'| L_{\alpha_l} .
\end{displaymath}
Finally, define for $y_0 \ge 0$:
\begin{displaymath}
L(\alpha_l) (y_0) \quad := \quad \sup \left\lbrace \left| 
\frac{{\cal F} \alpha_l (y)  - {\cal F} \alpha_l (y') }{ y - y'} \right| : 
y > y' \ge y_0 \mbox{ or } y < y' \le -y_0  \right\rbrace
\end{displaymath}
and assume that the function $L (\alpha_l) $ decays exponentially, 
\begin{displaymath}
L(\alpha_l ) (y) \quad \le \quad K_{L(\alpha_l)} e^{-c_{L(\alpha_l)} y}
\end{displaymath}
with constants $ K_{L (\alpha_l) }, c_{L(\alpha_l)} > 0$.

Examples of functions possessing all properties assumed about $\alpha_l$
are lorentzians, gaussians, but also convolutions of measurable and bounded functions
with compact support with lorentzians or gaussians.

\underline{Definition:}
Let $f_l (x)$ be the Fermi-Dirac distribution at chemical potential $\mu_l$ 
and temperature $T$,
\begin{displaymath}
f_l (x) \quad = \quad f \left( \frac{x - \mu_l}{k_B T} \right) 
\end{displaymath} 
with $f (y) = (1 + e^y)^{-1}$, the normalized distribution. Let the function
$\alpha_l^+$ be defined as $\alpha_l^+ : = \alpha_l f_l $.

\underline{Statement:}                        
The Fourier transform of $\alpha_l^+ $ decays exponentially. The constants
$ K_{\alpha_l^+}, c_{\alpha_l^+} > 0$ in an estimate  
\begin{displaymath}
| {\cal F} \alpha_l^+ (y) | \quad \le \quad K_{\alpha_l^+} e^{- c_{\alpha_l^+} |y|}
\end{displaymath}  
can be chosen locally independent of $\mu_l $.

\underline{Additional statement (decay of $\alpha_l$):}
There are $c > 0 $ and 
$x_0 > 0$ such, that for all $x \in \mathbb{R}$:
\begin{displaymath}
|\alpha_l (x) | \quad \le \quad \frac{c}{x_0 + |x|}.
\end{displaymath}

\subsection*{\textbf{Proof:}}

Note that 
\begin{equation}
{\cal F} ( \alpha_l f_l ) (x) \quad = \quad k_B T e^{-ix\mu_l} \quad {\cal F}
(\gamma f) (x k_B T) 
\label{connection between F (alpha_l f_l) and F (f gamma) }
\end{equation}
with
\begin{displaymath}
\gamma (x) \quad := \quad (T_{-\mu_l} \alpha_l ) ( k_B T x) .
\end{displaymath}

\underline{Remark:}
The Fourier transform of $\gamma$ is 
\begin{displaymath}
( {\cal F} \gamma ) (x) \quad =\quad \frac{1}{k_B T} e^{i \mu_l \frac{x}{k_B T}}
\quad ({\cal F} \alpha_l ) \left(\frac{x}{k_B T}\right) ,
\end{displaymath}
so ${\cal F \gamma }$ decays exponentially, 
\begin{displaymath}
|{\cal F \gamma } (y) | \quad \le \quad K_{\gamma} e^{- c_{\gamma} |y|}, 
\end{displaymath}
and it is Lipschitz continuous, 
\begin{displaymath}
|{\cal F} \gamma (y) - {\cal F} \gamma (y') | \quad \le \quad 
|y - y'| L_{\gamma},
\end{displaymath}
with constants
\begin{eqnarray*}
K_\gamma \quad & = & \quad \frac{K_{\alpha_l}}{k_B T}, \\
c_\gamma \quad & = & \quad \frac{c_{\alpha_l}}{k_B T}, \\
L_\gamma \quad & = & \quad \frac{1}{(k_B T)^2} 
           \left( L_{\alpha_l} + |\mu_l| c_{\alpha_l} \right).  
\end{eqnarray*}
Finally, if for $x_0 > 0$
\begin{displaymath}
L(\gamma) (x_0) \quad := \quad \sup \left\lbrace \left| 
\frac{{\cal F} \gamma (x)  - {\cal F} \gamma (x') }{ x - x'} \right| : 
x > x' \ge x_0 \mbox{ or } x < x' \le -x_0  \right\rbrace ,
\end{displaymath} 
then the function $L(\gamma) $ decays exponentially, 
\begin{displaymath}
L(\gamma ) (x) \quad \le \quad K_{L(\gamma)} e^{-c_{L(\gamma)} x} ,
\end{displaymath}
with constants
\begin{eqnarray*}
K_{L(\gamma)} \quad &:=& \quad  \frac{1}{(k_B T)^2} 
\left(   K_{L(\alpha_l)} + K_{\alpha_l} |\mu_l| \right) ,\\ 
c_{L(\gamma)} \quad &:=& \quad  \frac{1}{k_B T} 
                      \min ( c_{\alpha_l} , c_{L(\alpha_l)} )    .
\end{eqnarray*}

Define for $\lambda \in ]0, 1[$ the function 
$f_\lambda$ by
\begin{displaymath}
 f_\lambda (x) \quad := \quad  f (x) e^{ \lambda x} .
\end{displaymath}
For any fixed value of $\lambda$, $f_\lambda$ decays exponentially. Moreover, 
there is the pointwise convergence
\begin{displaymath}
 f_\lambda \gamma \to f \gamma \quad (\lambda \to 0), 
\end{displaymath}
and the convergence has an integrable upper bound, since $\gamma$ is 
integrable. Hence: 
\begin{displaymath}
 {\cal F} ( \gamma f ) (y) \quad = \quad  \lim_{\lambda \to 0} 
    \quad                   {\cal F} ( f_\lambda \gamma ) (y) 
\end{displaymath}
for every single value of $y$.

Apply now the rule $({\cal F} g)(y) = \frac{1}{(iy)^n} {\cal F} 
\left( g^{(n)} \right) (y)$ \cite{Hackenbroch} to the case 
$g = f_\lambda$: The function $f_\lambda$ is smooth, $f_\lambda, f_\lambda' $ 
as well as $f_\lambda''$ decay exponentially. Hence, it follows with integration
by parts, that 
\begin{displaymath}
({\cal F} f_\lambda ) (x) \quad = \quad \frac{1}{(ix)^2} ({\cal F} f_\lambda'') (x)
\end{displaymath}
decays quadratically. (Let the latter property be defined through the existence of 
$a, b > 0$ such, that for all $x \in \mathbb{R}$: $ |({\cal F} f_\lambda ) (x)|
\le \frac{a}{1 + (x/b)^2} $.)

The Fourier transform of $\gamma$  decays even exponentially. The convolution 
of two quadratically decaying functions decays quadratically, so the convolution 
theorem for Fourier transforms can be applied (let $(Sg)(x) := g (-x)$ as in Sec. 
\ref{theory for diagrams}): 
\begin{displaymath}
{\cal F}^{-1} \left[ ({\cal F} \gamma ) * (S {\cal F} f_\lambda ) \right]
\quad = \quad 2 \pi \quad \gamma f_\lambda,
\end{displaymath}
hence
\begin{eqnarray}
{\cal F} (\gamma f_\lambda) \quad &=& \quad \frac{1}{2 \pi} \quad  
({\cal F} \gamma ) * (S {\cal F} f_\lambda ) \nonumber \\
&=& \quad ({\cal F} \gamma ) * ({\cal F}^{-1} f_\lambda ) .
\label{convolution}
\end{eqnarray}

At the end of this appendix the equality             
\begin{displaymath}
 \left( {\cal F}^{-1} f_\lambda \right)(y) \quad = \quad \frac{i}{2}
  \quad \frac{1} {\sinh (\pi(y - i\lambda))}
\end{displaymath}
will be shown. Using this information and the notation:
\begin{displaymath}
 H(z) := \frac{\pi z}{\sinh (\pi z)} \quad (\mbox{holomorphic on } 
\left\lbrace Im \in ]-1,1[ \right\rbrace ,
\mbox{ with } H(0) = 1),
\end{displaymath}
 the limit $\lambda \to 0 $ of equation (\ref{convolution}) turns into: 
\begin{displaymath}
 {\cal F} (\gamma f) (y) \quad = \quad  \frac{-1}{2\pi} \quad  \lim_{\lambda \to 0} 
 \int_\mathbb{R} dx  \quad \frac{({\cal F} \gamma)(x+y) \quad H(x-i\lambda)}
 {\lambda + ix}.
\end{displaymath}

Integrate like
\begin{displaymath}
 \int_{-\infty}^\infty dx \quad function(x) \quad = \quad  \int_0^\infty dx \quad 
 \left( function (-x) + function (x) \right);
\end{displaymath}
then, multiply numerator and denominator of the fractions 
with the complex conjugate of their denominators. Upon grouping the contributions 
according to their prefactors two summands are obtained. The first summand reads after  
an elementary integral transformation, omitting the prefactor $\frac{-1}{2 \pi} $:
\begin{displaymath}
 \int_0^\infty dx \frac{\left( {\cal F} \gamma \right) (-\lambda x +y) 
		    H(\lambda(-x-i))  
	   +   \left( {\cal F} \gamma \right) (\lambda x +y) 
		    H(\lambda(x-i)) }    {1+x^2} ,
\end{displaymath}
\begin{displaymath}
 \to  \pi \left( {\cal F} \gamma \right) (y)  \quad (\lambda \to 0) \quad 
\mbox{ with Lebesgue, Lebesgue.}
\end{displaymath}
The corresponding contribution in an additive decomposition 
\begin{displaymath}
{\cal F} (f \gamma ) (y) \quad = \quad \sum_i F_i (y)
\end{displaymath}
of ${\cal F} (f \gamma ) (y)$ is 
\begin{displaymath}
F_1 (y) \quad = \quad \frac{-1}{2} ({\cal F} \gamma ) (y).
\end{displaymath}

The second integral without prefactor is
\begin{displaymath}
i \int_0^\infty dx \frac{x}{\lambda^2 + x^2} \left\lbrace 
\left( {\cal F} \gamma \right) (-x + y) H ( -x-i \lambda)     - 
\left( {\cal F} \gamma \right) (x + y) H ( x-i \lambda) \right\rbrace 
\end{displaymath}
 \begin{eqnarray} \label{decay}
  &=& i \int_0^\infty  dx \frac{x}{\lambda^2 + x^2}  \left( {\cal F} \gamma 
\right) (-x + y)
	\left\lbrace H(-x-i\lambda) - H(x-i\lambda) \right\rbrace +  \nonumber \\
&&   i \int_0^\infty dx  \frac{x}{\lambda^2 + x^2} H(x-i\lambda) 
\left\lbrace \left( {\cal F} \gamma \right) (-x + y) -  \left( {\cal F} \gamma 
\right) (x + y) \right\rbrace. 
 \end{eqnarray}
To the first line of the right hand side of Eq. (\ref{decay}),  the theorem of 
Lebesgue can be applied; because of the symmetry $H(x) = H(-x)$, it is  
zero. For the treatment of the second line use that ${\cal F}
\gamma$ satisfies the Lipschitz condition noted above. Hence, the convergence 
theorem can be applied also to the second line and one obtains in the limit 
$\lambda \to 0$: 
\begin{displaymath}
 -i\pi \int_0^\infty dx \frac{\left({\cal F} \gamma \right) (y+x) - 
\left({\cal F} \gamma \right) (y-x)}{\sinh (\pi x)}.
\end{displaymath}

So far, the Fourier transform of $f \gamma$ has been additively decomposed 
according to 
\begin{equation}
{\cal F} (f \gamma ) (y) \quad = \quad  (F_1 + F_2 )(y)
\label{additive decomposition of f gamma}
\end{equation}
with 
\begin{displaymath}
F_1 (y) \quad   =  \quad \frac{-1}{2} ({\cal F} \gamma ) (y),
\end{displaymath}
\begin{displaymath}
F_2 (y) \quad   =  \quad  \frac{i}{2} \int_0^\infty dx \frac{\left({\cal F} 
\gamma \right) (y+x) - \left({\cal F} \gamma \right) (y-x)}{\sinh (\pi x)} .
\end{displaymath}

Verify now that the latter integral decays exponentially as a function of $y$. 
Write $\int_0^\infty = \int_0^{x_0} + \int_{x_0}^{\infty} $ with arbitrary
$x_0 > 0$ and treat the two intervals separately. Note that 
\begin{eqnarray*}
  |i/2| \int_{x_0}^\infty dx \left| \frac{ \left( {\cal F} \gamma \right) (y+x) }
{ \sinh (\pi x)}  \right|
&\le  & \quad K_\gamma  \int_{x_0}^\infty dx \quad \frac{ e^{-c_\gamma |y + x|}}  
{ e^{ \pi x  } - e^{ -\pi x } } \\ 
& \le & \quad K_\gamma  \int_{x_0}^\infty dx 
\quad \frac{ e^{-c_\gamma |y + x|} e^{-\pi x} } { 1 - e^{ -2 \pi x_0 } } 
\\ & \le &
\quad \frac{K_\gamma } {1 - e^{ -2 \pi x_0 }} \int_{-\infty}^{\infty} dx \quad  
e^{-c_\gamma |y + x|} e^{-\pi |x|}  \\ & \le &
\quad \frac{K_\gamma } {1 - e^{ -2 \pi x_0 }} \left( \frac{2}{c_\gamma} 
+ \frac{2}{\pi} \right) e^{-\frac{1}{2} \min(\pi, c_\gamma) |y| }  \\ 
& =: & \quad K_1   e^{- c_1 |y| }, \\
  |i/2| \int_{x_0}^\infty dx \left| \frac{ \left( {\cal F} \gamma \right) (y+x) }
{ \sinh (\pi x)}  \right| & \le &  \quad K_1   e^{- c_1 |y| }.
\end{eqnarray*}

Moreover, for $y \in \mathbb{R} $ with $|y| > x_0$:  
\begin{displaymath}
  \left| i/2 \right| \left| \int_0^{x_0} dx \quad \frac{\left({\cal F} 
  \gamma \right) (y+x) - \left({\cal F} \gamma \right) (y-x)} {\sinh (\pi x)} \right| 
\end{displaymath}
\begin{displaymath}
  \le \quad  \int_0^{x_0} dx \quad \left| \frac{\left({\cal F} \gamma \right) (y+x) - 
\left({\cal F} \gamma \right) (y-x)} {2x}   \frac{x}    {\sinh (\pi x)} \right|
\end{displaymath}
\begin{displaymath}
 \le \quad 1/\pi  \int_0^{x_0} dx \quad \left| \frac{\left({\cal F} \gamma \right) (y+x) - 
\left({\cal F} \gamma \right) (y-x)} {2x} \right|
\end{displaymath}
\begin{displaymath}
\le \quad \frac{x_0}{\pi}  L (\gamma ) ( |y| - x_0 ) 
\end{displaymath}
\begin{displaymath}
\le \quad  \frac{x_0}{\pi} K_{L(\gamma)} e^{c_{L(\gamma)} x_0 } \quad e^{-c_{L(\gamma)} |y|},
\end{displaymath}
while for $y \in \mathbb{R}$ with $ |y| \le x_0 $:
\begin{displaymath}
  \left| i/2 \right| \left| \int_0^{x_0} dx \quad \frac{\left({\cal F} 
  \gamma \right) (y+x) - \left({\cal F} \gamma \right) (y-x)} {\sinh (\pi x)} \right| 
\end{displaymath}
\begin{displaymath}
\le \quad \frac{x_0}{\pi} L_\gamma .
\end{displaymath}
In summary, for all $y \in \mathbb{R}$: 
\begin{displaymath}
\left| i/2 \right| \left| \int_0^{x_0} dx \quad \frac{\left({\cal F} 
  \gamma \right) (y+x) - \left({\cal F} \gamma \right) (y-x)} {\sinh (\pi x)} \right| 
\end{displaymath}
\begin{displaymath}
\le \quad K_2 e^{- c_2 |y|}
\end{displaymath}
with constants 
\begin{eqnarray*}
K_2 \quad &:=& \quad  \frac{x_0}{\pi} \quad \max ( K_{ L(\gamma) }), L_\gamma ) \quad
             e^{c_{L(\gamma) x_0}}     ,\\
c_2 \quad &:=& \quad   c_{L(\gamma)}.
\end{eqnarray*}

Recalling at this stage the additive decomposition of the Fourier transform of 
$f \gamma$, Eq. (\ref{additive decomposition of f gamma}), the estimate 
\begin{displaymath}
 | {\cal F} (f \gamma ) (y)| \quad \le \quad  K_{f\gamma} e^{- c_{f_\gamma} |y|}
\end{displaymath}
is obtained, with constants
\begin{eqnarray*}
K_{f \gamma} \quad &:=& \quad 1/2 K_\gamma + 2 K_1 + K_2  ,\\
c_{f \gamma} \quad &:=& \quad  \min ( c_\gamma, c_1, c_2 )  .
\end{eqnarray*}
Inserting this into Eq. (\ref{connection between F (alpha_l f_l) and F (f gamma) }),
an estimate for the Fourier transform of $\alpha_l^+$ by an exponential decay 
is obtained. A review of the two constants - prefactor and factor in the exponent -
shows that they depend continuously on the chemical potential $\mu_l$, so the 
constants can be chosen locally independent of $\mu_l$. This independence is 
useful for deriving diagrams with respect to chemical potentials.

\subsection*{Calculation of the Fourier back transform of $f_\lambda$:}  
Integration by parts yields:
\begin{displaymath}
 \left( {\cal F}^{-1} f_\lambda \right)( y) \quad  = \quad 
 \frac{-1}{2\pi (\lambda + iy) } F (\lambda + iy), 
\end{displaymath}
with  
\begin{displaymath}
 F(z):= \int dx \quad  e^{zx} f'(x). 
\end{displaymath}
The function $F(z)$ is holomorphic on the stripe 
$\left\lbrace Re \in ]-1, 1[ \right\rbrace$ within the complex plane. For purely 
imaginary arguments, its calculation is equivalent to the calculation of the 
Fourier transform of $f'$. This I let be performed by the software
{\em Mathematica} \cite{Mathematica} and obtained the equation
\begin{displaymath}
 F(z) = \frac{-i\pi z}{\sinh ( -i\pi z)},
\end{displaymath}
at first only for purely imaginary arguments. (I did not or not yet find a way 
to reproduce this result, nor do I have another reference.) However, the two 
functions are holomorphic on $\left\lbrace Re \in ]-1,1[ \right\rbrace$, and so 
the equality holds also on this larger set. Implicitly: 
\begin{displaymath}
 \left( {\cal F}^{-1} f_\lambda \right) (y) \quad  = \quad 
 \frac{i}{2} \quad \frac{1}{\sinh  (\pi(y - i\lambda))}.
\end{displaymath}

\textbf{Proof of the additional statement:}
Let 
\begin{displaymath}
F \quad := \quad {\cal F} \alpha_l, 
\end{displaymath}
and represent $\alpha_l$ as the Fourier back transform of $F$. Since $\alpha_l$ is 
bounded by a constant, it suffices to show that   
\begin{displaymath}
2 \pi \quad \alpha_l (x) \quad = \quad \int_\mathbb{R} dt \quad F(t) \quad e^{ixt}
\end{displaymath}
has an upper bound which decays like $1/|x|$. Assume $x \neq 0$ and write with the 
convergence theorem
\begin{displaymath}
 \left| \int_\mathbb{R} dt \quad F(t) \quad e^{ixt} \quad \right|  
\end{displaymath}
\begin{displaymath}
= \quad \left| \lim_{n \to \infty } \quad \int_\mathbb{R} dt 
        \quad F(t) \quad e^{ixt} \quad \textbf{1}_{\left[ - n \frac{2\pi}{|x|}  , 
        n \frac{ 2\pi }{|x|} \right] } (t) \right| 
\end{displaymath}
\begin{displaymath}
= \quad \left| \sum_{n=0}^\infty \quad \int_\mathbb{R} dt 
        \quad F(t) \quad e^{ixt} \quad \{ \textbf{1}_{\left[ - (n+1) \frac{2\pi}{|x|}  , 
        (n+1) \frac{ 2\pi }{|x|} \right] }   -
        \textbf{1}_{\left[ - n \frac{2\pi}{|x|}  , 
        n \frac{ 2\pi }{|x|} \right] } \} (t) \right|   
\end{displaymath}
\begin{displaymath}
\le \quad \sum_{n=0}^\infty \quad \left| \int_{-(n+1) 
       \frac{2\pi}{|x|}}^{-n \frac{2\pi}{|x|}}  dt \quad F(t) e^{ixt} \right| \quad +
\quad \left| \int_{n 
       \frac{2\pi}{|x|}}^{(n+1) \frac{2\pi}{|x|}}  dt \quad F(t) e^{ixt} \right|
\end{displaymath}
\begin{displaymath}
\le \quad \sum_{n=0}^\infty \quad \left| \int_{-(n+1) 
       \frac{2\pi}{|x|}}^{-(n+1) \frac{2\pi}{|x|} + \frac{\pi}{|x|}}  dt
       \left\lbrace 
        F(t) e^{ixt}  \quad + \quad F \left( t + \frac{\pi}{|x|} \right)   
       e^{ix \left(t + \frac{\pi}{|x|} \right) } \right\rbrace      
       \right|        
\end{displaymath}
\begin{displaymath}
\quad \quad \quad  + \quad \left| \int_{n \frac{2\pi}{|x|}}^{ n \frac{2\pi}{|x|} + 
       \frac{\pi}{|x|}}  dt
       \left\lbrace 
        F(t) e^{ixt}  \quad + \quad F \left( t + \frac{\pi}{|x|} \right)   
       e^{ix \left(t + \frac{\pi}{|x|} \right) } \right\rbrace      
       \right|        
\end{displaymath}

\begin{displaymath}
\le \quad \sum_{n=0}^\infty \quad  \int_{-(n+1) 
       \frac{2\pi}{|x|}}^{-(n+1) \frac{2\pi}{|x|} + \frac{\pi}{|x|}}  dt
       \quad \left|
        F(t)   \quad - \quad F \left( t + \frac{\pi}{|x|} \right)   
       \right|        
\end{displaymath}
\begin{displaymath}
\quad \quad \quad \quad  + \quad  \int_{n \frac{2\pi}{|x|}}^{ n \frac{2\pi}{|x|} + 
       \frac{\pi}{|x|}}  dt
       \quad \left|
        F(t)  \quad - \quad F \left( t + \frac{\pi}{|x|} \right)   
       \right|        
\end{displaymath}
\begin{displaymath}
\le \quad \sum_{n=0}^{\infty} \quad \frac{\pi^2}{x^2} \quad L (\alpha_l ) 
\left( n \frac{2 \pi}{|x|} \right) \quad 2
\end{displaymath}
\begin{displaymath}
= \quad \frac{2 \pi^2}{x^2} \left\lbrace L(\alpha_l) (0) \quad + \quad 
\sum_{n=1}^{\infty}   \quad L (\alpha_l ) 
\left( n \frac{2 \pi}{|x|} \right) \right\rbrace
\end{displaymath}
\begin{displaymath}
\le \quad \frac{2 \pi^2}{x^2} \left\lbrace L(\alpha_l) (0) \quad + \quad 
\int_0^\infty dt     \quad L (\alpha_l ) 
\left( t \frac{2 \pi}{|x|} \right) \right\rbrace
\end{displaymath}
\begin{displaymath}
\le \quad \frac{2 \pi^2}{x^2} \left\lbrace L(\alpha_l) (0) \quad + \quad 
\int_0^\infty dt     \quad K_{L (\alpha_l )} 
\exp \left[ -c_{L(\alpha_l)}\left( t \frac{2 \pi}{|x|} \right) \right] 
\right\rbrace
\end{displaymath}
\begin{displaymath}
= \quad 2 \pi^2 L(\alpha_l) (0) \quad \frac{1}{x^2} \quad + \quad 
\pi K_{L(\alpha_l)} c_{L(\alpha_l)}^{-1} \quad \frac{1}{|x|}.
\end{displaymath}

\section*{Appendix B: Lemma for deriving RTA and DSO rates with respect to the 
          coupling parameter}

Consider the function 
\begin{displaymath}
f(w) =  \int_\mathbb{R} d t \frac{ g(t) }{ \gamma^2 (t) + ( t/w + p(t) )^2 }
     = w\int_\mathbb{R} d x \frac{ g(wx) }{ \gamma^2 (wx) + ( x + p(wx) )^2 } ,
\end{displaymath}
where $g, \gamma, p$ are smooth and bounded real functions of the real variable $t$, 
$\gamma$ strictly positive. The argument $w$ of the function $f(w)$ be real and
positive. Define then $f_0 := f$ and for $n = 0,1,2, \dots $: 
\begin{displaymath}
f_{n+1}  (w) =  ( f_n (w) - f_n (0) ) /w,
\end{displaymath}  
as long as the limits $w \rightarrow 0$ exist. Then: $f_0 (0) = 0, f_1 (0 ) =  
\pi g(0)/ \gamma (0)$. Moreover, 
\textcolor{blue}{
\begin{eqnarray}
      f_2 (0) &=& \int_0^\infty \frac{dt}{t^2}  \left( g(t)  + g (-t) - 2 g(0) \right) 
      		\nonumber \\  && -  \pi \left(  \frac{gp}{\gamma}  \right)' (0),    
    \label{fourth_order_derivative}
\end{eqnarray}}
and
\textcolor{red}{
\begin{eqnarray}
      f_3 (0) &=& -2 \int_0^\infty \frac{dt}{t^3}  
      			\left( (gp)(t)  - (gp) (-t) - 2t (gp)'(0) \right)  \nonumber \\&&  
           + \frac{\pi}{2} \left(  \frac{ g (p^2  - \gamma^2 ) }{\gamma}  
           \right)'' (0). 
    \label{sixth_order_derivative}
\end{eqnarray}}
The derivatives of the fractions in $f_2(0) $ and $f_3(0)$ depend only on the behaviour 
of the functions $g, \gamma, p$ locally around zero. The integrals, on the other hand,
are in this sense non-local contributions. 

During the following proof the results for $f_2(0) $ and $f_3(0)$ will be obtained only 
as the sum of many contributions. I used the colours blue and red (grey) to mark those
contributions.

\subsection*{Proof:}
With the terminology
\begin{eqnarray*}
\left[ \dots \right]_1  &=& \left[ w^2 \gamma^2(0) + ( t + w p(0) )^2  \right],  \\
\left[ \dots \right]_2  &=& \left[ w^2 \gamma^2(t) + ( t + w p(t) )^2  \right]
\end{eqnarray*}
write 
\begin{displaymath}
f_2 (w) = \int dt \frac {g(t) \left[ \dots \right]_1  - g(0) \left[ \dots \right]_2 }
                       { \left[ \dots \right]_1 \left[ \dots \right]_2 }    
\end{displaymath}
\begin{equation}
        =  \int dt \frac{g(t) - g (0) }{\left[ \dots \right]_2}
        		+ g(0) \int \frac{ \left[ \dots \right]_1 - \left[ \dots \right]_2}
        			{\left[ \dots \right]_1  \left[ \dots \right]_2}.               
\label{group I and II}
\end{equation}
The function $f_2 (w)$ will be written as a sum of several contributions, 
the contributions are grouped in such a way that an overview is possible.

\subsection*{Group I}
The first summand on the right-hand side of Eq. (\ref{group I and II}).
With 
\begin{displaymath}
( \delta g ) (t) := (g (t) - g (0))/ t
\end{displaymath} 
and with the terminology
\begin{eqnarray*}
\left[ \dots \right]_a  &=& \left[ w^2 \gamma^2 (-t) + (-t + wp(-t))^2 \right] , \\
\left[ \dots \right]_b  &=& \left[ w^2 \gamma^2 (t) + (t + wp(t))^2\right]
\end{eqnarray*}
this contribution turns into 
\begin{eqnarray}
&& \int_0^\infty dt \quad t \frac{ \delta g (t) \left[ \dots \right]_a  
				  -     \delta g (-t) \left[ \dots \right]_b }
				{ \left[ \dots \right]_a  \left[ \dots \right]_b} \nonumber  \\  &=&
		\int_0^\infty dt \quad t \frac{ \delta g (t) - \delta g (-t) }
					{\left[ \dots \right]_b }   +
	 \int_0^\infty dt \quad t \delta g (-t) \frac{\left[ \dots \right]_a - 
	 	\left[ \dots \right]_b}{\left[ \dots \right]_a   \left[ \dots \right]_b} 
\label{subgroup I.1 and I.2}
\end{eqnarray}

\subsection*{Subgroup I.1}
The first additive contribution to the right-hand side of Eq. (\ref{subgroup I.1 and I.2}).
With 
\begin{displaymath}
(  r \delta g ) (t) := ( \delta g (t) -  \delta g (-t) )/ t
\end{displaymath} 
this contribution turns into
\begin{displaymath}
\int dt \frac{t^2}{w^2 \gamma ^2(t) + ( t + wp(t) )^2} r \delta g (t).
\end{displaymath}
The factor in front of $r \delta g (t)$ is bounded by a constant independent of $w$ and
$t$. Hence, the convergence theorem yields
\textcolor{blue}{
\begin{displaymath}
\int_0^\infty dt (r\delta g) (t) = \int_0^\infty \frac{dt}{t^2}  
                                             \left( g(t)  + g (-t) - 2 g(0) \right) 
\end{displaymath}}
as the non-local contribution to the right-hand side of Eq. (\ref{fourth_order_derivative}).

The corresponding contribution to $f_3(0) = \lim_{w \to 0} (f_2(w) - f_2(0) )/w $ is 
\begin{displaymath}
  \frac{1}{w} \int_0^\infty dt \quad r \delta g (t) \left[  \frac{t^2}
   {w^2 \gamma ^2 (t) + ( t + w p (t) )^2 }   - 1    \right]  
\end{displaymath}
\textcolor{red}{
\begin{displaymath} 
 \equiv  
 - (r \delta g) (0) ( \gamma ^2 + p^2 ) (0)  \int_0^\infty dx 
 \frac{1}{\gamma ^2 (0) + ( x + p(0) )^2 }  
 \end{displaymath}  }
 \begin{equation} 
  - \int_0^\infty dx  \frac{2x ((r \delta g)p) (wx) }{\gamma ^2 (0) + ( x + p(0) )^2 } ,
 \label{decomposition of subgroup I.1}
\end{equation} 
where I used the congruence 
\begin{displaymath}
 \int_0^\infty dx  \frac{2x ((r \delta g) p) (wx) }{\gamma ^2 (wx) + ( x + p(wx) )^2 } \equiv 
  \int_0^\infty dx  \frac{2x ((r \delta g) p) (wx) }{\gamma ^2 (0) + ( x + p(0) )^2 }
\end{displaymath}
modulo the limit $w \to 0$: The difference between the two functions of $w$ converges to 
zero as $w \to 0$. With the notation 
\begin{displaymath}
\delta p (t) := ( p (t) - p(0) )/ t,
\end{displaymath}
the second term on the right-hand side of Eq. (\ref{decomposition of subgroup I.1}) 
can be written as
\begin{equation}
 \frac{-2}{w} p(0) \int_0^\infty dx  \frac{ 1 }{\gamma ^2 (0) + ( x + p(0) )^2 }  
 \left( \delta g (wx) - \delta g (-wx ) \right)
\label{contribution with later partner}
\end{equation} 
\begin{displaymath}
-2  \int_0^\infty dt  \frac{ t^2 }{w^2 \gamma ^2 (0) + ( t + w p(0) )^2 } \quad
\left[ (r \delta g) (\delta p) \right] (t),
\end{displaymath} 
where the second line converges to 
\begin{displaymath}
-2  \int_0^\infty dt   \quad
\left[ (r \delta g) (\delta p) \right] (t)
\end{displaymath}
\textcolor{red}{
\begin{displaymath}
= -2 \int_0^\infty \frac{dt}{t^3} \quad ( p(t) - p(0) )\quad ( g(t) + g (-t) -2 g (0) )
\end{displaymath} }
with Lebesgue. The possibly divergent term (\ref{contribution with later partner})
will later be added to another possibly divergent term; their sum is not divergent.

\subsection*{Subgroup I.2}
The second summand on the right-hand side of Eq. (\ref{subgroup I.1 and I.2}). 

\subsection*{Subgroup I.2.a}
The latter can be decomposed into two qualitatively different contributions to $f_2(w)$, 
subgroup I.2.a be defined by the first one of these contributions: 
\begin{displaymath}
- \int_0^\infty dt \quad t \delta g (-t) \frac{w^2 \left(
      (\gamma^2 + p^2 ) (t) - (\gamma^2 + p^2 ) (-t)  \right)}
      {\left[ \dots \right]_a   \left[ \dots \right]_b}
\end{displaymath}
\begin{displaymath}
= - \int_0^\infty dt \quad t^2 \delta g (-t) \frac{w^2 ( r [\gamma^2 + p^2 ] ) (t) }
      {\left[ \dots \right]_a   \left[ \dots \right]_b}
\end{displaymath}
\begin{displaymath}
=  -w \int_0^\infty dx \frac{\delta g (-wx) ( r [\gamma^2 + p^2 ] ) (wx)  x^2}
     {\left[ \gamma ^2 ( -wx) + ( -x + p(-wx) )^2  \right]
      \left[ \gamma ^2 (  wx) + (  x + p( wx) )^2  \right]} 
\end{displaymath}
\textcolor{blue}{
\begin{displaymath}
 \quad \to 0 \quad ( w \to 0).
\end{displaymath}}

The corresponding contribution to $\lim_{w \to 0} f_3(w) = 
\lim_{w \to 0} (f_2 (w) - f_2(0))/w $ is 
\textcolor{red}{
\begin{displaymath}
- \delta g (0) ( r [\gamma^2 + p^2 ] ) (0) \int_0^\infty dx \frac{ x^2}
     {\left[ \gamma ^2 ( 0 ) + ( -x + p(0) )^2  \right]
      \left[ \gamma ^2 ( 0) + (  x + p(0) )^2  \right] }. 
\end{displaymath}}
The final integral is obtained from the residue calculus, a list of the values
of the appearing integrals of this kind is found at the end of this appendix. 
Moreover, 
\begin{displaymath}
\delta g (0) = g'(0) , \quad r [ \gamma^2 + p^2 ] (0) = 
2 ( \gamma ^2 + p^2 )' (0).
\end{displaymath}

\subsection*{Subgroup I.2.b}
As noted above, the second summand on the right-hand side of Eq. (\ref{subgroup I.1 and I.2}) 
can be decomposed into two qualitatively different contributions to $f_2(w)$,
let the subgroup I.2.b be defined by the second one of these contributions. Using the
terminology 
\begin{displaymath}
\eta ( t) := p (-t) + p (t),
\end{displaymath}
this second contribution reads:
\begin{displaymath}
\int_0^\infty dt \quad t \delta g (-t) \frac{-2tw \eta (t) }
                                {\left[ \dots \right]_a   \left[ \dots \right]_b}
\end{displaymath}
\begin{displaymath}
= \int_0^\infty dx \frac{-2x^2 \delta g ( -wx) \eta (wx) }
{ \left[ \gamma^2 (-wx) + ( -x + p(-wx) )^2   \right]
  \left[ \gamma^2 (wx) + ( x + p(wx) )^2   \right]} 
\end{displaymath}
\textcolor{blue}{
\begin{displaymath}
\equiv -2 \eta (0) \delta g (0) \int_0^\infty dx \frac{ x^2  }
{ \left[ \gamma^2 (0) + ( -x + p(0) )^2   \right]
  \left[ \gamma^2 (0) + ( x + p(0) )^2   \right]} .
\end{displaymath}}

In order to determine the corresponding contribution to $\lim_{w \to 0} f_3(w) = 
\lim_{w \to 0} (f_2 (w) - f_2(0))/w $ use the abbreviations
\begin{eqnarray*}
\left[ \dots \right]_{v,1} &=& \gamma ^2 (-wx) + ( -x + p (-wx) )^2, \\
\left[ \dots \right]_{v,2} &=& \gamma ^2 (wx) + ( x + p (wx) )^2, \\
\left[ \dots \right]_{c,1} &=& \gamma ^2 (0) + ( -x + p (0) )^2, \\
\left[ \dots \right]_{c,2} &=& \gamma ^2 (0) + ( x +  p(0) )^2. 
\end{eqnarray*}
Consider 
\begin{displaymath}
\frac{1}{w} \left\lbrace  \int_0^\infty dx \frac{-2x^2 \delta g ( -wx) \eta (wx) }
{ \left[  \dots  \right]_{v,1} \left[ \dots   \right]_{v,2} }  - 
\int_0^\infty dx \frac{-2x^2 \delta g ( 0) \eta (0) }
{ \left[ \dots   \right]_{c,1} \left[ \dots  \right]_{c,2} }   \right\rbrace
\end{displaymath}
\begin{displaymath}
= \frac{1}{w} \int_0^\infty dx \frac{-2 x^2  
\left[ \delta g ( -wx ) \eta (wx) - \delta g (0) \eta (0) \right]  }
  { \left[  \dots  \right]_{v,1} \left[ \dots   \right]_{v,2}} 
\end{displaymath}
\begin{equation}
+  \frac{\delta g (0) \eta (0) }{w}   \int_0^\infty dx \quad -2 x^2
\frac{\left[  \dots  \right]_{c,1} \left[  \dots   \right]_{c,2}
  - \left[  \dots  \right]_{v,1} \left[ \dots   \right]_{v,2}}
{\left[  \dots  \right]_{c,1} \left[  \dots   \right]_{c,2}
  \left[  \dots  \right]_{v,1} \left[ \dots   \right]_{v,2}}.
  \label{decomposition of subgroup I.2.b}
\end{equation}

\subsection*{Subgroup I.2.b.i}
The group of contributions to $\lim_{w \to 0} f_3 (w)$ originating from the 
the first summand on the right-hand side of Eq. (\ref{decomposition of 
subgroup I.2.b}). It is 
\begin{displaymath}
\frac{1}{w} \int_0^\infty dx \frac{-2 x^2  
\left[ \delta g ( -wx ) \eta (wx) - \delta g (0) \eta (0) \right]  }
  { \left[  \dots  \right]_{v,1} \left[ \dots   \right]_{v,2}}
\end{displaymath}
\begin{displaymath}
\equiv  \frac{1}{w} \int_0^\infty dx \frac{-2 x^2  
\left[ \delta g ( -wx ) \eta (wx) - \delta g (0) \eta (0) \right]  }
  { \left[  \dots  \right]_{c,1} \left[ \dots   \right]_{c,2}}
\end{displaymath}
\begin{displaymath}
= \frac{-2}{w} \int_0^\infty dx \frac{ x^2} { \left[  \dots  \right]_{c,1} 
   \left[ \dots   \right]_{c,2}}  ( \quad \left[ \eta (wx) - \eta (0) \right] 
   \left[ \delta g ( -wx) - \delta g (0) \right]  
\end{displaymath}
\begin{displaymath}
 \quad + \quad \eta (0) \left[ \delta g ( -wx) - \delta g (0) \right]
\end{displaymath}
\begin{equation}
 \quad + \quad  \delta g (0) \left[ \eta  ( wx) - \eta  (0) \right] \quad ).
\label{decomposition of subgroup I.2.b.i}
\end{equation}
The first line of the right-hand side of Eq. (\ref{decomposition of subgroup I.2.b.i})
yields the contribution 
\textcolor{red}{
\begin{displaymath}
 -2  \int_0^\infty dt \quad \frac{\eta (t) - \eta (0) }{t} \quad   
 \frac{\delta g (-t) - \delta g (0) }{t} 
\end{displaymath}}
to $f_3(0)$.
The second summand (second line) of the right-hand side of Eq. 
(\ref{decomposition of subgroup I.2.b.i}) can, upon adding and subtracting
\begin{displaymath}
p^2 (0) + \gamma ^2 (0) - 2x p(0), 
\end{displaymath}
in the numerator of the fraction, and by the use of the function
\begin{displaymath}
(\delta \delta g ) (t) := (\delta g (t) - \delta g (0) )/t,
\end{displaymath} 
be rewritten as
\begin{equation}
\frac{-4}{w} p (0) \int_0^\infty dx \quad 
      \frac{1}{\gamma ^2 (0) + ( x + p(0) )^2 } 
       \quad ( \delta g (-wx) - \delta g (0) )
\label{the later contribution}
\end{equation} 
\textcolor{red}{                                              
\begin{displaymath}                                             
-2 ( \gamma ^2 + p^2 ) (0)  (  \delta g )' (0) \quad \int_0^\infty dx 
\frac{ 2 p(0)x }{\left[  \gamma ^2 (0) + ( -x + p (0) )^2  \right] 
 \left[ \gamma ^2 (0) + ( x + p (0) )^2   \right] }
\end{displaymath}
\begin{displaymath}
+ 4 p  (0) \eta (0) ( \delta g )' (0) \quad \int_0^\infty dx 
\frac{x^2}{\left[  \gamma ^2 (0) + ( -x + p (0) )^2  \right] 
 \left[ \gamma ^2 (0) + ( x + p (0) )^2   \right] }.
\end{displaymath}}
(Note: $\delta \delta g (0) = ( \delta g ) ' (0) = 1/2 g '' (0)$.)
The sum of the present term (\ref{the later contribution}) and the earlier 
contribution (\ref{contribution with later partner}) is 
\begin{displaymath}
\frac{-2}{w} p(0) \int_0^\infty dx  \frac{ 1 }{\gamma ^2 (0) + ( x + p(0) )^2 }  
 \left( \delta g (wx) + \delta g (-wx ) - 2 \delta g (0) \right ).
\end{displaymath}
The function $ \delta g ( t) + \delta g (-t) - 2 \delta g (0)$ vanishes 
quadratically in $t = 0$. Multiply and divide through $t^2$, then 
integrate with respect to $t = wx$ instead of $x$. The convergence theorem 
yields the contribution
\textcolor{red}{
\begin{displaymath}
-2 p(0) \int_0^\infty dt \frac{\delta g (t) + \delta g (-t) - 2 \delta g (0) }
               { t^2 }
\end{displaymath} }
to $f_3 (0)$. The treatment of the third term (third line) on the 
right-hand side of Eq. (\ref{decomposition of subgroup I.2.b.i}) is analogous: 
$\eta (t) = p (t) + p (-t)$ is even, hence $\eta (t) - \eta (0)$ 
vanishes even quadratically in $t=0$. Multiply and divide through $t^2$, 
then integrate with respect to $t = wx$ instead of $x$. The convergence theorem
yields the contribution
\textcolor{red}{
\begin{displaymath}
-2 \delta g (0) \int_0^\infty dt \frac{p (t) + p(-t) - 2 p (0) }{t^2 }
\end{displaymath}}
to $f_3 (0)$.

\subsection*{Subgroup I.2.b.ii}
The group of contributions to $\lim_{w \to 0} f_3 (w)$ originating from the 
the second summand on the right-hand side of Eq. (\ref{decomposition of 
subgroup I.2.b}). It is
\begin{displaymath}
  -2 \delta g (0) \eta (0)    \int_0^\infty dx \quad  
\frac{x^3} {\left[  \dots  \right]_{c,1} \left[  \dots   \right]_{c,2}
  \left[  \dots  \right]_{v,1} \left[ \dots   \right]_{v,2}}  \quad  [
\end{displaymath}
\begin{displaymath}
\frac{(\gamma ^2 + p^2 )^2 (0) - ( \gamma ^2 + p^2 ) (-wx) 
                ( \gamma ^2 + p^2 ) (wx)}{wx}
\end{displaymath}
\begin{displaymath}
 + \quad x^2 \quad \frac{ 2 (\gamma ^2 + p^2 ) (0) - ( \gamma ^2 + p^2 ) ( wx) 
               - ( \gamma ^2 + p^2 ) (- wx)}{wx}
\end{displaymath}
\begin{displaymath}
- \quad 4 x^2 \quad  \frac{  p^2  (0) -  p (wx) p(-wx) }  {wx}
\end{displaymath}
\begin{displaymath}
 + \quad 2 x  \quad \frac{ p (-wx) (\gamma ^2 + p^2 ) (wx)  - 
    p (wx) ( \gamma ^2 + p^2 ) ( -wx) }{wx}
\end{displaymath}
\begin{displaymath}
 - \quad 2 x^3  \quad \frac{ p (wx) - p (-wx) }{wx} \quad  ] .
\end{displaymath}
Of these altogether five contributions to $\lim_{w \to 0}f_3 (w)$,
the first three vanish: The functions
of $wx$ which are found in the numerators of the fractions are even
and vanish quadratically in $wx = 0$. Again, the convergence theorem
can be applied. For the treatment of the fourth and fifth of the above 
contributions define
\begin{displaymath}
Q (t) := \frac{p (-t) (\gamma ^2 + p^2 ) (t)  - 
                                 p (t) ( \gamma ^2 + p^2 ) ( -t)}{t}
\end{displaymath}
and 
\begin{displaymath}
r p (t) :=  \frac{ p (t) - p (- t) }{t}
\end{displaymath}
to obtain
\textcolor{red}{
\begin{displaymath}
 - \quad 2 \quad Q(0) \delta g (0) \eta (0)    \int_\mathbb{R} dx \quad  
\frac{x^4} {\left[  \gamma ^2 (0) + ( -x + p (0) )^2 \right]^2 
\left[  \gamma ^2 (0) + ( x + p (0) )^2   \right]^2 }
\end{displaymath} }
and 
\textcolor{red}{
\begin{displaymath}
 + \quad  2 \quad rp (0) \delta g (0) \eta (0)    \int_\mathbb{R} dx \quad  
\frac{x^6} {\left[  \gamma ^2 (0) + ( -x + p (0) )^2  \right]^2 
\left[  \gamma ^2 (0) + ( x + p (0) )^2   \right]^2 },
\end{displaymath} }
respectively. Note: $ \delta g (0) = g'(0), rp (0) = 2 p'(0),$ and 
\begin{displaymath}
Q(0) = -2 ( \gamma ^2 + p^2 ) (0) p' (0) + 2 p(0) ( \gamma ^2 + p^2 )' (0). 
\end{displaymath}

\subsection*{Group II}
Let this be the second summand on the right-hand side of Eq. 
(\ref{group I and II}), which was a sum of contributions to $f_2(w)$. 
This second summand is the sum of two qualitatively different 
contributions: 
\begin{displaymath}
g(0) \int_\mathbb{R} dt \frac{w^2 \left[ ( \gamma ^2 + p^2 ) (0) - 
    ( \gamma ^2 + p^2 ) (t) \right] } 
     { [\dots ]_1 [ \dots ]_2 }   
\end{displaymath}
\begin{equation}
+ \quad g(0) \int_\mathbb{R} dt \frac{2tw \left[ p  (0) -  p (t) \right] }
      { [\dots ]_1 [ \dots ]_2 },   
\label{decomposition of group II}
\end{equation}
where 
\begin{eqnarray*}
\left[ \dots \right]_1  &=& \left[ w^2 \gamma^2(0) + ( t + w p(0) )^2  \right],  \\
\left[ \dots \right]_2  &=& \left[ w^2 \gamma^2(t) + ( t + w p(t) )^2  \right].
\end{eqnarray*}

\subsection*{Subgroup II.1}
The first contribution in the sum (\ref{decomposition of group II}). By the use 
of the function
\begin{displaymath}
\delta ( \gamma ^2 + p^2 ) (t) := 
       [ ( \gamma ^2 + p^2 ) (t) - ( \gamma ^2 + p^2 ) (0) ]/t
\end{displaymath}
this first contribution turns into
\begin{displaymath}
- g(0) \int dx \frac{x \delta ( \gamma ^2 + p^2 ) (wx) }
         {  [\gamma ^2 (0) + (x + p(0)) ^2] [ \gamma ^2 (wx) + (x + p (wx))^2  ]  }
\end{displaymath}
\textcolor{blue}{
\begin{displaymath}
\equiv - g(0) \delta ( \gamma ^2 + p^2 ) (0) \int_\mathbb{R} dx \frac{x  }
     {  [\gamma ^2 (0) + (x + p(0)) ^2]^2  }.
\end{displaymath}}
In order to determine the corresponding contribution to $\lim_{w \to 0} f_3 (w) $,
use the abbreviations
\begin{eqnarray*}
\left[ \dots \right]_{c} &=& \gamma ^2 (0) + ( x + p(0))^2, \\
\left[ \dots \right]_{v} &=& \gamma ^2 (wx) + ( x + p(wx))^2,
\end{eqnarray*}
and consider
\begin{displaymath}
 \frac{g(0)}{w} \int dx \frac{   x   }{[ \dots ]_{c}^2 [ \dots ]_{v}} \quad 
  [ \quad \delta ( \gamma ^2 + p^2 ) (0) [ \gamma ^2 (wx) + (x + p (wx) )^2 ]
\end{displaymath}  
\begin{displaymath}
\quad \quad -  \delta (\gamma ^2 + p^2 ) (wx) [ \gamma ^2 (0) 
                + ( x + p(0) )^2 ]  \quad ]
\end{displaymath}
\begin{displaymath}
 = g (0) \int dx \frac{x}{ [ \dots]_{c}^2  [ \dots]_{v} } \quad [
\end{displaymath}
\begin{displaymath}
\quad x \quad \left\lbrace \delta ( \gamma ^2 + p^2 ) (0) \delta (\gamma ^2 + p^2 ) (wx)
\quad - \quad ( \gamma ^2 + p^2 ) (0) \delta \delta ( \gamma ^2 + p^2 ) (wx) \right\rbrace  
\end{displaymath}
\begin{displaymath}
+ \quad 2 x^2 \quad  \left\lbrace  \delta (\gamma ^2 + p^2 ) (0) \delta p ( wx) 
 \quad - \quad p (0) \delta \delta ( \gamma ^2 + p^2 )  (wx) \right\rbrace
\end{displaymath}
\begin{displaymath}
- \quad x^3 \quad  \delta \delta ( \gamma ^2 + p^2 ) ( wx) \quad ]
\end{displaymath}

\textcolor{red}{
\begin{displaymath}
\equiv \left\lbrace g(0) \delta ( \gamma ^2 + p^2 ) (0)^2  \quad - \quad  
       g(0) (\gamma ^2 + p^2 ) (0)   \delta \delta ( \gamma ^2 + p^2 ) (0) 
       \right\rbrace  
\end{displaymath}
\begin{displaymath}
\quad \quad \int_\mathbb{R} dx \frac{x^2}
   { [ \gamma ^2 (0) + ( x + p(0))^2 ]^3 }
\end{displaymath}
\begin{displaymath}
+ \quad \left\lbrace 2 g(0) \delta (\gamma ^2 + p^2 ) (0) \delta p (0) \quad  
- \quad   2 g (0) p (0) \delta \delta ( \gamma ^2 + p ^2 ) (0) \right\rbrace 
\end{displaymath}
\begin{displaymath}
\quad \quad \int_\mathbb{R} dx \frac{x^3}
   { [ \gamma ^2 (0) + ( x + p(0))^2 ]^3 }
\end{displaymath}
\begin{displaymath}
 -  \quad g(0) \delta \delta ( \gamma ^2 + p ^2 ) (0) \quad \int_\mathbb{R} 
 dx \frac{x^4}{[ \gamma ^2 (0) + ( x + p(0))^2 ]^3 } .
\end{displaymath} }

\subsection*{Subgroup II.2}
The second contribution in the sum (\ref{decomposition of group II}) of 
contributions to $f_2 (w)$. It is
\begin{displaymath}
-2 g(0) \int_\mathbb{R} dx \frac{x^2 \delta p (wx)}{[\gamma ^2 (0) + ( x + p(0))^2]
[\gamma ^2 (wx) + ( x + p(wx))^2]}
\end{displaymath}
\textcolor{blue}{
\begin{displaymath}
\equiv -2 g (0) \delta p (0) \int_\mathbb{R} dx \frac{x^2 }
  {[\gamma ^2 (0) + ( x + p(0))^2]^2}.
\end{displaymath} }
The corresponding contribution to $ f_3 (w)$ is
\begin{displaymath}
\int_\mathbb{R} dx \frac{x^2}{\left[ \dots \right]_c^2 \left[ \dots \right]_v}
\quad [
\end{displaymath} 
\begin{displaymath}
x \quad \left\lbrace 2 g(0) \delta p (0) \delta (\gamma ^2 + p^2)  (wx) \quad 
   -  \quad 2 g(0) \delta \delta p (wx) (\gamma ^2 + p^2 ) (0)     \right\rbrace
\end{displaymath}
\begin{displaymath}
x^2 \quad \left\lbrace 4 g(0) \delta p (0) \delta p (wx) \quad 
   -  \quad 4 g(0) p(0)  \delta \delta p (wx)   \right\rbrace   
\end{displaymath}
\begin{displaymath}
- x^3 \quad \left\lbrace 2 g(0) \delta \delta p (wx)   \right\rbrace \quad  ]   
\end{displaymath}
\textcolor{red}{
\begin{displaymath}
\equiv \left\lbrace 2 g(0) \delta p (0) \delta (\gamma ^2 + p^2)  (0) \quad 
   -  \quad 2 g(0) \delta \delta p (0) (\gamma ^2 + p^2 ) (0)     \right\rbrace
\end{displaymath}
\begin{displaymath}
\quad \quad \quad \quad \int_\mathbb{R} dx \frac{x^3}
         { \left[ \gamma ^2 (0) + ( x + p(0))^2   \right]^3}
\end{displaymath}
\begin{displaymath}
 + \quad \left\lbrace 4 g(0) \delta p (0) \delta p (0) \quad 
   -  \quad 4 g(0) p(0)  \delta \delta p (0)   \right\rbrace   
\end{displaymath}
\begin{displaymath}
\quad \quad \quad \quad \int_\mathbb{R} dx \frac{x^4}
         { \left[ \gamma ^2 (0) + ( x + p(0))^2   \right]^3}
\end{displaymath} }
\begin{displaymath}
-\quad 2 g(0) \int_\mathbb{R} dx \frac{x^5}
         { \left[ \gamma ^2 (0) + ( x + p(0))^2   \right]^3}
         \delta \delta p (wx) .
\end{displaymath}

Write the contribution of the last line as 
\begin{displaymath}
-2g (0) \int_0^\infty dx \quad \frac{x^5}
      {\left[ \gamma ^2 (0) + ( x + p(0))^2   \right]^3} \quad 
     [ \delta \delta p (wx) - \delta \delta p (-wx)   ]
\end{displaymath}
\begin{displaymath}
-2g (0) \int_0^\infty dx \quad \left\lbrace \frac{1}
      {\left[ \gamma ^2 (0) + ( x + p(0))^2   \right]^3} -  
      \frac{1}  {\left[ \gamma ^2 (0) + ( x - p(0))^2   \right]^3} 
      \right\rbrace  
\end{displaymath}
\begin{displaymath}
 \quad \quad \quad \quad \quad \quad   x^5 \delta \delta p (-wx) \quad \quad   =
\end{displaymath}
\textcolor{red}{
\begin{displaymath}
  - \quad 2 g(0) \quad \int_0^\infty dt \quad \frac{p (t) - p (-t) - 2 p'(0) t}
	{t^3}
\end{displaymath}
\begin{displaymath}
+  \quad 2 g(0) p''(0) \quad \left\lbrace 4 p^3 (0) \quad + \quad 
                                              6 p(0) (\gamma ^2 + p^2 )(0) \right\rbrace
\end{displaymath}
\begin{displaymath}
\quad \quad \quad \quad \int_\mathbb{R} dx \quad \frac{x^8}{\left[ \gamma ^2 (0) + 
         ( x + p(0))^2   \right]^3 \left[ \gamma ^2 (0) + ( x - p(0))^2   \right]^3}
\end{displaymath}
\begin{displaymath}
+ \quad 6 g(0) p''(0) p(0) 
\end{displaymath}
\begin{displaymath}
\quad \quad \quad \quad \int_\mathbb{R} dx \quad \frac{x^{10}}{\left[ \gamma ^2 (0) + 
         ( x + p(0))^2   \right]^3 \left[ \gamma ^2 (0) + ( x - p(0))^2   \right]^3}
\end{displaymath}
\begin{displaymath}
+ \quad 6 g(0) p''(0) p(0) ( \gamma ^2 + p^2 ) ^2 (0) 
\end{displaymath}
\begin{displaymath}
\quad \quad \quad \quad \int_\mathbb{R} dx \quad \frac{x^{6}}{\left[ \gamma ^2 (0) + 
         ( x + p(0))^2   \right]^3 \left[ \gamma ^2 (0) + ( x - p(0))^2   \right]^3} ,
\end{displaymath}
}
with which all contributions to $f_2 (0) $ and to $f_3 (0)$ are determined.

\newpage

\subsection*{List of required integrals}
With the definitions
\begin{displaymath}
I_n^{(k)} := \int_\mathbb{R} dx \quad \frac{x^k}{\left[ \gamma^2(0) + ( x - p (0) )^2 
   \right]^n \left[ \gamma^2(0) + ( x + p (0) )^2 \right]^n},
\end{displaymath}
\begin{displaymath}
J_n^{(k)} := \int_\mathbb{R} dx \quad \frac{x^k}{  [ \gamma^2(0) + ( x + p (0) )^2 ]^n},
\end{displaymath}
the following equations hold: 
\begin{eqnarray*}
I_1^{(0)} &=& \quad \pi /2 \quad \frac{1}{\gamma (0)} \quad \frac{1}{( \gamma ^2 + p^2 ) (0) } , \\
I_1^{(2)} &=& \quad \pi /2 \quad \frac{1}{\gamma (0)} , 
\end{eqnarray*}
\begin{eqnarray*}
I_2^{(4)} &=& \quad \pi  \quad \frac{1}{\gamma (0) ^3} \quad \frac{1}{ 2^4 } , \\
I_2^{(6)} &=& \quad \pi  \quad \frac{1}{\gamma (0) ^3} \quad \frac{1}{ 2^4 } \quad 
               ( 5 \gamma ^2  + p^2 ) (0) . 
\end{eqnarray*}
Furthermore, with the abbreviation $\quad c := \quad p(0) /\gamma (0)$:
\begin{eqnarray*}
J_2^{(1)} &=& \quad -\pi /2  \quad \frac{1}{\gamma (0) ^2} \quad c , \\
J_2^{(2)} &=& \quad \pi /2   \quad \frac{1}{\gamma (0) } \quad ( 1 + c^2 ) , 
\end{eqnarray*}
\begin{eqnarray*}
J_3^{(2)} &=& \quad \pi /8  \quad \frac{1}{\gamma (0) ^3} \quad ( 1 + 3 c^2 ) , \\
J_3^{(3)} &=& \quad -3 \pi /8   \quad \frac{1}{\gamma (0) ^2 } \quad ( c + c^3 ) , \\
J_3^{(4)} &=& \quad 3 \pi /8   \quad \frac{1}{\gamma (0) } \quad ( 1 + 2 c^2 + c^4 ) .
\end{eqnarray*}

Finally, 
\begin{eqnarray*}
I_3^{(6)} &=& \quad 3 \pi /2^8  \quad \frac{1}{\gamma (0) ^5}  , \\
I_3^{(8)} &=& \quad  \pi /2^8   \quad \frac{1}{\gamma (0) ^3 } \quad ( 7  + 3 c^2 ) , \\
I_2^{(6)} &=& \quad \pi / 2^4   \quad \frac{1}{\gamma (0) } \quad ( 5 + c^2  ) ,
\end{eqnarray*}
and 
\begin{displaymath}
( \gamma ^2 + p^2 )^2 (0) \quad I_3^{(6)} \quad + \quad I_3^{(10)} \quad = \quad 
I_2^{(6)} \quad -  \quad 2 ( \gamma ^2 - p^2 ) (0) \quad I_3^{(8)}.
\end{displaymath}

\end{document}